\documentclass[twocolumn,trackchanges]{aastex701}

\usepackage{color}
\usepackage{amsmath}
\usepackage{CJK}

\newcommand{\OI}{\makebox{[O\,{\sc I}]}}

\newcommand{\HeI}{He\,{\sc i}}
\newcommand{\HeII}{He\,{\sc ii}}

\newcommand{\hbeta}{H{$\beta$}}
\newcommand{\halpha}{H{$\alpha$}}

\newcommand{\pbeta}{Pa$\beta$}

\newcommand{\OII}{[O{\sevenrm\,II}]}
\def \OIII {[O\,{\sc iii}]}

\newcommand{\OIIIab}{[O{\sevenrm\,III}]\,$\lambda\lambda$4959,5007}

\newcommand{\NII}{[N{\sevenrm\,II}]}

\newcommand{\NIIab}{[N{\sevenrm\,II}]\,$\lambda\lambda$6548,6584}
\newcommand{\SII}{[S{\sevenrm\,II}]}

\newcommand{\SIIab}{[S{\sevenrm\,II}]\,$\lambda\lambda$6717,6731}
\newcommand{\NeIII}{[Ne{\sevenrm\,III}]}

   \font\sevenrm=cmr7 scaled 1000

\newcommand{\SIII}{[S{\sevenrm\, III}]}
\newcommand{\FeII}{Fe{\sevenrm\, II}}

\newcommand{\mbh}{$M_{\rm BH}$}

\usepackage{soul}
\usepackage{changepage}
\usepackage{multirow}

\begin{document}

\title{NEXUS: Spectral Variability of Little Red Dots and Blue Active Galactic Nuclei at $2 \lesssim z \lesssim 6$}

\author[0000-0002-8501-3518]{Zachary Stone}
\affiliation{Department of Astronomy, University of Illinois at Urbana-Champaign, Urbana, IL 61801, USA}
\email[show]{stone28@illinois.edu}  

\author[0000-0003-1659-7035]{Yue Shen}
\affiliation{Department of Astronomy, University of Illinois at Urbana-Champaign, Urbana, IL 61801, USA}
\affiliation{National Center for Supercomputing Applications, University of Illinois at Urbana-Champaign, Urbana, IL 61801, USA}
\email{shenyue@illinois.edu}

\author[0000-0001-5105-2837]{Ming-Yang Zhuang}
\affiliation{Department of Astronomy, University of Illinois at Urbana-Champaign, Urbana, IL 61801, USA}
\email{mingyang@illinois.edu}

\author[0000-0002-1605-915X]{Junyao Li}
\affiliation{Department of Astronomy, University of Illinois at Urbana-Champaign, Urbana, IL 61801, USA}
\email{}

\author[0000-0003-0230-6436]{Zhiwei Pan}
\affiliation{Department of Astronomy, University of Illinois at Urbana-Champaign, Urbana, IL 61801, USA}
\email{}

\author[0000-0002-5612-3427]{Jenny E. Greene}
\affiliation{Department of Astrophysical Sciences, 4 Ivy Lane, Princeton University, Princeton, NJ 08540}
\email{}

\author[0000-0002-7633-431X]{Feige Wang}
\affiliation{Department of Astronomy, University of Michigan, 1085 S University, Ann Arbor, MI 48109, USA}
\email{fgwang@umich.edu}


\begin{abstract}
We present spectral measurements for 17 Little Red Dots (LRDs) and 14 blue broad-line active galactic nuclei (AGNs) at $2\lesssim z \lesssim 6$ using multi-epoch JWST NIRSpec MSA spectra from the NEXUS program, sampling rest-frame timescales of $\sim 1-3$~months. Overall, the LRD population shows significantly enhanced Balmer decrement compared with both blue JWST AGNs at similar redshifts and 56 low-redshift broad-line AGNs matched in \halpha\ luminosity. The rest-optical continua of LRDs show little ensemble variability (rms~$\lesssim 3\%$), and the total \halpha\ emission also shows weaker ensemble variability compared with low-redshift AGNs matched in \halpha\ luminosity and rest-frame timescales. Based on the flux uncertainties, we constrain the intrinsic \halpha\ rms variability to be $\lesssim 4\%$ for the LRD population over these timescales. Combining our results with recent broad-line variability measurements of LRDs over yearly to decade timescales reveals a low-level white-noise pattern across all timescales, in stark contrast to the variability amplitude ($\sim 6\%$ over monthly timescales) and red-noise pattern observed in normal AGNs. These results add to the growing observational studies that suggest population-wise, LRDs have weak variability both in optical continuum and broad-line emission. Furthermore, the distinct white-noise broad-line variability pattern suggests different production mechanisms of broad-line emission in LRDs as opposed to normal AGNs, and/or different properties of the driving ionizing flux from the central engine.    
\end{abstract}



\section{Introduction} 
A key observational characteristic of active galactic nuclei (AGNs) is multi-wavelength and multi-timescale variability. Studies of low-redshift AGNs suggest that rest-frame UV/optical continuum variability traces the variability of a central accretion disk surrounding the supermassive black hole (SMBH). Continuum emission from the accretion disk in the UV/optical will reverberate to the broad emission lines in the broad-line region \citep[BLR;][]{BlandfordMcKee1982a, Peterson2001a}. Joint analyses of variability from both the continuum and the response of broad emission lines thus probe the inner regions of AGNs near the SMBH, providing estimates of AGN properties such as the BLR size and SMBH mass \citep[e.g.,][]{Peterson2001a}.

Over the past two decades, studies of photometric and spectroscopic variability of AGNs have mostly been limited to the $z\lesssim 2$ Universe. Characteristic timescales and amplitudes of variability have been obtained by fitting multi-band continuum light curves to continuous auto-regressive moving average (CARMA) models, which have been shown to describe AGN UV/optical variability well \citep{KellyEtAl2014}. Such metrics of variability have also been shown to correlate with AGN properties, including \mbh, Eddington ratio $\lambda_{\rm Edd}$, and luminosity \citep[e.g.,][]{KellyEtAl2009, MacLeodEtAl2010, BurkeEtAl2021, YuEtAl2025}. Hence, variability presents a unique way to probe properties of AGNs.

A major focus of recent JWST observations has been high-redshift broad-line AGNs \citep{FurtakEtAl2023, MaiolinoEtAl2024, UblerEtAl2023, CarnallEtAl2023, GreeneEtAl2024, HarikaneEtAl2023, KocevskiEtAl2023a, KokorevEtAl2023, LarsonEtAl2023, BarroEtAl2024, MattheeEtAl2024}, as most are identified via broad \halpha\, emission, shifted into the near-infrared at $z \gtrsim 2$. Meanwhile, the variability of low-luminosity AGNs at high redshifts (e.g., $z \gtrsim 2$) is relatively unexplored. Large multi-epoch spectro-photometric reverberation mapping (RM) surveys have recently allowed for estimates of rest-UV photometric and spectroscopic variability in a number of sources out to $z\sim3$, probing rest-frame timescales $\lesssim$~years \citep[e.g.,][]{ShenEtAl2015,ShenEtAl2019}. Correlated photometric and spectroscopic variability in low-$z$ AGNs have allowed for \mbh\, measurements \citep[uncertainty $\sim$0.3~dex;][]{ShenEtAl2024_rm}, and tight correlations between AGN parameters \citep[e.g., broad-line region radius and luminosity; ][]{BentzEtAl2013}. However, the evolution of these relations towards high-$z$ is unknown, and may depend on physical properties of the system \citep{DuEtAl2016}.

JWST has revealed an exciting new population of high-redshift ($z \gtrsim 4$) broad-line objects dubbed Little Red Dots \citep[LRDs;][]{MattheeEtAl2024}. LRDs are compact, red (in rest-optical) sources displaying steep Balmer breaks and unique V-shaped spectral energy distributions (SEDs), appearing red in the rest-frame optical and blue in the rest-frame UV \citep{FurtakEtAl2024, BarroEtAl2024, FujimotoEtAl2024, GreeneEtAl2024, AkinsEtAl2025, KocevskiEtAl2025, TaylorEtAl2025, LabbeEtAl2023, KokorevEtAl2024, MattheeEtAl2024, LabbeEtAl2025}. Their compactness and the presence of both broad emission lines and Balmer breaks stronger than normal stellar populations can produce \citep{WangEtAl2024b, AkinsEtAl2025, InayoshiIchikawa2024, DeGraaffEtAl2025a, MaEtAl2025} suggest that LRDs harbor SMBHs and display significant AGN emission. Empirical measurements of LRD luminosity functions find that LRDs are more common than normal AGNs in the early Universe (e.g., $z\gtrsim 4$), constituting $\sim$a few percent of the galaxy population at $z \approx 5$, up to 100 times that of UV-selected SMBHs \citep{BarroEtAl2024, KokorevEtAl2024, LabbeEtAl2025, GreeneEtAl2024, MattheeEtAl2024, ZhuangEtAl2025, HarikaneEtAl2023, KocevskiEtAl2025,PanEtAl2026}. Such a high frequency at early times have brought LRDs to the forefront of SMBH growth and seeding at high-$z$.

However, LRDs contain many puzzling features counter to the AGN narrative. Firstly, LRDs have little-to-no observed X-ray emission, even using stacked observations \citep{AnannaEtAl2024, YueEtAl2024, MaiolinoEtAl2025}. Additionally, although they exhibit red optical colors, many LRDs display little warm \citep{WilliamsEtAl2024, SettonEtAl2025, WangEtAl2025} and cold \citep{SettonEtAl2025, XiaoEtAl2025, CaseyEtAl2025, LabbeEtAl2025} dust emission, although recent results show significant mid-infrared emission in some LRDs \citep{LinEtAl2026, JiEtAl2026, DelvecchioEtAl2025, BrazziniEtAl2026}. Interestingly, LRDs display little-to-no photometric variability on rest-frame time-scales $\lesssim$ months \citep{JiEtAl2025, WangEtAl2024, LabbeEtAl2024, LabbeEtAl2025, FurtakEtAl2025, ZhangEtAl2025c, StoneEtAl2025a}. In fact, \citet{ZhangEtAl2025c} found that out of a sample of 300 LRDs, only eight potentially have significant variability, granting a rate of $\sim 3\%$ at best. Many LRDs also display Balmer absorption $\sim$hundreds of km s$^{-1}$ offset from the emission line center \citep{MaiolinoEtAl2024, KokorevEtAl2025, JuodzbalisEtAl2024, KocevskiEtAl2025, MaiolinoEtAl2025,ZhuangEtAl2025,HarikaneEtAl2023, KocevskiEtAl2023a, MattheeEtAl2024, MattheeEtAl2026}, a rare occurrence for low-$z$ BLAGNs \citep[e.g.,][]{HutchingsEtAl2002, WangXu2015}.

The combination of strong Balmer breaks, Balmer absorption, a lack of dust, and a scarcity of X-ray detections suggest a dense gas atmosphere encasing the SMBH. Several models have been invoked to explain the observed properties of LRDs utilizing regions of dense gas -- these models are likely not mutually exclusive: super-Eddington AGNs \citep{InayoshiEtAl2025a, LambridesEtAl2024, MadauHaardt2024, PacucciNarayan2024, TrincaEtAl2024, LiuEtAl2025b, KidoEtAl2025, ScholtzEtAl2026, MadauMaiolino2026, MadauEtAl2026, Liu_etal_2026}, quasi-stars / direct-collapse black holes \citep{PacucciEtAl2026, BegelmanDexter2026}, enshrouded ``cocooned" SMBHs \citep[e.g.,][]{SneppenEtAl2026}, and ``black hole stars" \citep[BH*s;][]{NaiduEtAl2025}. While the dense gas explains the optical continuum, the blue UV continuum may be produced by host-galaxy emission \citep{ChenEtAl2025, ZhuangEtAl2025, LinEtAl2026, NaiduEtAl2025, RinaldiEtAl2025}. 

Each of these above models has a separate prescription for the lack of observed continuum variability: super-Eddington accretion typically displays low intrinsic optical continuum variability \citep{LuEtAl2019, SecundaEtAl2025b}, quasi-stars experience pulsations similar to Cepheids on decades-long timescales in the rest-frame \citep{CantielloEtAl2025, PacucciEtAl2026}, the photosphere of the dense gas envelope surrounding the BH* experience little variability \citep{SneppenEtAl2026}. If LRDs are powered by a normal AGN, both low-$z$ variability relations and physically-based models suggest significant variability will be observed with a multi-year baseline, especially for low-mass systems \citep{SecundaEtAl2025b, ZhouEtAl2025b}. Super-Eddington hydrodynamical simulations suggest even lower variability, hampering prospects of a significant detection \citep{SecundaEtAl2025b}. Additionally, observed LRD variability may be dampened by the host galaxy, depending on the fraction of continuum emission originating from the AGN \citep{BurkeEtAl2023}. Reproducing the observed low LRD continuum variability with typical low-$z$ AGN variability amplitudes requires a low AGN-to-total emission fraction of $\lesssim$ 30\% \citep{KokuboHarikane2025, ZhangEtAl2025c, TeeEtAl2025}, which is difficult to reconcile with the low to little emission from the host galaxy in LRDs \citep[e.g.,][]{HarikaneEtAl2023, ChenEtAl2025}.

Spectroscopic broad-line variability is a complementary line of evidence to test proposed models. For low-$z$ AGNs, the broad-line region (BLR) is illuminated by the ionizing continuum, echoing the variability from the UV/optical continuum from the accretion disk. While X-rays from the LRD will not be detectable due to supposed Compton thickness in the dense gas surrounding the SMBH \citep{MadauMaiolino2026}, if BLR emission is powered by leaked ionizing continuum from within the envelope, it will reverberate the variability from the central engine. Therefore, measurements of broad emission line variability can place constraints on the different scenarios for LRDs and rule out non-AGN models. For example, the lensed LRD A2744-QSO1 ($z = 7.04$) displays equivalent-width variations over rest-frame years \citep{FurtakEtAl2025, JiEtAl2025}. However, systematics induced from lens modeling complicate the significance of these detections. 

\begin{figure}[]
    \centering
    \includegraphics[width=0.48\textwidth]{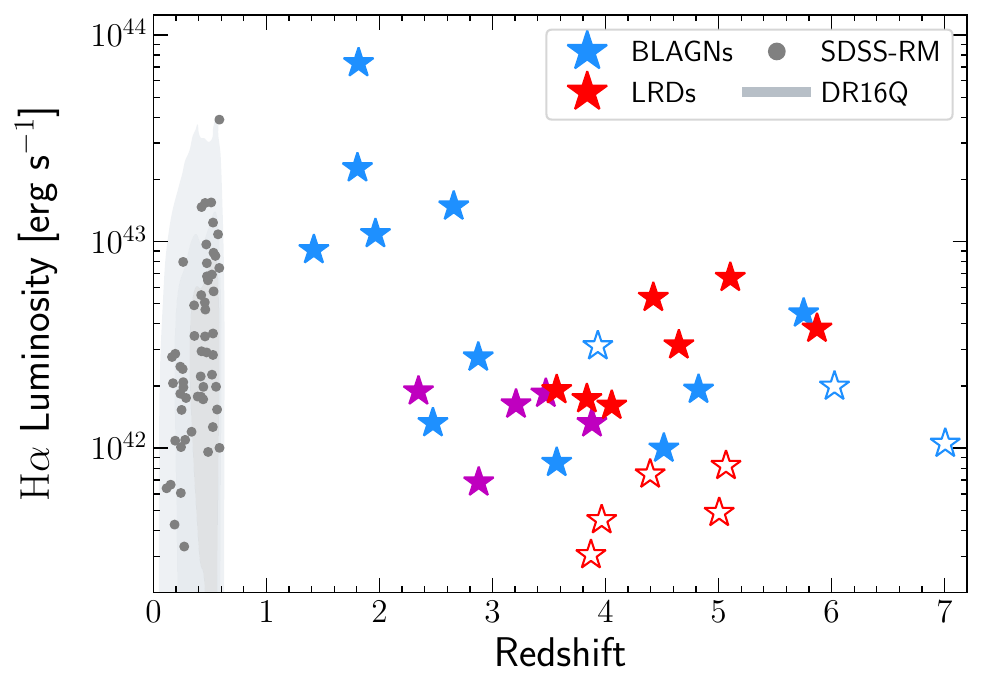}
    \caption{The NEXUS BLAGNs (blue)/LRDs (red/purple) in $\rm H\alpha$ luminosity-redshift space The SDSS-RM low-$z$ AGN comparison sample (gray points) and SDSS DR16Q comparison sample (gray contours) are also shown. Objects represented by filled stars show the subsample considered for our \halpha~variability analysis. The purple stars represent the subset of LRDs that have ``cooler'' temperatures (see Section~\ref{subsec:15499})}
    \label{fig:Lz}
\end{figure}

To investigate the possibility of (rest-optical) broad-line AGN reverberation mapping in the early Universe with JWST, and to place some of the first constraints on broad-emission-line variability in LRDs, we analyze multi-epoch JWST NIRSpec spectroscopy from the NEXUS survey \citep{ShenEtAl2024a}. NEXUS is a multi-cycle (Cycles 3-5) Treasury JWST program, obtaining repeated NIRCam imaging and WFSS spectroscopy over a $\sim$0.1 deg$^2$ field within the North Ecliptic Pole (NEXUS-Wide). The central $\sim$50 arcmin$^2$ region of the NEXUS field is visited with a higher cadence with joint NIRCam imaging and NIRSpec MSA spectroscopy (NEXUS-Deep). NEXUS-Wide data are obtained with a yearly cadence, while NEXUS-Deep data are obtained with a bi-monthly cadence, both over the course of 2024-2028. In this study, we utilize the first six epochs of NIRSpec MSA spectroscopy from NEXUS-Deep, spanning nearly six months in the observed-frame. 

This paper is organized as follows: in Section~\ref{sec:data}, we describe the spectroscopic data obtained through NEXUS, its reduction, and the samples of BLAGNs and LRDs. In Section~\ref{sec:analysis}, we describe our emission line modeling procedure and other metrics of sample classification and variability. In Section~\ref{sec:results}, we present the results of our line ratio and variability analyses. In Section~\ref{sec:discussion}, we contextualize our observed lack of variability in reference to local AGNs and recent JWST variability work. Finally, we conclude in Section~\ref{sec:conclusion}. By default flux variations are measured in fractional units relative to the mean flux. 

\section{Data}\label{sec:data}

\subsection{JWST Observations}

NEXUS Deep epochs were conducted with a cadence of approximately two months. We make use of the first six Deep epochs executed between June 1 2025 and March 28 2026. All primary NIRSpec observations used the MSA PRISM mode, providing a spectral resolution of $R \sim 100-300$ over 0.6--5.3\,\micron. The observations employed the \texttt{NRSIRS2RAPID} readout pattern with 43 groups and one integration, together with a two-shutter slitlet nodding pattern for each pointing, yielding an effective exposure time of 21.4 minutes per target.

Each Deep epoch comprises four MSA pointings and is capable of observing $\sim$800--900 targets. Targets are assigned different priorities according to their brightness and photometry-based science classifications \citep{nexus_edr,nexus-qdr}. The highest-priority, Class 1 targets include all broad-line objects identified in DESI DR1 optical spectra \citep{DESI_DR1} and in NIRCam WFSS spectra from the NEXUS Early Data Release \citep[EDR;][]{nexus_edr}. Photometrically selected LRD candidates from \citet{nexus_edr} and \citet{PanEtAl2026} are also assigned to Class 1. All spectra analyzed in this work correspond to Class 1 targets. Further details on the observing strategy and data reduction are provided in \citet{nexus_edr} and \citet{nexus-qdr}. 


\subsection{NIRSpec Reduction}

The MSA spectra were obtained with the PRISM/CLEAR configuration and reduced following the NEXUS Quick Release notes \citep{nexus-qdr}. We started from the stage 1 rate.fits products retrieved from MAST. The spectra were processed using msaexp v0.9.12 \citep{msaexp}, the JWST calibration pipeline v1.16.1 and the CRDS reference context jwst\_1298.pmap. In addition to executing the default pipeline, msaexp applies corrections for 1/f noise, detector bias, rescales the readout noise using source-free regions of each exposure and extends the wavelength limits to $0.54 - 5.5\ \rm \micron$. The sky background was estimated by differencing the nodded exposures. The background-subtracted 2D spectra were then coadded, and the final 1D spectra were obtained through optimal extraction.  

Our custom reduction performs equally as well as the DAWN JWST Archive (DJA)\footnote{https://dawn-cph.github.io/dja/} reduction, with all resulting spectroscopic light curves matching within uncertainty. We visually inspect all spectra to ensure that the reduction did not induce any artifacts or systematics. During this procedure, NX10749 Deep-02 and NX10835 Deep-02 were removed due to poor reduction (both DJA and NEXUS reductions).

\subsection{BLAGNs and LRDs from NEXUS}

\begin{deluxetable*}{lccccccccc}
\tablecaption{Properties of the BLAGN/LRD Sample\label{tab:sample}}
\tablehead{
\colhead{Name} & \colhead{DR1 ID} & \colhead{R.A.} & \colhead{Decl.} & \colhead{$z$} & \colhead{$\log_{10}(L_{\rm H\alpha})$} & \colhead{$\log_{10}(\rm H\alpha / H\beta)$} & \colhead{Normalization} & \colhead{max$\left( |\Delta F_{\rm H\alpha}| \right)$} & \colhead{$N_{\rm reliable}$} \\
\multicolumn{1}{c}{(1)} & (2) & (3) & (4) & (5) & (6) & (7) & (8) & (9) & (10)
}
\startdata
\\[-10pt]
\multicolumn{9}{c}{\large LRDs} \\[2pt]
\hline
\textbf{NX15499}\tablenotemark{c} & 93986 & 268.486879 & 65.187836 & 2.345\tablenotemark{a} & 42.28 & 0.64 $\pm$ 0.02 & F200W & 10.42 & 4/4 \\ 
\textbf{NX35791}\tablenotemark{c} & 92199 & 268.440149 & 65.197540 & 2.878 & 41.83 & 0.89 $\pm$ 0.05 & F200W & 9.31 & 2/2 \\ 
\textbf{NX7607}\tablenotemark{c} & 31215 & 268.432888 & 65.147653 & 3.208 & 42.21 & 1.10 $\pm$ 0.05 & F444W & 27.04 & 2/2 \\ 
\textbf{NX28630}\tablenotemark{c} & 101522 & 268.505631 & 65.233142 & 3.472 & 42.26 & 1.03 $\pm$ 0.04 & F444W & 5.13 & 2/2 \\ 
\textbf{NX34350} & 90371 & 268.484841 & 65.207206 & 3.568 & 42.28 & 0.96 $\pm$ 0.04 & Spec. & 9.27 & 4/4 \\ 
\textbf{NX27049} & 86363 & 268.392363 & 65.243382 & 3.835 & 42.24 & 1.10 $\pm$ 0.05 & Spec. & 3.55 & 2/2 \\ 
NX16624 & 93206 & 268.432919 & 65.191916 & 3.871 & 41.48 & 0.81 $\pm$ 0.08 & Spec. & \nodata & 0/2 \\ 
\textbf{NX25108}\tablenotemark{c} & 153244 & 268.394589 & 65.254180 & 3.880 & 42.12 & 0.78 $\pm$ 0.03 & F200W & 10.07 & 3/3 \\ 
NX8515 & 30608 & 268.416011 & 65.152857 & 3.966 & 41.65 & 0.92 $\pm$ 0.06 & Spec. & \nodata & 0/2 \\ 
\textbf{NX33289} & 89443 & 268.404027 & 65.212111 & 4.055 & 42.21 & 0.96 $\pm$ 0.05 & Spec. & 12.32 & 2/2 \\ 
NX10868 & 98249 & 268.441027 & 65.167628 & 4.395 & 41.87 & 0.79 $\pm$ 0.06 & Spec. & \nodata & 0/3 \\ 
\textbf{NX5732} & 44927 & 268.541892 & 65.138807 & 4.423 & 42.73 & 1.23 $\pm$ 0.03 & Spec. & 8.19 & 3/3 \\ 
\textbf{NX10835} & 98270 & 268.437673 & 65.167487 & 4.649 & 42.50 & 1.13 $\pm$ 0.05 & Spec. & 5.38 & 2/2 \\ 
NX27135 & 86487 & 268.381841 & 65.242666 & 5.006 & 41.69 & 1.01 $\pm$ 0.08 & Spec. & \nodata & 0/2 \\ 
NX12300 & 96854 & 268.472997 & 65.174102 & 5.066 & 41.91 & 1.05 $\pm$ 0.07 & Spec. & \nodata & 0/3 \\ 
\textbf{NX27270} & 86558 & 268.408439 & 65.242195 & 5.104 & 42.83 & 1.00 $\pm$ 0.03 & Spec. & 3.32 & 2/2 \\ 
\textbf{NX12349} & 96784 & 268.568555 & 65.174341 & 5.871 & 42.58 & 1.02 $\pm$ 0.05 & Spec. & 3.19 & 2/2 \\ 
\hline \\[-9pt]
\multicolumn{9}{c}{\large BLAGNs} \\[2pt]
\hline
\textbf{NX29467} & 102382 & 268.454907 & 65.228638 & 1.419 & 42.96 & 0.73 $\pm$ 0.01 & F200W & 35.29 & 3/3 \\ 
\textbf{NX33414} & 106038 & 268.327197 & 65.211474 & 1.805 & 43.36 & 0.73 $\pm$ 0.01 & F200W & 14.99 & 4/4 \\ 
\textbf{NX6456} & 31036 & 268.372523 & 65.148987 & 1.815 & 43.86 & \nodata & F200W & 14.28 & 2/2 \\ 
\textbf{NX10749} & 80824 & 268.584760 & 65.169367 & 1.964 & 43.04 & 0.89 $\pm$ 0.02 & F200W & 12.05 & 2/2 \\ 
\textbf{NX27038} & 88600 & 268.461037 & 65.216262 & 2.473 & 42.12 & 0.86 $\pm$ 0.02 & F200W & 43.70 & 5/5 \\ 
\textbf{NX30350} & 103173 & 268.441897 & 65.222079 & 2.657 & 43.17 & 0.68 $\pm$ 0.01 & F444W & 28.05 & 5/5 \\ 
\textbf{NX34911} & 90949 & 268.428706 & 65.203777 & 2.874 & 42.44 & 0.62 $\pm$ 0.02 & F444W & 48.38 & 4/4 \\ 
\textbf{NX7168} & 43317 & 268.483153 & 65.147721 & 3.568 & 41.93 & 0.87 $\pm$ 0.04 & F200W & 47.51 & 3/5 \\ 
NX6877\tablenotemark{b} & 44006 & 268.442547 & 65.144322 & 3.932 & 42.49 & 0.72 $\pm$ 0.02 & F444W & \nodata & 1/1 \\ 
\textbf{NX23770} & 151906 & 268.542056 & 65.260274 & 4.516 & 42.00 & 0.85 $\pm$ 0.04 & F444W & 26.04 & 3/3 \\ 
\textbf{NX26545} & 154588 & 268.520316 & 65.246468 & 4.823 & 42.28 & 1.02 $\pm$ 0.04 & F444W & 8.21 & 3/3 \\ 
\textbf{NX10347} & 98798 & 268.520051 & 65.164873 & 5.753 & 42.65 & 0.58 $\pm$ 0.01 & F444W & 7.72 & 3/3 \\ 
NX8409 & 30689 & 268.353201 & 65.152326 & 6.026 & 42.30 & 0.74 $\pm$ 0.04 & F444W & \nodata & 1/2 \\ 
NX7680\tablenotemark{b} & 43433 & 268.393567 & 65.147946 & 7.006 & 42.02 & 0.53 $\pm$ 0.07 & F444W & \nodata & 0/2 \\ 
\enddata
\vspace{5pt}
(1--5): Object name (EDR IDs from \citealt{nexus_edr}), DR1 ID, R.A., Decl., and spectroscopic redshift. (5): Mean total integrated \halpha\, luminosity (erg s$^{-1}$) over all available MSA epochs. (6): Weighted mean of Balmer decrements across all observed epochs. (7): Normalization scheme used for \halpha. (8): Maximum fractional change in \halpha\, flux (\%), after removing fluxes with uncertainty $\geq$~5\%. (9): Number of reliable epoch \halpha~fluxes passing our 5\% uncertainty cut.
\tablecomments{Objects included in our \halpha~variability analysis are highlighted in bold. ($a$): $z$ for NX15499 was incorrectly obtained in \citet{ZhuangEtAl2025}, and has since been corrected. ($b$): These sources were declared LRD candidates in \citet{ZhuangEtAl2025} and later confirmed to be BLAGNs \citep{PanEtAl2026}. ($c$): Classified as cool LRDs in \citet{PanEtAl2026}. }
\end{deluxetable*}

We construct a sample of BLAGNs and LRDs, using the procedures outlined in \citet{ZhuangEtAl2025}, subsequently adopted to produce source classifications in \citet{nexus-qdr}. For a full description of the criteria used to identify broad \halpha\, emitters and select LRDs (i.e., Class 1 objects from \citealt{nexus-qdr}), see Section~2 of \citet{ZhuangEtAl2025}. We initially construct our sample to include all sources currently classified as broad \halpha\, emitters -- i.e., broad-line AGNs (BLAGNs) and LRDs. We then restrict the sample to include sources with more than one Deep spectroscopic epoch, leaving a sample of 31 sources with 17 (14) LRDs (BLAGNs). All sources are visually inspected to ensure they are bona fide BLAGNs and LRDs. We confirm all of our LRDs are present in the sample from \citet{PanEtAl2026}. Six of our LRDs exhibit V-shaped upturns that do not coincide with the Balmer break, which may be indicative of a blackbody component with temperature $T \lesssim 3000$~K. Such ``cool" LRDs have been identified in prior studies \citep[e.g.,][]{DeGraaffEtAl2025}, and we form a LRD subsample composed of solely these objects.

\begin{figure}
    \centering
    \includegraphics[width=\linewidth]{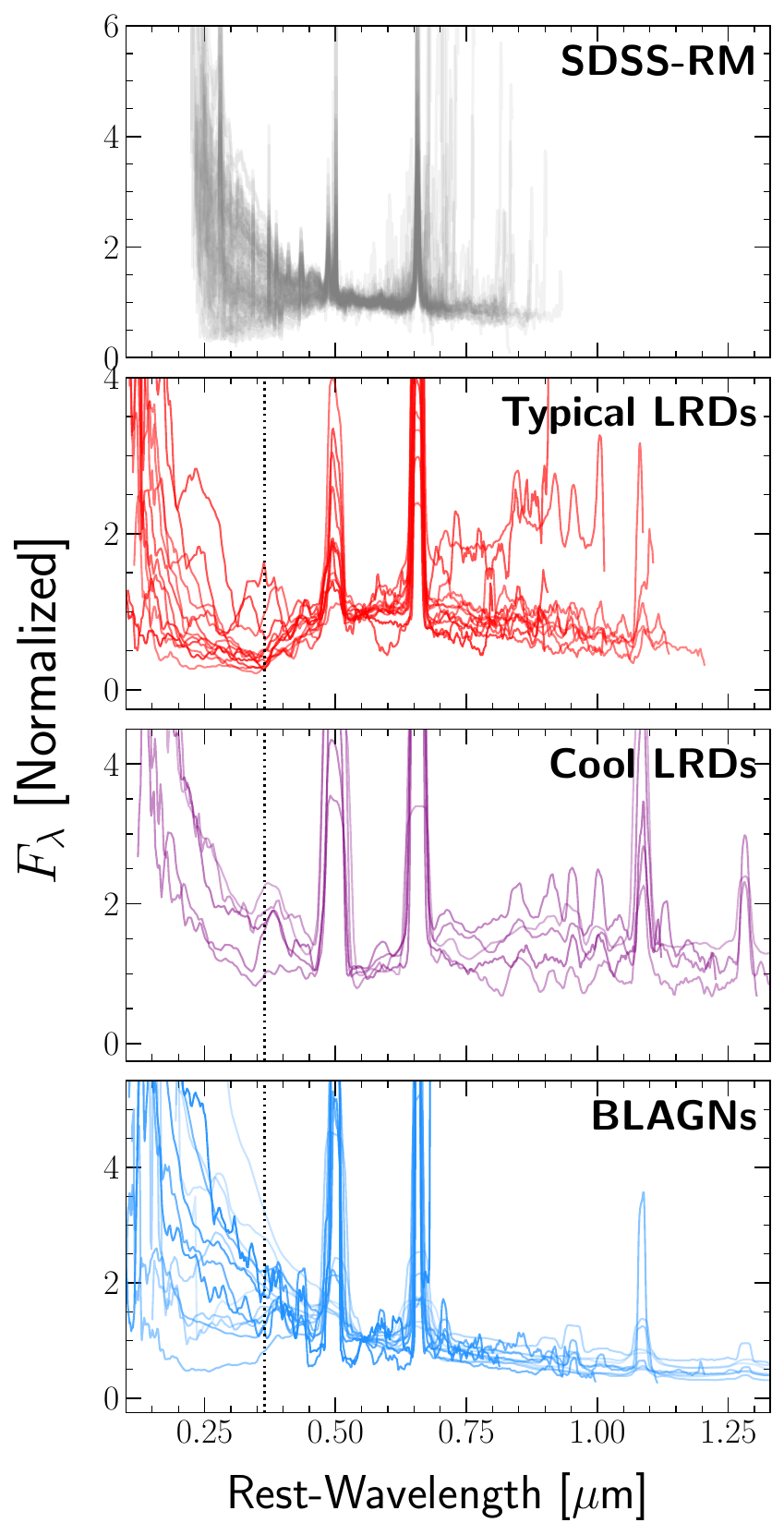}
    \caption{SEDs for our NEXUS BLAGNs (\emph{bottom})/LRDs (\emph{middle}), and the SDSS-RM comparison sample (\emph{top}). All spectra are shifted to the rest-frame of the source, normalized within the $\rm 5400-5700\,\AA$ continuum window, and smoothed with a Gaussian kernel of two pixels. The Balmer break, signifying the supposed location of the ``V" within the LRD, is shown with a vertical dotted line.} 
    \label{fig:SpecPlot}
\end{figure}

We construct a comparison sample of low-$z$ AGNs from the SDSS-RM program \citep{ShenEtAl2015, ShenEtAl2019}, with spectral measurements from \citet[][DR16Q]{WuShen2022}, and similar $L_{\rm H\alpha}$ to the NEXUS sample. In total, the low-$z$ SDSS-RM comparison sample consists of 56 AGNs with $L_{\rm H\alpha} > 10^{41}$ erg s$^{-1}$. When useful, we compare to all DR16Q AGNs with the same luminosity constraint. We show our sample of BLAGNs and LRDs in Fig.~\ref{fig:Lz}, and their SEDs in Fig.~\ref{fig:SpecPlot}. We refer to our sample by names corresponding to their Early Data Release (EDR) IDs. The upcoming first NEXUS data release (DR1, Zhuang et~al., in prep) will use different IDs, which we list in Table~\ref{tab:sample} as well.

\section{Analysis}\label{sec:analysis}

\subsection{Emission Line Modeling}


We first fit each of the emission lines present in each epoch spectrum. The emission line complexes considered and the models used for each are described in Table~\ref{tab:lines}. For each line complex, we consider a $\rm 1000\, \AA$ (rest-frame) region around the complex center, and adopt a straight line as the local continuum model. A straight line continuum suffices for the narrow wavelength ranges considered. In general, these spectral fitting windows are chosen per-line, as to not be contaminated by other broad lines or significant continuum variations. We use a larger spectral window around \halpha~ ($\sim2000\, \rm\AA$), as this line is the basis of our variability analysis and proper sampling of the continuum is essential. 

We fit all emission line complexes jointly for a single spectrum. All complexes are fit with multiple Gaussians, using {\tt LMFIT} \citep{NewvilleEtAl2025}. The width and offset of narrow emission lines (if they can be constrained) are all tied to each other, assuming all narrow-line emission is of similar origin. In particular, the \NIIab, \SIIab, and \OIIIab\, doublets, \OI\, $\lambda\lambda$6300,6363, and all \HeI\,  transitions are all treated as narrow lines. This greatly constrains the profiles, especially in lower-resolution portions of the spectra for low-redshift sources. A majority of the significant Balmer and Paschen lines, as well as \HeII $\lambda$4686, are modeled with a broad Gaussian component as well. We include the \FeII\, ``bump" near the ninth Paschen transition (Pa9), implementing it as two broad Gaussians. Due to the MSA resolution of the spectra, \OII\, $\lambda 3727$ and \NeIII\, $\lambda 3869$ are blended, so we fit them as a complex. Similarly, H$\gamma$ and [O \textsc{iii}] $\lambda4364$ are heavily blended, so we fit this complex to a single narrow Gaussian, obtaining an upper limit on H$\gamma$. We restrict the narrow component of a complex to have a full-width half-maximum (FWHM) at least 200 km s$^{-1}$ less than that of the broad component. We further constrain that narrow component has a FWHM $< 700$ km s$^{-1}$, and the broad component has a FWHM $\in [1500, 15000]$ km s$^{-1}$. All components must have an offset from the line center $<$ 1000 km s$^{-1}$. None of our LRDs displays apparent Balmer absorption, mostly likely due to PRISM's poor spectral resolution, so we exclude absorption modeling from this analysis.

\begin{figure*}
    \centering
    \includegraphics[width=\linewidth]{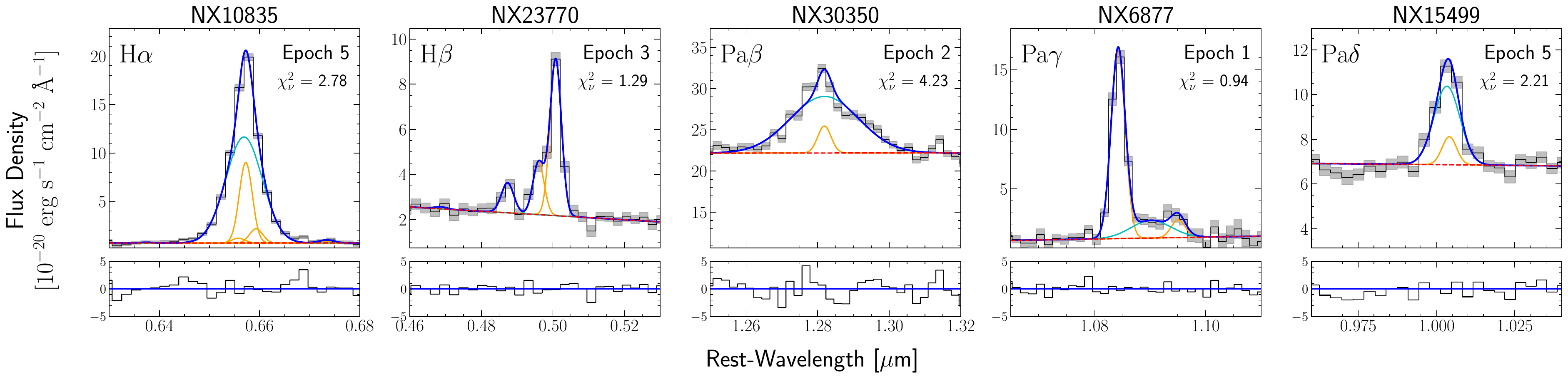}
    \caption{Example line model fits for some objects in our sample. For each source (panel), an example emission line is shown for a particular Deep epoch spectrum. For each panel, the top sub-panel shows the emission line in black, with the $1\sigma$ uncertainty shaded in gray, the emission line fit in blue, broad components in cyan, narrow components in orange, and the continuum in red. Some line complexes (e.g., \halpha) have multi-component fits shown within the profile. The bottom sub-panel displays the residuals (relative to error) of the fits. The reduced $\chi^2$ value for each fit is shown in the corresponding panel.}
    \label{fig:line_fits}
\end{figure*}

To account for instrumental broadening, we convolve the model with the publicly-available line-spread function (LSF) for NIRSpec prism\footnote{https://jwst-docs.stsci.edu/jwst-near-infrared-spectrograph/nirspec-instrumentation/nirspec-dispersers-and-filters}. We note that the line-spread function in reality is different than that provided by the JWST documentation due to both the morphology of the source \citep{DeGraaffEtAl2024} and the data reduction procedure \citep{ShajibEtAl2025}. The measured spectral resolution for point sources can be $>$~60\% higher than the fiducial dispersion function due to point source morphology alone, so we increase the LSF resolution by 60\% across all wavelengths. This rough and conservative increase in resolution produces reasonable model fits for all lines (Fig.~\ref{fig:line_fits}), even at low spectral resolution. We are primarily concerned with the total integrated flux of the line complexes and relative changes in line properties, not absolute line profiles, so we forsake the accuracy of model parameters for better fits. We discard any emission lines close to the edge of the spectra to mitigate the effect of systematics in our analysis.

A recent focus with LRDs has been to determine the shape of the broad component of Hydrogen and Helium lines, as many models predict exponential broad components from scattering in a dense gas medium \citep[e.g.,][]{NaiduEtAl2025, BegelmanDexter2026, SneppenEtAl2026, ChangEtAl2025}. However, the low-resolution MSA spectra in this study are inadequate to distinguish between different models. Additionally, instrumental broadening causes the broad lines to approach Gaussians, especially for the low-$z$ low-resolution sources. Therefore, we only utilize a single Gaussian to extract the broad-line flux for each emission line. Future dedicated follow-up will be able to determine the origin and shapes of these broad components.


To estimate the uncertainty in the emission line models, we utilize a Markov Chain Monte Carlo (MCMC) approach, using {\tt emcee} \citep{emcee}. We adopt the same model as used previously, but introduce priors using our initial {\tt LMFIT}-based fit. For each parameter, we utilize a uniform prior using 50\% to 150\% of the initial fit value as the prior bounds. This allows for {\tt emcee} to properly sample the likelihood distribution around the maximum obtained by {\tt LMFIT}. We run our MCMC with 1000 burn-in samples, 1000 used samples, and 1000 walkers, as necessitated by the large number of parameters in the model. Because of the complexity of the model's likelihood-space, some walkers fail to migrate from their starting position. Due to the large number of walkers and MCMC samples, we discard these walkers and still obtain well-sampled parameter distributions, discarding $< 50\%$ walkers for each spectrum. We utilize the distribution of model parameters to determine the upper and lower errors (16th--84th percentiles) on the best-fit parameter value. For each line, we measure the integrated flux and rest-frame FWHM. Most lines are either too low signal or unresolved to obtain reliable measurements of integrated flux or line profile properties. High-SNR lines such as \halpha\, and the Paschen series are among the most reliable in this analysis. We acknowledge that the uncertainty in our measurements is likely underestimated; \citet{GlimmIr} find a $\sim30\%$ systematic uncertainty in multi-epoch G395M spectra. Stacking analysis or highly-resolved follow-up spectroscopy will allow for more detailed line profile modeling in the future.

\begin{table}
    \caption{Measured Emission Line Complexes}

    \centering
    \begin{tabular}{|l|r|l|}
        \hline
        Emission Line Complex & $\lambda_{\rm rest} \, [\rm \AA]$ & Model \\
        \hline
         \OII\, + \NeIII\, $\lambda 3869$ & 3727.42 & narrow + narrow \\
         H$\delta$ & 4101.74 & narrow \\
         H$\gamma$ & 4340.46 & narrow \\
         \hbeta\, + \OIII\, + \HeII & 4861.32 & complex \\[3pt]
         \halpha\, + \NII\, + \SII\, + & \multirow{2}{*}{6562.80} & \multirow{2}{*}{complex} \\
          \;\;\;\; \OI\, + \HeI\, & \, & \, \\[3pt]
         Pa9 + \FeII & 9229.01 & complex \\
         \SIII\, $\lambda 9531$ & 9531.00 & narrow \\
         Pa$\delta$ & 10049.40 & broad + narrow \\
         Pa$\gamma$ + \HeI & 10938.10 & complex \\
         Pa$\beta$ & 12818.10 & broad + narrow\\
         Pa$\alpha$ & 18751.00 & broad + narrow \\
         \hline
    \end{tabular}
    \label{tab:lines}
\end{table}

A prominent issue throughout our analysis involves instrument systematics and observational effects within NIRSpec MSA spectra. Slit losses especially cause fluxes to vary significantly epoch-to-epoch. The effect of these slit losses is dependent on both wavelength and slit position, making it difficult to account for. Seeing as the line flux may vary, we can in principle normalize the spectral fluxes using simultaneous broad-band photometry, commonly adopted in reverberation mapping studies \citep{ShenEtAl2015}. However, utilizing aperture photometry on the NEXUS NIRCam imaging data of our LRD sample, we observe $\sim$6\% apparent, epoch-to-epoch continuum variability in F200W on average. When using photometry from more sophisticated difference imaging techniques \citep[i.e.,][]{StoneEtAl2025a}, the F200W epoch-to-epoch continuum variability decreases to $\sim$2\% on average, with all sources displaying $\lesssim$10\% variability. Results in F444W are similar. Therefore, we caution using standard aperture photometry in future JWST variability studies, as it may be insufficient for accurate spectro-photometric calibrations, especially for low-variability sources observed with JWST.

Because we do not have accompanying difference imaging for each spectrum, we devise a normalization scheme utilizing the local continuum. LRDs are compact by definition, so both the host-galaxy emission (if present) and LRD emission of the SED will be affected by slit losses in the same manner. Our initial difference-image photometry shows minimal photometric variability in our NEXUS LRD sample, which allows us to use the local spectral continuum to normalize. Subsequently, we divide the emission line profile for each MCMC sample by the corresponding continuum model, pixel-by-pixel. We then rescale the epoch spectrum using the continuum model from the spectroscopic epoch with the largest continuum value at the line center -- i.e., the epoch least affected by slit loss -- to recover the absolute flux scale of the spectrum. 

Our spectral normalization procedure for BLAGNs is more complex, as the more extended host-galaxy emission and compact AGN emission will be affected by slit losses differently. Furthermore, some BLAGNs show clear signs of a varying continuum, violating the assumption for LRDs. In this case, we normalize by the aperture photometry, using the broad-band filter (either F200W or F444W) closest to the emission line. Specifically, we obtain a multiplicative factor to scale the synthetic broad-band photometry to the measured photometry. This photometry will contain emission from both the AGN and host galaxy components, allowing us to compensate for slit losses better than with the local continuum normalization. While this may introduce additional scatter present in the photometric light curves, in all cases the aperture photometry is less variable than the local SED continuum. We add the photometric uncertainty in quadrature to all flux estimates with this normalization. Three of our BLAGNs (NX6456, NX10749, NX23770) do not have corresponding aperture photometry for each MSA epoch. For these epochs, we normalize with the aperture photometry taken nearest to the observed spectrum. While this may induce an additional $\sim$6\% scatter, it reduces all \halpha~variability estimates significantly relative to the local continuum normalization. 

\begin{figure*}
    \includegraphics[width=\textwidth]{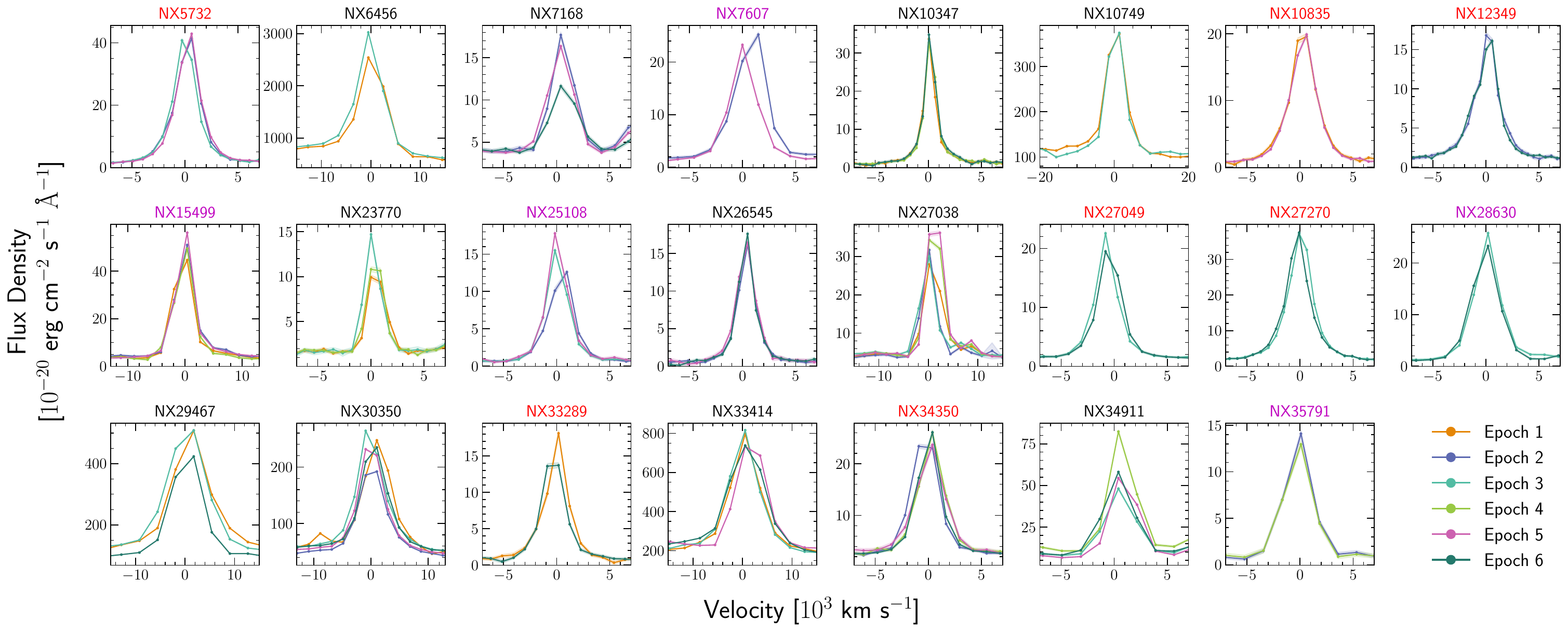}
    \caption{Normalized multi-epoch \halpha\, line profiles for the broad-line variability sub-sample (Section~\ref{subsec:agn_var}), after imposing a $<5\%$ integrated flux uncertainty cut on each epoch. Typical LRDs are labeled in red, and cool LRDs are labeled in purple.}
    \label{fig:spec_mosaic}
\end{figure*}

For the six cool LRDs, we utilize the aperture photometry to normalize instead of the local continuum. Importantly, this reduces the scatter in \halpha~variability significantly within each object as opposed to using the local continuum. If the cool LRDs represent an intermediate stage between BLAGNs and LRDs, they may exhibit AGN-like spectral variability.
Fig.~\ref{fig:spec_mosaic} displays a mosaic of normalized multi-epoch \halpha~profiles used for variability analysis.

\begin{figure}
    \centering
    \includegraphics[width=0.48\textwidth]{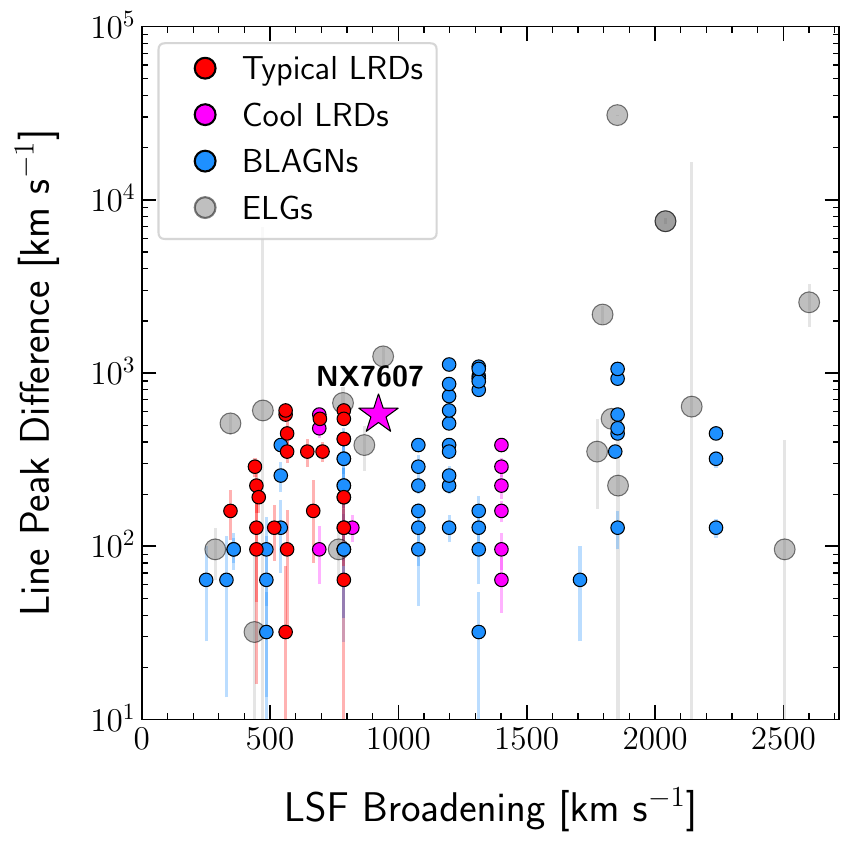}
    \caption{Multi-epoch offsets in the peak model \halpha\, flux from our line fitting procedure for the sample of BLAGNs/LRDs (blue/red), as well as a sample of ELGs (gray). Offsets between two epochs are shown as a function of the LSF broadening at \halpha. Large offsets can be caused by systematics in MSA spectral reduction, as evidenced by the results for the ELGs.}
    \label{fig:offsets}
\end{figure}

Within our normalized line profiles, we find that some sources exhibit significant velocity offsets between epochs (e.g., NX7607), reaching $\gtrsim 1000$ km s$^{-1}$. While emission line profile variability is commonly seen in local AGNs \citep[e.g.,][]{FriesEtAl2023}, systematics in NIRSpec reduction may also lead to artificial profile variations. To investigate this, we utilize a sample of 46 emission line galaxies (ELGs) with repeated MSA spectra within the first six NEXUS Deep epochs. We expect the narrow emission lines present within the ELGs to be constant in shape, so any observed variations suggest unaccounted-for systematics. All available epochs for each object are fit using the same procedure described earlier in this section. For a proper comparison to the sample we utilize for variability analysis (Section~\ref{subsec:agn_var}), we only consider the \halpha\, model fits. Only 18 of our ELGs display \halpha\, emission in multiple spectra, spanning $0 \lesssim z \lesssim 7$. We then calculate the velocity offset between the model line fits in each available epoch. Our results are shown in Fig.~\ref{fig:offsets} as a function of the amount of broadening from the LSF at \halpha\, for each source. Most NEXUS LRDs and BLAGNs display offsets $\sim 10^{2-3}$ km s$^{-1}$ with uncertainties $\sim$50 km s$^{-1}$. The ELGs show similar results, reaching $\sim 10^4$ km s$^{-1}$ for large LSF broadening (i.e., low redshifts). Seeing as the ELGs are able to reproduce the offsets seen in our BLAGN/LRD sample, any observed offset likely stems from issues in wavelength calibration within the MSA reduction coupled with an inability to resolve the profile. Based on this test, we cannot claim detection of significant profile variability in NX7607 {or other BLAGNs/LRDs with similar offsets measured from multi-epoch MSA spectra}. More work must be done in the future to mitigate such systematics.

\subsection{Line Ratios}

\begin{figure*}
    \includegraphics[width=\textwidth]{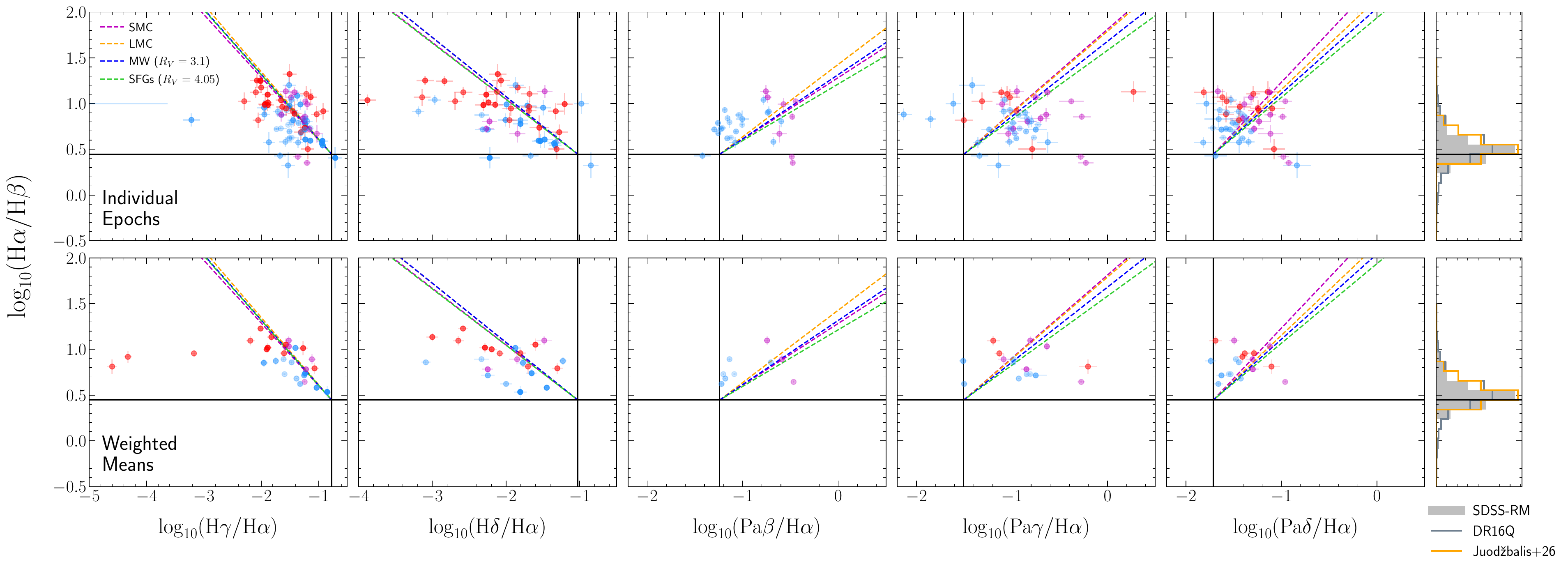}
    \caption{Ratios between different Balmer and Paschen lines from our sample. The top row shows ratios from individual deep epoch spectra, while the bottom row shows the weighted mean over all epochs for each object. The colors and transparency of the sources match those in Fig.~\ref{fig:SpecPlot}. Black lines mark line ratios assuming Case B recombination. Dashed colored lines represent line ratios from different extinction laws (SMC, LMC, Milky Way, star-forming galaxies), evolving with increasing $A_V$. The rightmost panels show histograms of Balmer decrements from the SDSS-RM and DR16Q comparison samples, and high-$z$ BLAGNs from \citet{JuodzbalisEtAl2026}.}
    \label{fig:line_ratios}
\end{figure*}

We calculate emission line ratios for all observed lines from the Balmer and Paschen series in order to compare to theoretical predictions for the physical situations within high-$z$ BLAGNs and LRDs. While observed line ratios are normally obtained with respect to \hbeta, we opt to utilize line ratios with respect to \halpha\, due to its ubiquity among our sample and the significantly higher signal-to-noise ratio (SNR). We calculate line ratios for each individual spectrum, but also the weighted mean of all line ratios across epochs. Many of the emission line fits straddle the resolution limit, especially low-SNR lines like $\rm H\gamma$. Such measurements may cause the weighted measurement to drop significantly. For comparison, we include multiple extinction curves within {\tt dust\_extinction} \citep{Gordon2024} and {\tt dust\_attenuation}: the Small Magellanic Cloud \citep[SMC;][]{GordonEtAl2024}, Large Magellanic Cloud \citep[LMC;][]{GordonEtAl2003}, Milky Way \citep{GordonEtAl2023}, and star-forming galaxies \citep[SFGs;][]{CalzettiEtAl2000}. We also calculate Balmer decrements from the SDSS-RM and DR16Q comparison samples using the total integrated line flux. Observed line ratios are shown in Fig.~\ref{fig:line_ratios}.

We convert the measured Balmer decrement (\halpha/\hbeta\, ratio) into an extinction $A_V$, assuming the decrement is due to dust reddening and the fiducial line ratio is from Case B recombination -- though this is probably not true for LRDs. Using individual line ratios, we obtain estimates for the dust extinction $A_V$ for each source: 

\begin{equation}
A_V = \frac{2.5}{\kappa(H\beta) - \kappa(H\alpha)}\log_{10}\left[ \frac{(H\alpha /{H\beta})_{\rm obs}}{(H\alpha / H\beta)_{\rm int}} \right]
\end{equation}

where $A_V = R_V E(B-V)$, $A_\lambda = \kappa(\lambda) E(B-V)$, $(H\alpha / H\beta)_{\rm obs}$ is the observed line ratio, and $(H\alpha / H\beta)_{\rm int}$ is the intrinsic line ratio. To obtain estimates for dust attenuation $A_V$, we utilize the average extinction curve for the SMC, with $R_V = 3.02$, and assume intrinsic line ratios from Case B recombination. For each source, we obtain multiple estimates of $A_V$ from each line ratio for a given epoch. We then utilize the weighted mean of all $A_V$ across all epochs as our final estimate (Fig.~\ref{fig:dust}). We choose to exclude $\rm H\gamma$ and $H\delta$, as their measurements are highly uncertain. It should also be noted that we only use \halpha/\hbeta\, for the SDSS-RM sample, using line flux values from \citet{ShenEtAl2019}.

The calculated dust attenuation only serves as a comparison for prior studies of broad \halpha-emitters. Low-$z$ analyses indicate that local extinction laws, or ones similar, may apply to high-$z$ BLAGNs \citep{GaskellBenker2007}. However, many models proposed for LRDs expect their Balmer decrements to be severely affected by different physical mechanisms than dust attenuation, such as collisional de-excitation and resonant scattering in BH*s \citep{NaiduEtAl2025, DeGraaffEtAl2025a}. Additionally, careful consideration of dust re-emission and mid-to-far infrared upper-limits within LRDs decreases simple $A_V$ estimates by $\sim$2.5 mag \citep{ChenEtAl2025c}. In our discussion, we only utilize our obtained $A_V$ as a means to discuss the large Balmer decrements within LRDs and compare with other samples of LRDs.

\subsection{Broad-line Variability}\label{subsec:agn_var}

We now compare emission-line variability between our NEXUS sample and the SDSS-RM comparison sample. To do so, we restrict our analysis to \halpha\, measurements, as they are present in almost all NEXUS spectra, by construction, and most SDSS-RM spectroscopic measurements. Furthermore, our \halpha\, measurements have the highest quality among all emission lines and are the least affected by the MSA spectral resolution. Still, many of the integrated \halpha\, fluxes have large measurement uncertainties of $\gtrsim 10\%$, rendering variability constraints nearly useless. In order to mitigate the influence of low-SNR fluxes within our analysis, we impose a cut of individual epoch fluxes with fractional uncertainty $< 5\%$ before any variability analysis. Importantly, SDSS-RM quasars have shown to display intrinsic variability $\sim 5-10\%$ on the rest-frame timescales probed by NEXUS (Fig.~\ref{fig:timescale}). In order to observe 5\% intrinsic variability within our data, we must restrict the flux uncertainty accordingly. This quality cut reduces our LRD and BLAGN \halpha\ variability sample size to {12} and {11}, respectively (see Table~\ref{tab:ks} for details), but largely preserves the global properties of these populations.

We expect \halpha\, in normal, broad-line AGNs to exhibit an intrinsic variability, originating from the central engine, as well as white noise from flux measurement uncertainties. The intrinsic rms variability of an AGN light curve can be measured with a maximum likelihood estimator $\sigma_0$ \citep[e.g.,][]{ShenEtAl2019}. This $\sigma_0$ metric presents a statistically rigorous estimate of the intrinsic flux variability, taking into account measurement errors in each individual data points. It bypasses many of the issues present in empirical methods, such as subtracting a mean flux error in variability amplitude for traditional structure function measurements.

In brief, the $\sigma_0$ estimator assumes the observed sample distribution is the combination of a single intrinsic Gaussian\footnote{In the current analysis, we assume Gaussian distributions for all underlying population-wide intrinsic variability. This is not a perfect assumption since individual objects may have different rms variability amplitudes, but necessary given the small sample statistics and limited epochs per object for the LRD/BLAGN sample.} variability (with dispersion $\sigma_0$) around the mean, and contributions from individual measurement uncertainties. The latter uncertainties do not have to be the same for individual data points. The intrinsic $\sigma_0$ represents a sample-averaged rms variability around the mean of the distribution. The estimator then evaluates $\sigma_0$ using a maximum-likelihood approach \citep[fully detailed in][]{ShenEtAl2019}.
In our case, the distribution sample refers to the pairwise flux variations (referred to as $\sigma_{0,\Delta F}$, or simply $\sigma_0$), or the collection of fluxes from the light curve (referred to as $\sigma_{0,\rm lc}$). 


{We investigate our sample's emission-line flux variations by utilizing pairwise fractional flux differences $\Delta F_{\rm H\alpha}$. For plotting purposes, we take their absolute values $|\Delta F_{\rm H\alpha}|$. These flux differences are obtained between each available spectral epoch, without duplication, relative to the mean of the full light curve after excluding epochs with $\geq 5\%$ flux uncertainty. Observed (error inflated) $|\Delta F_{\rm H\alpha}|$ can reach $\gtrsim 10\%$ for our sample, and $\gtrsim 30\%$ for individual SDSS-RM objects over similar rest-frame timescales ($\lesssim$ months). For reference, we list the mean flux uncertainty for each sub-sample ($\sigma_{\rm white\, noise}$) in Table~\ref{tab:ks}. Utilizing the flux uncertainties and cadence within current NEXUS data, we can simulate distributions of $|\Delta F_{\rm H\alpha}|$ generated from pure flux uncertainties, as well as from flux uncertainties combined with intrinsic variability measured for the SDSS-RM sample. A Kolmogorov-Smirnov (KS) test could then declare the similarity of the observed and simulated $|\Delta F_{\rm H\alpha}|$ distributions. However, current NEXUS data contains too few LRD/BLAGN flux pairs to perform this analysis, considering that the KS test is more subject to bias from flux uncertainties and small-sample statistics. Longer baselines and more flux pairs from future NEXUS data will be required to perform KS tests for a rigorous detection of variability (or lack thereof).}

Using the distribution of fractional pairwise flux differences, we obtain estimates of the intrinsic variability within the different considered subsamples. Because the $\sigma_0$ estimator assumes a Gaussian distribution around zero mean \citep{ShenEtAl2019}, we first create negative duplicates of all $|\Delta F_{\rm H\alpha}|$ flux differences. This enforces that the distribution has zero mean and a symmetric profile. We use the procedure outlined in \citet{ShenEtAl2019} to estimate $\sigma_{0, \Delta F}$ for a given subsample. We then divide by $\sqrt{2}$ to account for the fact we use pairwise flux differences instead of the fluxes themselves, as the intrinsic variability contributes equally to each flux in the pair, granting a sample-wide $\sigma_0$. To avoid confusion with the pairwise flux variations $\sigma_0$, we label this estimate $\sigma_{0,\rm lc}$ as it refers to the intrinsic rms variability around the light-curve mean. We list these estimates in Table~\ref{tab:ks}. We exclude NX7607 when obtaining $\sigma_{0, \rm lc}$, as its $\sim30\%$ variability significantly inflates the estimated intrinsic variability. Due to unknown systematics, we clarify that these estimates are upper limits to intrinsic variability.

For comparison, we construct a flux-uncertainty-corrected ensemble structure function (SF) of \halpha~variability using the SDSS-RM \halpha\, light curves \citep{ShenEtAl2024_rm} from our comparison sample. We first compile all fractional flux differences, and their uncertainties, between all flux measurements for every AGN in the SDSS-RM sample, noting its rest-frame time difference. We bin these flux differences by timescale, then perform the following:

\begin{equation}\label{eqn:sf}
    {\rm SF}(\Delta t) = \sqrt{{\rm med}[(\Delta F)^2] - ( k \cdot {\rm med}[\sigma_{\Delta F}] )^2}
\end{equation}

\noindent
where $k \approx 0.67$ is the ratio of the median absolute deviation (MAD) to the standard deviation for a Gaussian distribution. This SF is essentially the MAD of all SDSS-RM $\Delta F$ estimates for a given timescale, directly comparable to the individual $\Delta F$ values shown in Fig.~\ref{fig:timescale}. As in prior analyses \citep{MacLeodEtAl2012, Kozlowski2016}, we subtract the median $\sigma_{\Delta F}$ among all pairs in the timescale bin in quadrature from the raw SF to correct for flux uncertainties, but we scale these uncertainties to the MAD. Uncertainties in the SF are obtained via boostrapping, but are small $(\sim 0.1\%)$ given the large number of SDSS-RM pairs (Table~\ref{tab:ks}). We display this SF in open gray circles in Fig.~\ref{fig:timescale}. Most observed LRD and non-variable BLAGN flux pairs are consistent or less-variable than the SDSS-RM SF, {although the measured $|\Delta F|$ values are inflated from flux uncertainties.} We also recover a sample-wide $\sigma_{0, \rm lc}$ for the SDSS-RM sample (Table~\ref{tab:ks}), using all flux differences on rest-frame timescales $< 100$d with the same procedure as for the BLAGNs and LRDs. 

We additionally construct an intrinsic $\sigma_0$-based ensemble SF for SDSS-RM. For each rest-frame timescale bin, we obtain an estimate $\sigma_{0, \Delta F}$. We again scale this dispersion to the MAD for a fair comparison with the observed $|\Delta F|$ values of LRDs and BLAGNs. This structure function is shown as filled circles in Fig.~\ref{fig:timescale}, and roughly traces the traditional structure function. At the timescales probed by the current NEXUS data, the $\sigma_0$-based SF is consistent with the traditional SF, but grows slightly larger at longer timescales. The two likely diverge due to each SF method accounting for flux uncertainty differently, e.g., the standard SF method utilizes the average flux uncertainty in the bin. We compare to a few other variability results in Fig.~\ref{fig:timescale}, including TWINKLE \citep{LiuEtAl2026}. They report a deficit of $3\sigma$ flux difference detections, given the number of expected $3\sigma$ detections within SDSS-RM light curves. To paraphrase, they find LRD variability is lower than SDSS-RM variability on $\sim$6 month timescales, which we display as an upper limit on Figs.~\ref{fig:timescale} \& \ref{fig:timescale_cont} under the SDSS-RM traditional SF estimate.

\section{Results}\label{sec:results}

\subsection{Balmer Decrement and Extinction}

Our line ratio measurements of LRDs and BLAGNs are consistent with recent JWST studies. High-redshift LRDs and BLAGNs generally display larger Balmer decrement (3.4--16.7 with a median value of 7.64) compared with that for low-redshift AGNs (1.3--10 with a median value of 3.22). Larger Balmer decrements suggest either elevated dust content or collisionally excited emission lines \citep{BrooksEtAl2025, DEugenioEtAl2026, TorralbaEtAl2025, NikopoulosEtAl2025}, or radiative transfer effects of Balmer lines in extremely dense gas \citep[e.g.,][]{MadauMaiolino2026, ChangEtAl2025}. In fact, many of our sources exhibit prominent \OI $\lambda 8446$ emission, indicative of fluorescence from Ly$\beta$\, trapping in a dense gas medium \citep{KwanKrolik1981}. The same could be said for Paschen-to-Balmer line ratios, although observational evidence suggest different broad line widths for the Paschen and Balmer lines in AGNs \citep{LandtEtAl2008, LampertiEtAl2017}, possibly from Paschen lines existing on the outer parts of the BLR \citep{KimEtAl2010}. \citet{Kokubo2024} demonstrate that Raman scattering may produce differing line widths for different Hydrogen lines, although results from the Rosetta Stone find inconclusive evidence \citep{JuodzbalisEtAl2024}.

\begin{figure}
    \centering
    \includegraphics[width=\linewidth]{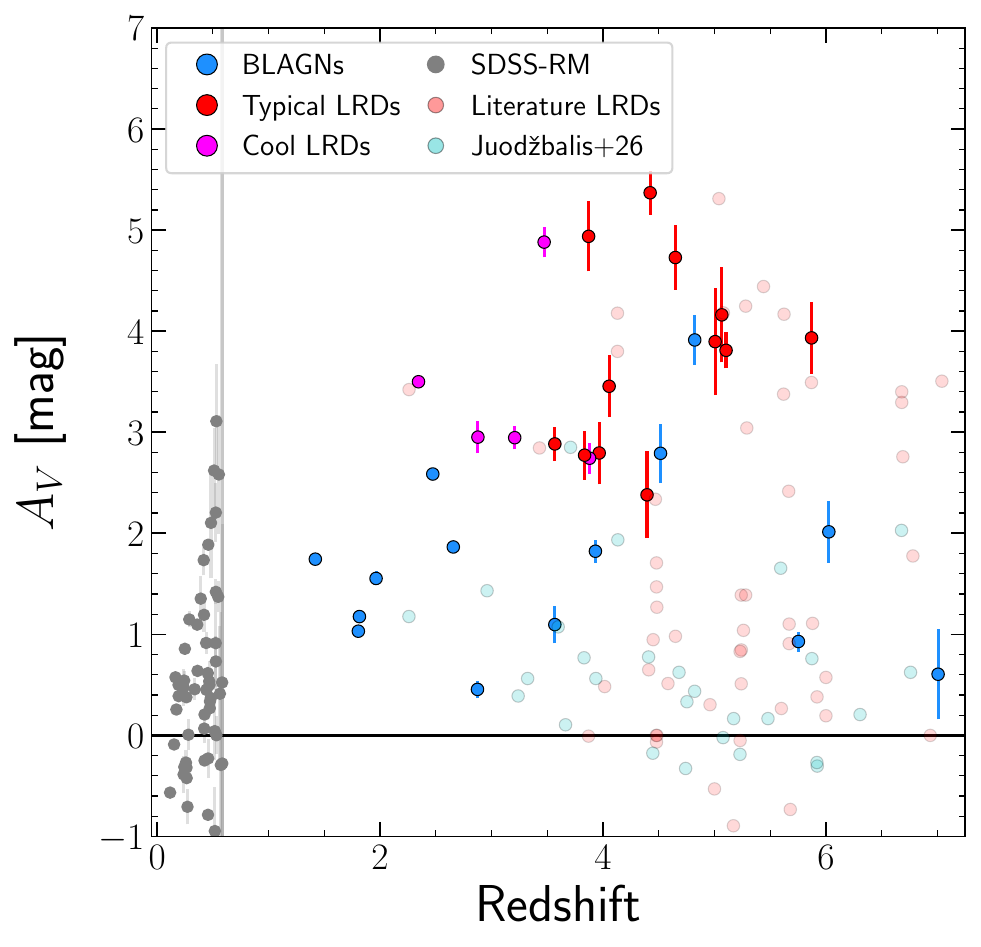}
    \caption{Estimates of dust extinction $A_V$ from the NEXUS BLAGN/LRD sample and SDSS-RM comparison sample. Estimates are obtained using the total integrated fluxes from each line. Faded red circles represent total Balmer decrement $A_V$ estimates for LRDs from the literature \citep{HarikaneEtAl2023, BrooksEtAl2024, JuodzbalisEtAl2024, LabbeEtAl2024, DEugenioEtAl2025, NikopoulosEtAl2025, DEugenioEtAl2026, DEugenioEtAl2026b}, and faded blue circles are high-$z$ BLAGNs from \citet{JuodzbalisEtAl2026}. A horizontal black line indicates no extinction ($A_V = 0$).}
    \label{fig:dust}
\end{figure}

Figure~\ref{fig:line_ratios} displays the measured line ratios for our LRDs and BLAGNs, and the comparisons with predictions from a dust-reddening model with various extinction laws. Under the dust-reddening assumption, none of our sources has line ratios significantly violating Case B recombination, as seen in a few LRDs \citep{NikopoulosEtAl2025, ZhangEtAl2025a}. Overall, the line ratios (and in particular the large Balmer decrement) in LRDs and BLAGNs can be explained by dust reddening, albeit with moderate scatter around the prediction ($\sim0.3$~dex). The required dust extinction $A_V$ values to produce these line ratios in LRDs are higher than typically seen in low-redshift AGNs, as shown in Figure~\ref{fig:dust}. However, the lack of dust thermal emission at longer wavelengths disfavors this interpretation for LRDs. More recent LRD models suggest the red rest-frame optical continuum emission originates from a quasi-blackbody from the thermalized atmosphere/envelope surrounding a central accreting BH \citep[e.g.,][]{NaiduEtAl2025, SneppenEtAl2026}. The same models could also explain the large Balmer decrement using collisional excitation and radiative transfer effects given the extreme gas densities.

\subsection{\halpha~Variability}

\begin{figure*}
    \centering
    \includegraphics[width=\textwidth]{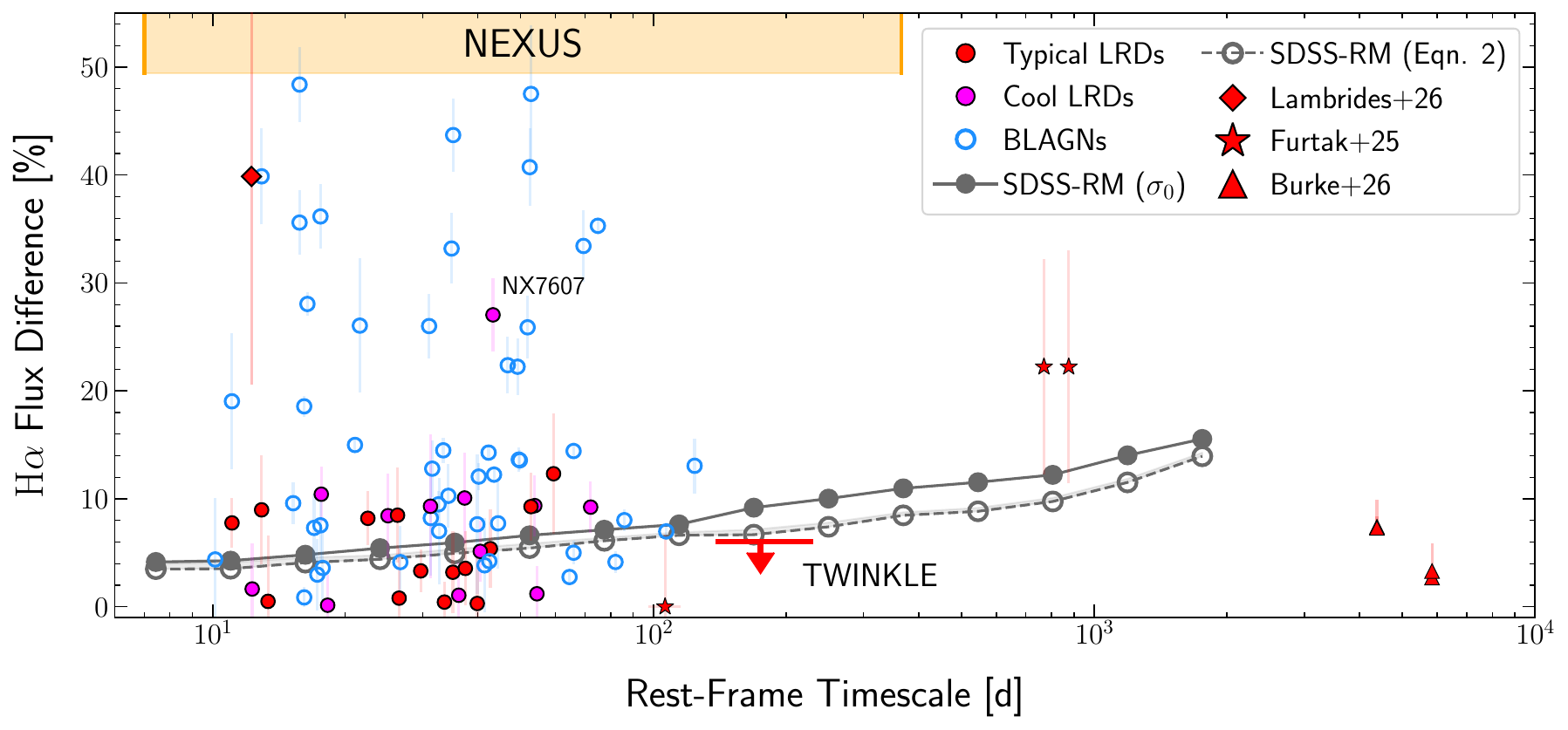}
    \caption{Pairwise \halpha\, flux variations as a function of rest-frame timescale. Estimates from LRDs are shown in red/purple, and BLAGNs are shown in blue. Note that these individual flux variations are error-inflated. The MAD-scaled traditional structure function (Eqn.~\ref{eqn:sf}) for the SDSS-RM comparison sample is shown as open gray circles and a dashed gray line. Bootstrapped uncertainties are shown shaded in light gray, but very small. The MAD-scaled $\sigma_0$-based SDSS-RM intrinsic structure function is shown with filled gray circles. A paraphrased result from TWINKLE \citep[][see Section~\ref{subsec:agn_var}]{LiuEtAl2026} and estimates from local LRD analogs \citep{BurkeEtAl2026}, GlimmIr \citep{GlimmIr} with systematic uncertainty included, and A2744-QSO1 \citep{FurtakEtAl2025} are also displayed. The range of timescales probed by future NEXUS data is shaded in orange. We caution the apparent significant variability in NX7607 among our sample due to potential systematic uncertainties (see Section~\ref{subsec:15499} for details).}
    \label{fig:timescale}
\end{figure*}

Our \halpha~flux variability for JWST LRDs and BLAGNs shown in Fig.~\ref{fig:timescale} shows a concentration of flux differences $\lesssim 15\%$, similar to SDSS-RM quasars on the same timescales, after accounting for uncertainty. There is a population of large flux differences within our sample, all produced by six BLAGNs (NX7168, NX23770, NX27038, NX29467, NX30350, NX34911). These objects produce both small and large flux differences across all observed timescales. These ``variable" BLAGNs mostly exist at lower redshifts ($z<3$), except for NX7168 and NX23770. They have moderate Balmer decrements (4.2--7.5) and dust attenuation ($A_V \in [0.5, 3.0]$), consistent with those from the SDSS-RM sample. NX23770 does not have accompanying photometry for all spectra, which may result in excess scatter. The higher flux differences in two of these objects result from a single MSA epoch with a lower integrated flux (NX7168, NX34911). Removing these epochs per-object reduces the range of observed variability to $\lesssim 15\%$. Additionally, even without removing these epochs, most fluxes fall within the 95th percentile of the SDSS-RM flux differences at the same timescales. The MSA spectra show no obvious signs of poor reduction and the aperture photometry displays minimal scatter, so we leave these fluxes in our analysis, but as a separate ``variable" BLAGN sample. Such variable fluxes exceed the expected variability from white noise and SDSS-RM-like variability from our procedure in Section~\ref{subsec:agn_var}, though residual systematics in MSA spectra may contribute to some of these apparent variations. 

Our sample-wide intrinsic rms variability estimates $\sigma_{0, \rm lc}$ agree with the low-amplitude variability seen in Fig.~\ref{fig:timescale}. The LRDs exhibit 4.02\% \halpha~variability on rest-frame $\sim$months, with consistent results seen for both the cool LRDs and more typical ``hot" LRDs. However, considering that the \halpha~flux estimates likely have underestimated uncertainty from systematics, this represents an upper limit of variability in LRDs. Only one of our cool LRDs (NX7607) exhibits tentative variability ($\sim$30\%), although the impact of systematics complicates its significance. We discuss this further in Section~\ref{subsec:15499}. 



The BLAGN sample produces 52 flux pairs from 14 objects, spanning $1.4 \leq z \leq 4.5$. The full and variable BLAGN samples produce $\sigma_{0, \rm lc} \sim 15\%$, higher than expected from SDSS-RM variability. Similar to the LRDs, the non-variable BLAGNs display 6.16\% mean intrinsic \halpha\, variability. Such variability produced flux difference pairs consistent with both LRD variability and SDSS-RM variability.

{We obtain a sample-wide $\sigma_{0, \rm lc}$ for the SDSS-RM sample using all flux differences over similar rest-frame timescales as the current NEXUS data ($< 100$d). The SDSS-RM sample exhibits $6.21\%$ intrinsic rms variability, higher than the LRDs by $3.8\sigma$ but consistent with the non-variable high-$z$ BLAGNs.} Our non-detection of \halpha~variability, and suppression relative to local quasars, adds to a wealth of evidence indicating a lack of variability in LRDs. Figure~\ref{fig:timescale} summarizes the \halpha\ flux variability results from different studies. \citet{LiuEtAl2026} observe a lack of LRD \halpha~variability on rest-frame $\sim$6 month timescales. Local LRD analogs monitored over the course of a decade only exhibit $\sim5\%$ \halpha~variability \citep{BurkeEtAl2026}. Significant short-timescale ($\sim$weeks) \halpha\, variability has only been reported in one LRD \citep{GlimmIr}, out of the dozens with multi-epoch spectroscopy. Year-scale \halpha~variability is reported in a lensed LRD, but with large uncertainties incurred via systematics \citep{FurtakEtAl2025, JiEtAl2025}. Such evidence points towards a relatively flat structure function of \halpha\ flux variability in LRDs, as also seen in the continuum variability for local LRDs \citep{BurkeEtAl2026}, discrepant with the variability properties and physical interpretation in normal AGNs. We further discuss the implications of the population-wide variability properties of LRDs in Section~\ref{sec:discussion}.

\subsection{Photometric Variability}

Photometric variability measurements for NEXUS data with robust difference imaging methods \citep[i.e.,][]{StoneEtAl2025a} have only been performed for the first three Deep imaging epochs. Only one BLAGN, NX27038 ($z=2.5$), has had a $3\sigma$ variability detection within these three epochs. Our current data is only able to constrain the majority of LRD and high-$z$ BLAGN variability to the photometric noise floor ($\sim2\%$). Future long-baseline NEXUS difference image light curves will place greater constraints on LRD/BLAGN continuum variability.

\begin{figure*}
    \centering
    \includegraphics[width=\textwidth]{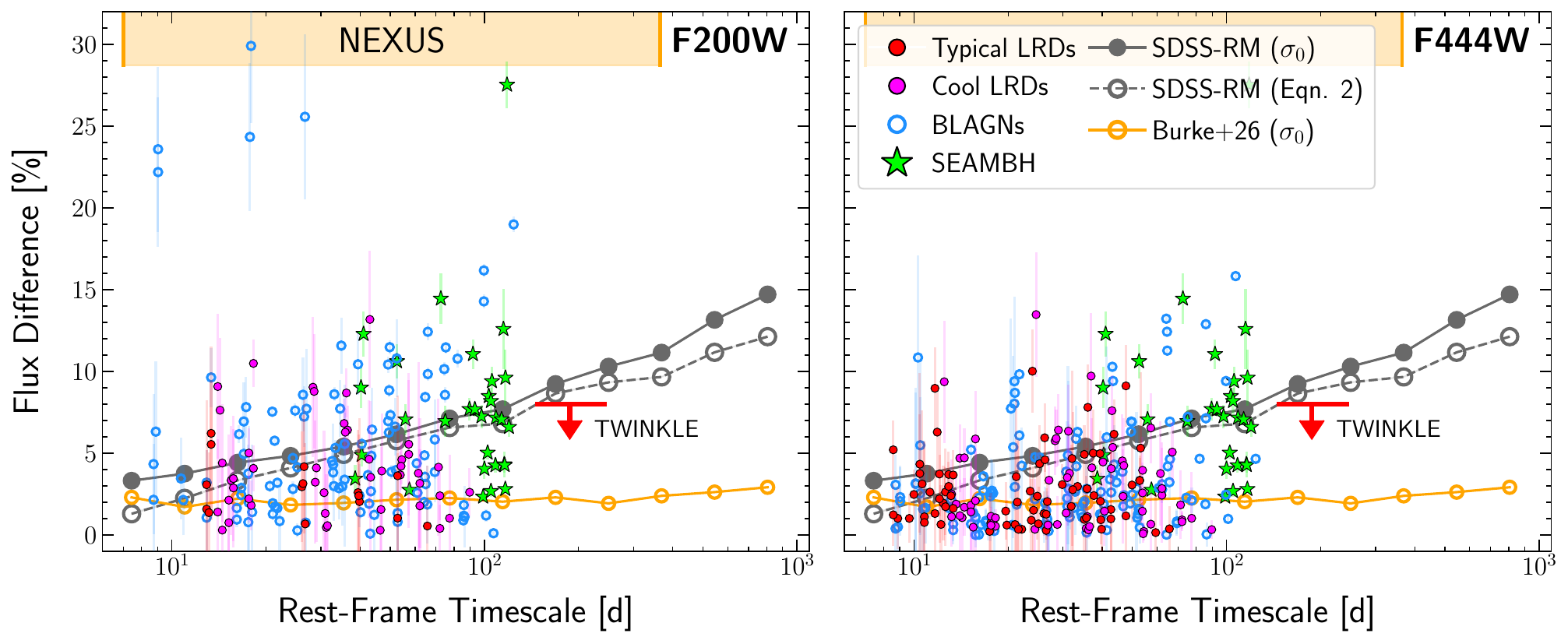}
    \caption{Same as Fig.~\ref{fig:timescale}, but using aperture photometry from F200W (\emph{left}) and F444W (\emph{right}) for JWST LRDs and blue AGNs. The open gray circles and shaded regions are for the low-redshift SDSS-RM comparison sample (sampling rest-frame continuum over $3900 - 5600 \, \rm \AA$). Filled gray circles show the $\sigma_0$-based SDSS-RM structure function. Orange points represent the ensemble $\sigma_0$-based SF using the local LRD analogs from \citet{BurkeEtAl2026}. Green stars display the continuum flux variability amplitudes of SEAMBH super-Eddington AGNs \citep{LuEtAl2019}. The upper-limit result from \citet{LiuEtAl2026} (paraphrased -- see Section~\ref{subsec:agn_var}) is shown in red. }
    \label{fig:timescale_cont}
\end{figure*}

Using the F200W and F444W aperture photometry available for all six MSA epochs, we perform the same analysis as described in Section~\ref{subsec:agn_var}, shown in Fig.~\ref{fig:timescale_cont} and Table~\ref{tab:ks}. While these measurements may present artificial variability from aperture photometry systematics, we obtain the most stringent estimates by again restricting to photometry with $< 5\%$ uncertainty. Indeed, a few BLAGNs exhibit particularly high variability $> 15\%$ in F200W, likely due to imaging systematics. We list these as tentative variability estimates in place of the upcoming difference imaging analyses. Generally, we find little-to-no LRD photometric continuum variability on the timescales probed by NEXUS ($|\Delta F| \lesssim 10\%$ with a sample-mean $\sigma_{0,\rm lc}\lesssim 3\%$). For comparison, we utilize the sample of super-Eddington AGNs from the SEAMBH campaign \citep{LuEtAl2019} with variability amplitude measurements. To compare with our sample, we convert their measured 5100$\rm \AA$ luminosity to the bolometric luminosity, using the luminosity-dependent bolometric correction ($\sim20$) from \citet{Netzer2019}. We then convert this to a broad \halpha\, luminosity using a factor of 130 from \citet{SternLaor2012}. Using a median broad-to-total \halpha\, ratio of 60\% from \citet{TaylorEtAl2025a}, we convert these broad \halpha~luminosities to total luminosities, and restrict the sample to sources with $41 < \log_{10}(L_{\rm H\alpha}) < 44$ and light curves with rest-frame temporal baselines $< 100$ days. This leaves a sample of 30 local super-Eddington sources. 

In addition, we utilize the SDSS-RM continuum light curves to construct an uncertainty-corrected SF for the comparison SDSS-RM sample, originating at rest-frame $3900 - 5600 \, \rm \AA$, blueward of the window probed by the NEXUS sample in F444W ($0.60 - 1.85 \, \mu \rm m$). For comparison, we assemble the $\sigma_0$-based intrinsic SF. Similar to the \halpha~SFs, the two are roughly consistent with each other. We also construct a SF using the \emph{gri} Zwicky Transient Facility \citep[ZTF;][]{ZTF} light curves for three local LRD analogs in \citet{BurkeEtAl2026}. We know a priori that the intrinsic variability is low ($\sigma_0 < 0.01$ mag), so we opt to use the intrinsic $\sigma_0$-based SF rather than the traditional one. The resulting SF for the local LRD analogs is relatively flat and low-amplitude ($\sim 3\%$) across all timescales, as opposed to the SDSS-RM SF which increases with timescale. The sample-wide $\sigma_{0, \rm lc}$ for these local LRD analogs across rest-frame timescales $< 100$d displays similar results (2.23\%), although well below the flux uncertainties. However, a recent study using ZTF light curves for additional local LRD analogs \citep{LinR_2026} suggests that some local LRDs may display similar continuum variability behaviors as normal AGNs.

Our constraints on continuum variability are much stricter than those for \halpha~due to the number of flux pairs. In F200W, we find $\sigma_{0, \rm lc}=2.70\%$ for LRDs and $\sigma_{0, \rm lc}=5.17\%$ for BLAGNs. We note that only one typical LRD has multi-epoch F200W coverage (NX34350), and cite F444W for significant LRD constraints. In F444W, LRDs (BLAGNs) display $\sigma_{0, \rm lc}=$1.78\% (3.18\%). The intrinsic photometric variability of LRDs and BLAGNs are generally consistent with each other, considering the comparable measurement uncertainties. The range of variability observed in our LRDs/BLAGNs is similar to that observed in local super-Eddington AGNs of $\sim$[3, 13]\%, although the definitions of variability are slightly different. Using the wavelength ranges probed by NEXUS and SDSS-RM and the relation between wavelength and variability amplitude from \citet{StoneEtAl2022b}, we estimate that NEXUS objects should exhibit $\sim$4\% additional continuum variability in the SDSS-RM wavelength range than the NEXUS wavelength range in F444W. After accounting for this wavelength difference, our NEXUS results are consistent with the SDSS-RM structure function and SEAMBH estimates. While the sample-wide $\sigma_{0, \rm lc}$ in F444W is higher for the SDSS-RM sample (5.72\%) than the LRDs and BLAGNs, after scaling the NEXUS estimate to the same rest-frame wavelengths probed by SDSS-RM, this difference becomes marginal ($< 3\sigma$).


\section{Discussion}\label{sec:discussion}

\begin{deluxetable*}{c|l|cccccc}
\tablecaption{Variability Statistics\label{tab:ks}}
\tablehead{
\colhead{} & \colhead{Sample} & \colhead{$N_{\rm obj}$} & \colhead{$N_{\rm pair}$} & \colhead{$\sigma_{\rm white \, noise}$ [\%]} & \colhead{$\sigma_{0, \rm lc}$ [\%]} \\
\multicolumn{1}{c}{} & \multicolumn{1}{c}{(1)} & (2) & (3) & (4) & (5)
}
    \startdata
        \multirow{7}{*}{\halpha} & All LRDs & 12 & 26 & 2.07 & $4.02_{-0.51}^{+0.59}$ \\[2pt]
        & Typical LRDs & 7 & 14 & 2.17 & $3.62_{-0.91}^{+1.17}$ \\[2pt]
        & Cool LRDs & 5 & 12 & 1.94 & $4.42_{-0.77}^{+0.96}$ \\[2pt]
        & All BLAGNs & 11 & 49 & 1.72 & $15.06_{-1.44}^{+1.71}$ \\[2pt]
        & ``Variable" BLAGNs & 6 & 35 & 1.62 & $17.48_{-1.94}^{+2.37}$ \\[2pt]
        & ``Non-variable" BLAGNs & 5 & 14 & 2.25 & $6.16_{-1.05}^{+1.43}$ \\[2pt]
        & SDSS-RM & 51 & $10^4$ & 2.88 & $6.29^{+0.02}_{-0.02}$ \\[2pt] 
        \hline
        \multirow{5}{*}{F200W} & All LRDs & 5 & 61 & 1.95 & $2.70_{-0.33}^{+0.38}$ \\[2pt]
        & Typical LRDs & 1 & 15 & 2.42 & $0.00_{-0.00}^{+1.10}$ \\[2pt]
        & Cool LRDs & 4 & 46 & 1.79 & $2.98_{-0.39}^{+0.46}$ \\[2pt]
        & BLAGNs & 8 & 92 & 1.10 & $5.17_{-0.38}^{+0.43}$ \\[2pt]
        & SDSS-RM & 55 & $10^6$ & 6.59 & $5.72^{+0.00}_{-0.00}$ \\[2pt]
        & Local LRDs & 3 & $10^4$ & 7.42 & $2.25^{+0.05}_{-0.05}$ \\[2pt]
        \hline
        \multirow{4}{*}{F444W} & All LRDs & 16 & 190 & 1.72 & $1.78_{-0.19}^{+0.21}$ \\[2pt]
        & Typical LRDs & 11 & 129 & 2.00 & $1.85_{-0.25}^{+0.27}$ \\[2pt]
        & Cool LRDs & 5 & 61 & 1.12 & $1.66_{-0.30}^{+0.33}$ \\[2pt]
        & BLAGNs & 12 & 124 & 1.05 & $3.18_{-0.20}^{+0.22}$ \\[2pt]
        & SDSS-RM & 55 & $10^6$ & 6.59 & $5.72^{+0.00}_{-0.00}$ \\[2pt]
        & Local LRDs & 3 & $10^4$ & 7.42 & $2.25^{+0.05}_{-0.05}$ \\[2pt]
    \enddata
\vspace{5pt}
(1): Sample used. (2): Number of objects in the sample. (3): The total number of flux difference pairs. (4): Average flux measurement uncertainty across the sample. (5): Sample-wide intrinsic rms flux variability (upper limits) relative to a constant mean flux, obtained from the flux difference distributions. We input the flux difference distributions duplicated with negated flux differences into the algorithm from \citet{ShenEtAl2019}, then divide by $\sqrt{2}$.
\end{deluxetable*}

\subsection{LRD Models Predict Little Variability}



We now discuss the implications of our variability measurements in the context of recent LRD models that feature a dense gas envelop surrounding a central accreting black hole. The origin of this dense gas region differs model-to-model. For a BH*, this dense gas medium is located outside of the ionized gas and accretion disk surrounding the SMBH, reaching $\sim 1000$AU in extent \citep{DeGraaffEtAl2025a}. Resonant scattering, electron scattering, and collisional de-excitation of Balmer photons produce large Balmer decrements, Balmer absorption, and large Balmer breaks - especially in BH*s, which contain a significant amount of $n = 2$ state Hydrogen \citep[e.g.,][]{NaiduEtAl2025, SneppenEtAl2026, BegelmanDexter2026, PacucciEtAl2026, ChangEtAl2025}. On the other hand, super-Eddington accretion in AGNs is believed to produce a slim disk, with a funnel of gas shielding much of the emission from the central engine, especially when this funnel intercepts most of our line of sight \citep[e.g.,][]{MadauMaiolino2026, MadauEtAl2026}. Both the dense gas funnel and perhaps the surrounding dusty torus (if existent) suppress high ionization emission from the continuum (i.e., UV), while preserving the strong, broad Hydrogen emission lines, shown in photoionization calculations with moderate covering factors \citep{MadauMaiolino2026}. Recent analytical works based on MHD simulations in super-Eddington accretion SMBHs also predict the formation of a photosphere that produces much of the observed optical continuum in LRDs \citep[e.g.,][]{LiuEtAl2025b, Liu_etal_2026}. 

However, the emission variability of this region surrounding the SMBH remains unclear. Recent analyses have shown that LRD photometric variability has much lower amplitude than typical AGNs \citep{KokuboHarikane2025, ZhangEtAl2025c, TeeEtAl2025, BrazziniEtAl2026,StoneEtAl2025a}, also supported by this work. Super-Eddington AGNs vary less than normal local quasars \citep[e.g.,][]{LuEtAl2019}, a consequence of variability amplitude decreasing with accretion rate \citep{Ai_etal_2010,MacLeodEtAl2010,LuEtAl2019, SimmEtAl2016, RakshitStalin2017, SanchezSaezEtAl2018}. Observations of local super-Eddington quasars find lower variability amplitudes in \hbeta\, than in the continuum \citep[by $\gtrsim 60\%$;][]{HuEtAl2021}, with broad-line variability $\lesssim 10\%$ \citep{DuEtAl2018}. Recent super-Eddington model studies focused on comparison with LRDs and JWST-discovered BLAGNs suggest low continuum variability amplitudes -- $\lesssim 10\%$ \citep{InayoshiEtAl2025a} or $\sim$0.06 mag \citep{SecundaEtAl2025b} -- and longer continuum variability timescales \citep[$\lesssim$ years;][]{KidoEtAl2025, SecundaEtAl2025b}. These constraints are also present within the current NEXUS continuum light curves.

In normal AGNs, the broad-line region reprocesses incident ionizing radiation from the central source, echoing its variability. Reverberation mapping campaigns show \halpha\, response times of months--years, with transfer function widths $\sim$tens of days \citep[e.g.,][]{GrierEtAl2017, ShenEtAl2024_rm, StoneEtAl2025b}. This reprocessed variability will be lower-amplitude and longer-timescale than that produced by the central engine, making the line variability more challenging to detect. \citet{SecundaEtAl2025b} propose that multi-epoch spectroscopy with a 3-month cadence will be able to detect LRD broad spectroscopic variability in both super-Eddington sources and sub-Eddington sources, assuming a sensitivity of 15\% in broad-line flux uncertainties. NEXUS reaches better flux sensitivity and cadence for the variability subsamples, but we do not detect such variability echoing a normal AGN driving ionizing flux for the LRD population over the current NEXUS baseline.

Most other considered models for LRDs predict very little variability, if any at all, that can be observed over the course of the NEXUS survey. Cocooned SMBHs with thick dense gas envelopes (e.g., quasi-stars, BH*s) will experience pulsations similar to Cepheids on timescales of $\sim$decades in the LRD rest-frame \citep{CantielloEtAl2025}. In the quasi-star scenario, as the system evolves, these pulsations will become more erratic and occur on shorter timescales. Direct-collapse black holes are predicted to experience continuum variability on similar timescales due to competing accretion and radiative feedback within the system \citep{PacucciEtAl2026}. However, in this scenario, most line emission originates below the photosphere, causing the broad-line components to primarily emerge from scattering \citep{BegelmanDexter2026}. Hydrodynamical simulations within cocooned systems predict variability on $\sim$year timescales, appearing mostly in the cores of emission lines, not the broad component \citep{SneppenEtAl2026}.


Sample-wide intrinsic variability estimates imply a striking ($>3\sigma$) difference between high-$z$ LRDs and local quasars. Additionally, because our measurements are upper limits, these estimates are likely more disparate. Only one LRD in our sample tentatively presents significant \halpha\, variability: NX7607 (but see Section~\ref{subsec:15499}). Both our variability results and others over rest-frame timescales extending to $\sim$decades \citep{BurkeEtAl2026, LiuEtAl2026} reveal a consistency with a white-noise pattern in both continuum and \halpha~measurements for LRDs and high-$z$ BLAGNs, if not at-most low-amplitude ($\lesssim 5\%$) intrinsic variability.



Current studies suggest variability will be difficult to measure in high-$z$ LRDs with current temporal baselines. \citet{FurtakEtAl2025} find only marginal $\sim$10\% broad-line variability in the lensed LRD A2744-QSO1 over rest-frame $\sim$years \citep[although it may be up to $\sim$50\%;][]{JiEtAl2025}, consistent with local quasar variability (Fig.~\ref{fig:timescale}). Significant $\sim$0.5 mag continuum variability is measured in the lensed R2211-RX1 over a rest-frame century \citep{ZhangEtAl2025d}, similar to timescales of pulsational instabilities in quasi/supermassive stars \citep{CantielloEtAl2025}. Assuming a rest-frame period of 41 yr \citep{ZhangEtAl2026} and an amplitude of 0.5 mag in F444W, R2211-RX1 exhibits $<1\%$ intrinsic variability over rest-frame years, consistent with NEXUS measurements over the same timescales.  If short-term variability is present within LRDs, it is exceedingly rare, most likely $\sim$a few percent of the total population at best \citep{ZhangEtAl2025c}. Indeed, only one case of short-term spectroscopic variability has been measured \citep{GlimmIr}, albeit with significant systematic uncertainty. Local LRD analogs present a unique avenue for probing long-timescale variability, with searches having been carried out recently utilizing numerous datasets \citep{LinEtAl2026b, BurkeEtAl2026, ParkEtAl2026, CaseyEtAl2026, LinEtAl2026}. However, the majority of these samples display little-to-no variability over decades \citep[$\sigma_0 \lesssim 0.01$ mag;][]{BurkeEtAl2026}, with only one source displaying $\sim$10\% variability \citep{LinEtAl2026b}. A viable path forward for LRD variability studies involves building large samples of local and high-$z$ LRDs with multi-epoch spectroscopy and imaging. While lensed systems are increasingly important to eke out long-term variability, samples of local LRD analogs with decades-long baselines are equally as useful. 

\begin{figure*}
    \centering
    \includegraphics[width=\textwidth]{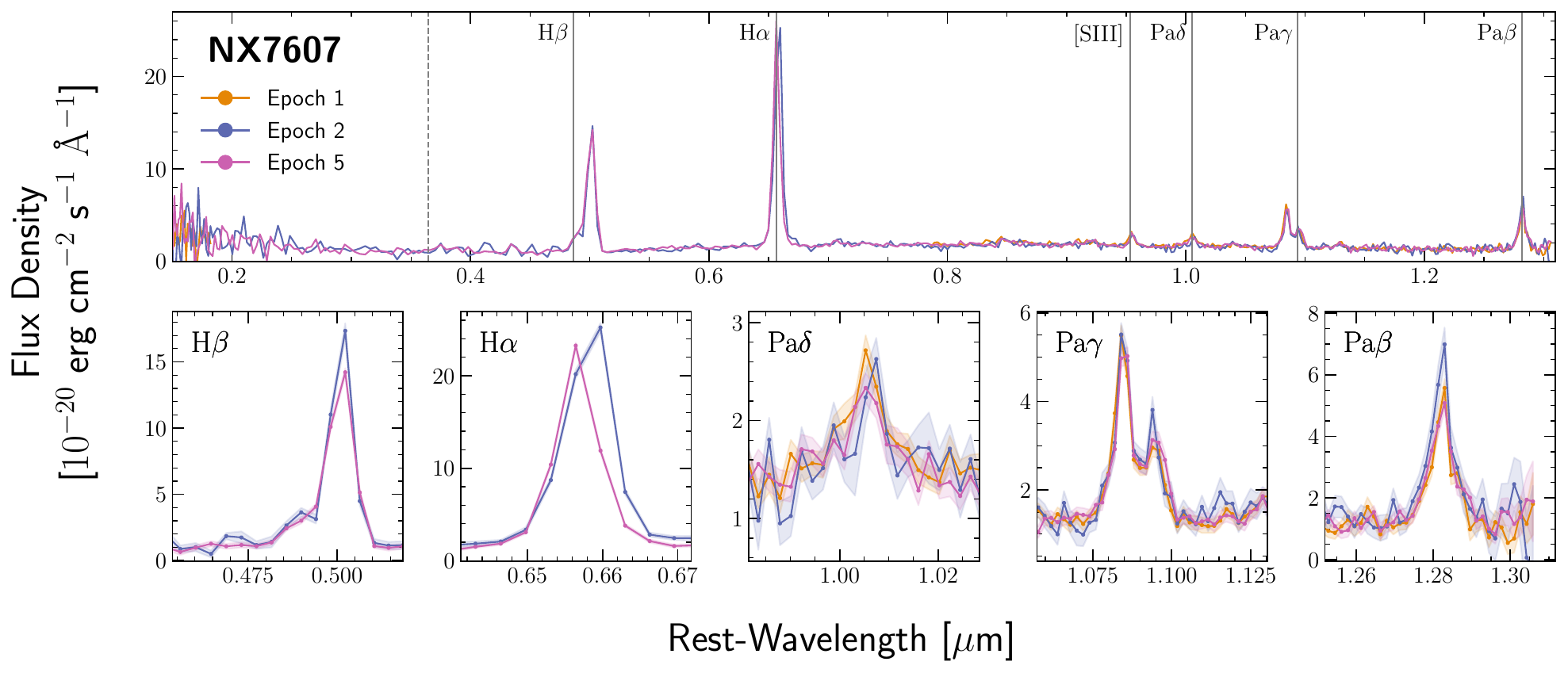}
    \caption{An overview of the cool LRD NX7607 ($z=3.2$). \emph{\ul{Top:}} The processed MSA prism spectra for each observed epoch. Prominent emission lines are indicated with vertical lines, and the Balmer break is shown as a dashed line. \emph{\ul{Bottom:}} Normalized emission line profiles for several emission line complexes. }
    \label{fig:NX7607}
\end{figure*}

\subsection{``Cool" LRDs}\label{subsec:15499}

Half of our LRD sample are defined as ``cool" LRDs, displaying UV upturns noncoincident with the Balmer break, and features mostly consistent with a cooler blackbody component in the rest-frame optical ($T \lesssim$ 3000K). This is somewhat common in JWST LRD/BLAGN samples -- \citet{SettonEtAl2025} find that only 50\% of their broad \halpha\, emitters have the UV upturn occur at the Balmer break. Our cool LRDs probe the low end of blackbody temperatures present in current LRD literature \citep[e.g.,][]{DeGraaffEtAl2025}, similar to five compiled in the low-$z$ DESI LRD sample \citep{LinEtAl2026b}. Using properties from \citet{PanEtAl2026}, these cool LRDs exhibit weaker Balmer breaks and smaller blackbody fractions (i.e., the fraction of the SED attributed to the blackbody) than the rest of the few dozen spectroscopically confirmed NEXUS LRDs. This agrees with correlations found in prior studies \citep{DeGraaffEtAl2025, PanEtAl2026, LinEtAl2026b}. It should be noted that all of the cool LRDs lie at lower redshifts $z \lesssim 4$ than the more typical LRDs, giving credence to cool LRDs being an intermediate stage between BLAGNs and LRDs.

In our sample, these cool LRDs exhibit somewhat larger Balmer decrements than JWST BLAGNs, but consistent with the typical LRDs in our sample. NX15499, however, exhibits a significantly lower Balmer decrement, suggestive of a possible reddened AGN. Interestingly, the cool LRDs display enhanced \pbeta/\halpha\, relative to the BLAGNs ($\sim$0.25), indicative of the higher equivalent widths seen in prior JWST work \citep{MattheeEtAl2024}. NX35791 displays \pbeta/\halpha\, consistent with extinction laws, but NX7607 and NX15499 do not. This suggests that NX35791 may be a transitionary object between BLAGNs and LRDs, or even a heavily reddened AGN. The agreement of other line ratios with extinction laws is less clear and varies by object. Previous works have attempted to reproduce LRD features with a reddened AGN \citep[e.g.,][]{SettonEtAl2025, StepneyEtAl2026}; these models will require more scrutiny with respect to the variety of sources seen within the LRD population.




Generally, the cool LRDs display intrinsic \halpha~variability consistent with the typical LRDs, less than the BLAGNs. One particular cool LRD (NX7607) displays significant $\sim$30\% variability over a rest-frame timescale of $\sim40$ days, well above $3\sigma$ flux uncertainties and the mean SDSS-RM variability. This object has a moderate Balmer decrement, and a shallower Balmer break (1.08), with a low blackbody temperature ($T=3210$). NX25108 and NX15499 also exhibit 10\% variability. Notably, one local LRD from \citet{LinEtAl2026b} with an upturn $\sim$1.1 $\mu$m and $T=2500$K displays a significant ($\sim 10\%$) dimming in the infrared over the course of nearly a rest-frame decade. 

We may be viewing a transitionary stage between LRDs and BLAGNs, where significant motion of dense gas causes significant broad/narrow \halpha\, variability. For example, \citet{MeridaEtAl2026} reported evidence for an evolution of blackbody temperature and stellar mass with redshift, suggesting LRDs cool overtime and become subdominant to stellar emission. Theories postulating the difference between LRDs and BLAGNs results from the line-of-sight gas content \citep[e.g.,][]{LinEtAl2026, ChenEtAl2026, SneppenEtAl2026c} will produce sub- or super-Eddington AGN \halpha\, variability, where the AGN's emission leaks through the poles of the system. In this case, the continuum and BLR emission would resemble SDSS-RM or SEAMBH variability; correlations between \halpha\, and UV/optical continuum emission complicate this interpretation \citep[e.g.,][]{AsadaEtAl2026, PangEtAl2026, ChenEtAl2026}. The answer may be in the middle -- \citet{PangEtAl2026} suggest a BH*-like model where the covering factor of the gas shell and temperature decrease with radius, housing a photoionized BLR. The exact geometry of these non-spherical systems is still in debate and will require future analysis.

While we cannot determine the precise line shifts of \halpha\, due to MSA calibration issues, observed profile changes will illuminate the origin of the variability. An observed blueward shift, such as seen in LRD Balmer absorption, suggests outflow reminiscient of Type IIn supernovae \citep{NaiduEtAl2026} and luminous blue variables \citep[LBVs;][]{Smith2013}, with recurring ``eruptions" occurring on timescales $\sim$days--centuries \citep{Smith2026}. As the BH* cools and its envelope expands, pulsations in the photosphere will become more erratic and outflows will begin to significantly affect the dense gas shell. Assuming opacity-driven pulsational instability expected in BH*s \citep[e.g.,][]{CantielloEtAl2025}, we estimate the dynamical time of the gas envelope. Following \citet{ZhangEtAl2025c}, we obtain a typical rest-frame timescale of 571 yr, assuming a $10^8 M_{\odot}$ SMBH and a modest Eddington ratio $\lambda_{\rm Edd} = 0.5$ -- requiring lensed systems for proper temporal sampling. Observing this variability for NX7607 ($z=3.2$) over the course of years would require a very low accretion rate ($\lambda_{\rm Edd} \sim 10^{-4}$). Super-Eddington outflows are a proposed origin for the observed blueshifted Balmer absorption in LRDs \citep{MadauEtAl2026}, seen in simulations of BH accretion \citep[e.g.,][]{JiangEtAl2014, SkadowskiEtAl2015, HuEtAl2022}. However, local super-Eddington AGNs display minimal ($\lesssim 10\%$) broad \hbeta\, variability \citep{DuEtAl2018}, similar to our noise floor.

We show NX7607's full MSA spectrum for all epochs in Fig.~\ref{fig:NX7607}. None of the its emission lines exhibit significant variability, except for \halpha. Some narrow-line variability is seen in \OIII, and slight broad-line variability is seen in \pbeta. LRD \OIII\, variability was recently reported in \citet{GlimmIr}, suggesting that the narrow-line region may also echo the central engine in an AGN scenario. If this short-term variability is real, it indicates a disparate origin or lack of correlation between the emission lines and the continuum, as the continuum has shown to vary little. Interestingly, the model introduced in \citet{PangEtAl2026} predict a lack of correlation between spectroscopic and continuum variability, with both occurring on short ($\sim$tens of days) timescales. This model also predicts an SED with a broader-than-expected blackbody feature, exactly as quantified in \citet{DeGraaffEtAl2025} for NX7607. However, our estimates may be affected by underestimated uncertainty and improper normalization. If the wavelength calibration is egregious enough, it may also affect total flux measurements. We therefore caution interpretation of significant \halpha\, variability in NX7607.

\section{Conclusion}\label{sec:conclusion}

We perform a spectral analysis for 31 broad-line objects (17 LRDs, 14 BLAGNs) at $2 \lesssim z \lesssim 6$ monitored within the NEXUS survey, each observed with JWST NIRSpec MSA bi-monthly for a maximum of 6 spectra over the course of one year. We focus our analysis on \halpha~and continuum variability, and emission line ratio diagnostics for both the LRDs and BLAGNs in comparison to a sample of local SDSS-RM AGNs with similar $L_{\rm H\alpha}$. Our main results are as follows:

\begin{itemize}
    \item Our BLAGNs exhibit Balmer decrements similar to or larger than the local sample of SDSS-RM AGNs. LRDs display Balmer decrements ($\sim 7.6$) on average higher than the BLAGNs and the local AGNs ($\sim 3.2$). 
    
    \item Utilizing other line ratios, our sample is somewhat consistent with local extinction laws, with the LRDs experiencing higher ``extinction" than the BLAGNs. Assuming an SMC extinction law, BLAGNs show $A_V$ similar to the SDSS-RM sample, while LRDs have larger $A_V \gtrsim 3$. These results are under the assumption of dust reddening in LRDs to interpret the observed rest-optical colors, which is likely incorrect. 
    
    \item Our LRDs and BLAGNs display intrinsic rms \halpha~flux variability of 4.02\% and 6.16\% on rest-frame timescales $\lesssim$100 days, though there are a population of JWST BLAGNs that are substantially more variable in \halpha\ flux. We consider these variability constraints upper limits given potential residual systematics in the flux measurements.
    
    
    

    \item Using accompanying aperture photometry in F200W and F444W, we constrain sample-wide rms intrinsic continuum variability for both our BLAGNs and LRDs to $\lesssim 3\%$ over rest-frame $\sim$months.

    \item Sample-wide measurements on $< 100$d rest-frame timescales show a $>3\sigma$ suppression of LRD broad-line variability relative to low-redshift SDSS-RM AGNs. High-redshift BLAGNs exhibit sample-wide broad-line variability consistent with these local AGNs.


    \item Combining with prior results over rest-frame timescales of $\sim$6 months \citep{LiuEtAl2026} to decades \citep{BurkeEtAl2026}, LRD variability (both \halpha\ and continuum) displays a white-noise pattern over all observed timescales, inconsistent with the structure functions for normal AGNs.


\end{itemize}


The variability measurements for LRDs from this work provide further insights on theoretical models of this enigmatic population. Combined with recent literature studies, LRDs as a population have weaker variability than normal AGNs across similar wavelengths and timescales. The average variability structure functions for the continuum and \halpha\ emission also deviate from the typical red-noise pattern of normal AGN variability. The origin of broad \halpha\ emission in LRDs is still debated. If \halpha\ is driven by the ionizing flux from the central engine in LRDs, the flat structure function of \halpha\ variability suggests different properties of the ionizing flux from central accretion. This is because while radiative transfer can modify the strengths of Balmer lines, it preserves the variability pattern from the driving light curve. Alternatively, if \halpha\ emission is mostly produced in the warm layer of the dense gas envelop surrounding the central BH, with major contributions from collisional excitation, then we would expect different variability patterns compared with \halpha\ variability in normal AGNs. Either way, the variability properties of LRDs are markedly different from those of normal AGNs, indicative of separate production mechanisms of the emission.  

Individual LRDs may still show significant variability consistent with normal AGN behaviors, as suggested in some recent studies \citep{ZhangEtAl2025c,FurtakEtAl2025,GlimmIr,LinR_2026}. If these reported variability incidents are real, they would provide valuable constraints on the diversity of LRDs. For example, some of these variable LRDs may be transitioning to normal AGNs or with different viewing angles. Nevertheless, we caution on two possible caveats on the reported variability detection. First, it is easy to produce apparent variability given nearly ubiquitous systematics (on the other hand, eliminating variability by systematics is much harder, if not impossible). Such systematics inclde improper photometry (e.g., aperture flux variations), instrumental systematics, or simply being statistical outliers. Second, not all ``LRDs'' are correctly classified or spectroscopically confirmed, which would require the full SED coverage to remove, e.g., contaminants from heavily dust-reddened normal AGNs. Given these considerations, continued monitoring of genuine LRDs with larger sample sizes and better systematics control would be necessary to robustly measure the variability properties of LRDs. Future NEXUS data, as well as other observing programs targeting high-z LRDs and their low-$z$ analogs, will provide crucial constraints on LRD variability, and thus the nature of this population.

\begin{acknowledgments}
Based on observations with the NASA/ESA/CSA James Webb Space Telescope obtained from the Barbara A. Mikulski Archive at the Space Telescope Science Institute, which is operated by the Association of Universities for Research in Astronomy, Incorporated, under NASA contract NAS5-03127. Support for Program number JWST-GO-05105 was provided through a grant from the STScI under NASA contract NAS5-03127. Some of the data products presented were retrieved from the Dawn JWST Archive (DJA). DJA is an initiative of the Cosmic Dawn Center (DAWN), which is funded by the Danish National Research Foundation under grant DNRF140.
\end{acknowledgments}





%
\facilities{JWST (NIRSpec/NIRCam)}

\software{ 
{\tt Astropy} \citep{2013A&A...558A..33A,2018AJ....156..123A,2022ApJ...935..167A},
{\tt emcee} \citep{emcee},
{\tt dust\_extinction} \citep{Gordon2024},
{\tt LMFIT} \citep{NewvilleEtAl2025},
{\tt Matplotlib} \citep{Hunter2007b},
{\tt Numba} \citep{LamEtAl2015a},
{\tt NumPy} \citep{HarrisEtAl2020b},
{\tt SciPy} \citep{scipy},
{\tt Seaborn} \citep{Waskom2021}
}






\bibliography{sample701, refs}{}

\begin{thebibliography}{}
\expandafter\ifx\csname natexlab\endcsname\relax\def\natexlab#1{#1}\fi
\providecommand{\url}[1]{\href{#1}{#1}}
\providecommand{\dodoi}[1]{doi:~\href{http://doi.org/#1}{\nolinkurl{#1}}}
\providecommand{\doeprint}[1]{\href{http://ascl.net/#1}{\nolinkurl{http://ascl.net/#1}}}
\providecommand{\doarXiv}[1]{\href{https://arxiv.org/abs/#1}{\nolinkurl{https://arxiv.org/abs/#1}}}

\bibitem[{Y.~L. {Ai} {et~al.}(2010){Ai}, {Yuan}, {Zhou}, {Wang}, {Dong}, {Wang}, \& {Lu}}]{Ai_etal_2010}
{Ai}, Y.~L., {Yuan}, W., {Zhou}, H.~Y., {et~al.} 2010, \bibinfo{title}{{Dependence of the Optical/Ultraviolet Variability on the Emission-line Properties and Eddington Ratio in Active Galactic Nuclei},} \apjl, 716, L31, \dodoi{10.1088/2041-8205/716/1/L31}

\bibitem[{H.~B. Akins {et~al.}(2025)Akins, Casey, Lambrides, Allen, Andika, Brinch, Champagne, Cooper, Ding, Drakos, Faisst, Finkelstein, Franco, Fujimoto, Gentile, Gillman, Gozaliasl, Harish, Hayward, Hirschmann, Ilbert, Kartaltepe, Kocevski, Koekemoer, Kokorev, Liu, Long, McCracken, McKinney, Onoue, Paquereau, Renzini, Rhodes, Robertson, Shuntov, Silverman, Tanaka, Toft, Trakhtenbrot, Valentino, \& Zavala}]{AkinsEtAl2025}
Akins, H.~B., Casey, C.~M., Lambrides, E., {et~al.} 2025, \bibinfo{title}{{{COSMOS-Web}}: {{The Overabundance}} and {{Physical Nature}} of "{{Little Red Dots}}"---{{Implications}} for {{Early Galaxy}} and {{SMBH Assembly}},} The Astrophysical Journal, 991, 37, \dodoi{10.3847/1538-4357/ade984}

\bibitem[{T.~T. Ananna {et~al.}(2024)Ananna, Bogd{\'a}n, Kov{\'a}cs, Natarajan, \& Hickox}]{AnannaEtAl2024}
Ananna, T.~T., Bogd{\'a}n, {\'A}., Kov{\'a}cs, O.~E., Natarajan, P., \& Hickox, R.~C. 2024, \bibinfo{title}{X-{{Ray View}} of {{Little Red Dots}}: {{Do They Host Supermassive Black Holes}}?} The Astrophysical Journal Letters, 969, L18, \dodoi{10.3847/2041-8213/ad5669}

\bibitem[{Y. Asada {et~al.}(2026)Asada, Inayoshi, Fei, Fujimoto, \& Willott}]{AsadaEtAl2026}
Asada, Y., Inayoshi, K., Fei, Q., Fujimoto, S., \& Willott, C. 2026, Origins of the {{UV}} Continuum and {{Balmer}} Emission Lines in {{Little Red Dots}}: Observational Validation of Dense Gas Envelope Models Enshrouding the {{AGN}}, arXiv

\bibitem[{ {Astropy Collaboration} {et~al.}(2013){Astropy Collaboration}, {Robitaille}, {Tollerud}, {Greenfield}, {Droettboom}, {Bray}, {Aldcroft}, {Davis}, {Ginsburg}, {Price-Whelan}, {Kerzendorf}, {Conley}, {Crighton}, {Barbary}, {Muna}, {Ferguson}, {Grollier}, {Parikh}, {Nair}, {Unther}, {Deil}, {Woillez}, {Conseil}, {Kramer}, {Turner}, {Singer}, {Fox}, {Weaver}, {Zabalza}, {Edwards}, {Azalee Bostroem}, {Burke}, {Casey}, {Crawford}, {Dencheva}, {Ely}, {Jenness}, {Labrie}, {Lim}, {Pierfederici}, {Pontzen}, {Ptak}, {Refsdal}, {Servillat}, \& {Streicher}}]{2013A&A...558A..33A}
{Astropy Collaboration}, {Robitaille}, T.~P., {Tollerud}, E.~J., {et~al.} 2013, \bibinfo{title}{{Astropy: A community Python package for astronomy},} \aap, 558, A33, \dodoi{10.1051/0004-6361/201322068}

\bibitem[{ {Astropy Collaboration} {et~al.}(2018){Astropy Collaboration}, {Price-Whelan}, {Sip{\H{o}}cz}, {G{\"u}nther}, {Lim}, {Crawford}, {Conseil}, {Shupe}, {Craig}, {Dencheva}, {Ginsburg}, {VanderPlas}, {Bradley}, {P{\'e}rez-Su{\'a}rez}, {de Val-Borro}, {Aldcroft}, {Cruz}, {Robitaille}, {Tollerud}, {Ardelean}, {Babej}, {Bach}, {Bachetti}, {Bakanov}, {Bamford}, {Barentsen}, {Barmby}, {Baumbach}, {Berry}, {Biscani}, {Boquien}, {Bostroem}, {Bouma}, {Brammer}, {Bray}, {Breytenbach}, {Buddelmeijer}, {Burke}, {Calderone}, {Cano Rodr{\'\i}guez}, {Cara}, {Cardoso}, {Cheedella}, {Copin}, {Corrales}, {Crichton}, {D'Avella}, {Deil}, {Depagne}, {Dietrich}, {Donath}, {Droettboom}, {Earl}, {Erben}, {Fabbro}, {Ferreira}, {Finethy}, {Fox}, {Garrison}, {Gibbons}, {Goldstein}, {Gommers}, {Greco}, {Greenfield}, {Groener}, {Grollier}, {Hagen}, {Hirst}, {Homeier}, {Horton}, {Hosseinzadeh}, {Hu}, {Hunkeler}, {Ivezi{\'c}}, {Jain}, {Jenness}, {Kanarek}, {Kendrew}, {Kern}, {Kerzendorf}, {Khvalko}, {King}, {Kirkby}, {Kulkarni},
  {Kumar}, {Lee}, {Lenz}, {Littlefair}, {Ma}, {Macleod}, {Mastropietro}, {McCully}, {Montagnac}, {Morris}, {Mueller}, {Mumford}, {Muna}, {Murphy}, {Nelson}, {Nguyen}, {Ninan}, {N{\"o}the}, {Ogaz}, {Oh}, {Parejko}, {Parley}, {Pascual}, {Patil}, {Patil}, {Plunkett}, {Prochaska}, {Rastogi}, {Reddy Janga}, {Sabater}, {Sakurikar}, {Seifert}, {Sherbert}, {Sherwood-Taylor}, {Shih}, {Sick}, {Silbiger}, {Singanamalla}, {Singer}, {Sladen}, {Sooley}, {Sornarajah}, {Streicher}, {Teuben}, {Thomas}, {Tremblay}, {Turner}, {Terr{\'o}n}, {van Kerkwijk}, {de la Vega}, {Watkins}, {Weaver}, {Whitmore}, {Woillez}, {Zabalza}, \& {Astropy Contributors}}]{2018AJ....156..123A}
{Astropy Collaboration}, {Price-Whelan}, A.~M., {Sip{\H{o}}cz}, B.~M., {et~al.} 2018, \bibinfo{title}{{The Astropy Project: Building an Open-science Project and Status of the v2.0 Core Package},} \aj, 156, 123, \dodoi{10.3847/1538-3881/aabc4f}

\bibitem[{ {Astropy Collaboration} {et~al.}(2022){Astropy Collaboration}, {Price-Whelan}, {Lim}, {Earl}, {Starkman}, {Bradley}, {Shupe}, {Patil}, {Corrales}, {Brasseur}, {N{\"o}the}, {Donath}, {Tollerud}, {Morris}, {Ginsburg}, {Vaher}, {Weaver}, {Tocknell}, {Jamieson}, {van Kerkwijk}, {Robitaille}, {Merry}, {Bachetti}, {G{\"u}nther}, {Aldcroft}, {Alvarado-Montes}, {Archibald}, {B{\'o}di}, {Bapat}, {Barentsen}, {Baz{\'a}n}, {Biswas}, {Boquien}, {Burke}, {Cara}, {Cara}, {Conroy}, {Conseil}, {Craig}, {Cross}, {Cruz}, {D'Eugenio}, {Dencheva}, {Devillepoix}, {Dietrich}, {Eigenbrot}, {Erben}, {Ferreira}, {Foreman-Mackey}, {Fox}, {Freij}, {Garg}, {Geda}, {Glattly}, {Gondhalekar}, {Gordon}, {Grant}, {Greenfield}, {Groener}, {Guest}, {Gurovich}, {Handberg}, {Hart}, {Hatfield-Dodds}, {Homeier}, {Hosseinzadeh}, {Jenness}, {Jones}, {Joseph}, {Kalmbach}, {Karamehmetoglu}, {Ka{\l}uszy{\'n}ski}, {Kelley}, {Kern}, {Kerzendorf}, {Koch}, {Kulumani}, {Lee}, {Ly}, {Ma}, {MacBride}, {Maljaars}, {Muna}, {Murphy}, {Norman},
  {O'Steen}, {Oman}, {Pacifici}, {Pascual}, {Pascual-Granado}, {Patil}, {Perren}, {Pickering}, {Rastogi}, {Roulston}, {Ryan}, {Rykoff}, {Sabater}, {Sakurikar}, {Salgado}, {Sanghi}, {Saunders}, {Savchenko}, {Schwardt}, {Seifert-Eckert}, {Shih}, {Jain}, {Shukla}, {Sick}, {Simpson}, {Singanamalla}, {Singer}, {Singhal}, {Sinha}, {Sip{\H{o}}cz}, {Spitler}, {Stansby}, {Streicher}, {{\v{S}}umak}, {Swinbank}, {Taranu}, {Tewary}, {Tremblay}, {de Val-Borro}, {Van Kooten}, {Vasovi{\'c}}, {Verma}, {de Miranda Cardoso}, {Williams}, {Wilson}, {Winkel}, {Wood-Vasey}, {Xue}, {Yoachim}, {Zhang}, {Zonca}, \& {Astropy Project Contributors}}]{2022ApJ...935..167A}
{Astropy Collaboration}, {Price-Whelan}, A.~M., {Lim}, P.~L., {et~al.} 2022, \bibinfo{title}{{The Astropy Project: Sustaining and Growing a Community-oriented Open-source Project and the Latest Major Release (v5.0) of the Core Package},} \apj, 935, 167, \dodoi{10.3847/1538-4357/ac7c74}

\bibitem[{G. Barro {et~al.}(2024)Barro, {P{\'e}rez-Gonz{\'a}lez}, Kocevski, McGrath, Trump, Simons, Somerville, Yung, Arrabal~Haro, Akins, Bagley, Cleri, Costantin, Davis, Dickinson, Finkelstein, Giavalisco, {G{\'o}mez-Guijarro}, Hathi, Hirschmann, Holwerda, {Huertas-Company}, Kartaltepe, Koekemoer, Lucas, Papovich, Pirzkal, Seill{\'e}, Tacchella, Wuyts, Wilkins, {de la Vega}, Yang, \& Zavala}]{BarroEtAl2024}
Barro, G., {P{\'e}rez-Gonz{\'a}lez}, P.~G., Kocevski, D.~D., {et~al.} 2024, \bibinfo{title}{Extremely {{Red Galaxies}} at z = 5--9 with {{MIRI}} and {{NIRSpec}}: {{Dusty Galaxies}} or {{Obscured Active Galactic Nuclei}}?} The Astrophysical Journal, 963, 128, \dodoi{10.3847/1538-4357/ad167e}

\bibitem[{M.~C. Begelman \& J. Dexter(2026)Begelman \& Dexter}]{BegelmanDexter2026}
Begelman, M.~C., \& Dexter, J. 2026, \bibinfo{title}{Little {{Red Dots}} as {{Late-stage Quasi-stars}},} The Astrophysical Journal, 996, 48, \dodoi{10.3847/1538-4357/ae274a}

\bibitem[{E.~C. {Bellm} {et~al.}(2019){Bellm}, {Kulkarni}, {Graham}, {Dekany}, {Smith}, {Riddle}, {Masci}, {Helou}, {Prince}, {Adams}, {Barbarino}, {Barlow}, {Bauer}, {Beck}, {Belicki}, {Biswas}, {Blagorodnova}, {Bodewits}, {Bolin}, {Brinnel}, {Brooke}, {Bue}, {Bulla}, {Burruss}, {Cenko}, {Chang}, {Connolly}, {Coughlin}, {Cromer}, {Cunningham}, {De}, {Delacroix}, {Desai}, {Duev}, {Eadie}, {Farnham}, {Feeney}, {Feindt}, {Flynn}, {Franckowiak}, {Frederick}, {Fremling}, {Gal-Yam}, {Gezari}, {Giomi}, {Goldstein}, {Golkhou}, {Goobar}, {Groom}, {Hacopians}, {Hale}, {Henning}, {Ho}, {Hover}, {Howell}, {Hung}, {Huppenkothen}, {Imel}, {Ip}, {Ivezi{\'c}}, {Jackson}, {Jones}, {Juric}, {Kasliwal}, {Kaspi}, {Kaye}, {Kelley}, {Kowalski}, {Kramer}, {Kupfer}, {Landry}, {Laher}, {Lee}, {Lin}, {Lin}, {Lunnan}, {Giomi}, {Mahabal}, {Mao}, {Miller}, {Monkewitz}, {Murphy}, {Ngeow}, {Nordin}, {Nugent}, {Ofek}, {Patterson}, {Penprase}, {Porter}, {Rauch}, {Rebbapragada}, {Reiley}, {Rigault}, {Rodriguez}, {van Roestel}, {Rusholme},
  {van Santen}, {Schulze}, {Shupe}, {Singer}, {Soumagnac}, {Stein}, {Surace}, {Sollerman}, {Szkody}, {Taddia}, {Terek}, {Van Sistine}, {van Velzen}, {Vestrand}, {Walters}, {Ward}, {Ye}, {Yu}, {Yan}, \& {Zolkower}}]{ZTF}
{Bellm}, E.~C., {Kulkarni}, S.~R., {Graham}, M.~J., {et~al.} 2019, \bibinfo{title}{{The Zwicky Transient Facility: System Overview, Performance, and First Results},} \pasp, 131, 018002, \dodoi{10.1088/1538-3873/aaecbe}

\bibitem[{M.~C. {Bentz} {et~al.}(2013){Bentz}, {Denney}, {Grier}, {Barth}, {Peterson}, {Vestergaard}, {Bennert}, {Canalizo}, {De Rosa}, {Filippenko}, {Gates}, {Greene}, {Li}, {Malkan}, {Pogge}, {Stern}, {Treu}, \& {Woo}}]{BentzEtAl2013}
{Bentz}, M.~C., {Denney}, K.~D., {Grier}, C.~J., {et~al.} 2013, \bibinfo{title}{{The Low-luminosity End of the Radius-Luminosity Relationship for Active Galactic Nuclei},} \apj, 767, 149, \dodoi{10.1088/0004-637X/767/2/149}

\bibitem[{R.~D. Blandford \& C.~F. McKee(1982)Blandford \& McKee}]{BlandfordMcKee1982a}
Blandford, R.~D., \& McKee, C.~F. 1982, \bibinfo{title}{Reverberation Mapping of the Emission Line Regions of {{Seyfert}} Galaxies and Quasars.,} The Astrophysical Journal, 255, 419, \dodoi{10.1086/159843}

\bibitem[{G. {Brammer}(2023){Brammer}}]{msaexp}
{Brammer}, G. 2023, {msaexp: NIRSpec analyis tools}, 0.6.17 Zenodo, \dodoi{10.5281/zenodo.8319596}

\bibitem[{M. {Brazzini} {et~al.}(2026){Brazzini}, {D'Eugenio}, {Maiolino}, {Lyu}, {DeCoursey}, {{\"U}bler}, {Ji}, {Juod{\v{z}}balis}, {Scholtz}, {Jones}, {Hainline}, {Dalla Bont{\`a}}, {{\'e}rez-Gonz{\'a}lez}, {Geris}, {Harshan}, {Feruglio}, {Bischetti}, {Mazzolari}, {Rieke}, {Alberts}, {Trefoloni}, {Carniani}, {Parlanti}, {Marconi}, {Risaliti}, {Ramos Almeida}, {Rinaldi}, {Perna}, {Zamora}, {Lamperti}, {Venturi}, {Cresci}, {Bunker}, \& {Ivey}}]{BrazziniEtAl2026}
{Brazzini}, M., {D'Eugenio}, F., {Maiolino}, R., {et~al.} 2026, \bibinfo{title}{{The Little Blue and Red Dots Rosetta Stones: Non-Gaussian broad lines, hot dust, and X-ray weakness},} arXiv e-prints, arXiv:2601.22214, \dodoi{10.48550/arXiv.2601.22214}

\bibitem[{M. Brooks {et~al.}(2024)Brooks, Simons, Trump, Taylor, Backhaus, Davis, Buat, Cleri, Finkelstein, Hirschmann, Holwerda, Kocevski, Koekemoer, Lucas, Pacucci, \& Seill{\'e}}]{BrooksEtAl2024}
Brooks, M., Simons, R.~C., Trump, J.~R., {et~al.} 2024, Here {{There Be}} ({{Dusty}}) {{Monsters}}: {{High Redshift AGN}} Are {{Dustier Than Their Hosts}}, arXiv, \dodoi{10.48550/arXiv.2410.07340}

\bibitem[{M. Brooks {et~al.}(2025)Brooks, Trump, Simons, Cole, Taylor, Bagley, Finkelstein, Davis, Amor{\'i}n, Backhaus, Cleri, Giavalisco, Grogin, Hirschmann, Holwerda, {Huertas-Company}, Kartaltepe, Kocevksi, Koekemoer, Lucas, Pacucci, \& Wang}]{BrooksEtAl2025}
Brooks, M., Trump, J.~R., Simons, R.~C., {et~al.} 2025, Beyond the {{Monsters}}: {{A More Complete Census}} of {{Black Hole Activity}} at {{Cosmic Dawn}}, arXiv, \dodoi{10.48550/arXiv.2511.19609}

\bibitem[{C.~J. Burke {et~al.}(2023)Burke, Shen, Liu, Natarajan, Caplar, Bellovary, \& Wang}]{BurkeEtAl2023}
Burke, C.~J., Shen, Y., Liu, X., {et~al.} 2023, \bibinfo{title}{Dwarf {{AGNs}} from Variability for the Origins of Seeds ({{DAVOS}}): {{Intermediate-mass}} Black Hole Demographics from Optical Synoptic Surveys,} Monthly Notices of the Royal Astronomical Society, 518, 1880, \dodoi{10.1093/mnras/stac2478}

\bibitem[{C.~J. {Burke} {et~al.}(2026){Burke}, {Stone}, {Shen}, \& {Jiang}}]{BurkeEtAl2026}
{Burke}, C.~J., {Stone}, Z., {Shen}, Y., \& {Jiang}, Y.-F. 2026, \bibinfo{title}{{Too Quiet for Comfort: Local Little Red Dots Lack Variability over Decades},} \apj, 1004, 16, \dodoi{10.3847/1538-4357/ae6b89}

\bibitem[{C.~J. {Burke} {et~al.}(2021){Burke}, {Shen}, {Blaes}, {Gammie}, {Horne}, {Jiang}, {Liu}, {McHardy}, {Morgan}, {Scaringi}, \& {Yang}}]{BurkeEtAl2021}
{Burke}, C.~J., {Shen}, Y., {Blaes}, O., {et~al.} 2021, \bibinfo{title}{{A characteristic optical variability time scale in astrophysical accretion disks},} Science, 373, 789, \dodoi{10.1126/science.abg9933}

\bibitem[{D. {Calzetti} {et~al.}(2000){Calzetti}, {Armus}, {Bohlin}, {Kinney}, {Koornneef}, \& {Storchi-Bergmann}}]{CalzettiEtAl2000}
{Calzetti}, D., {Armus}, L., {Bohlin}, R.~C., {et~al.} 2000, \bibinfo{title}{{The Dust Content and Opacity of Actively Star-forming Galaxies},} \apj, 533, 682, \dodoi{10.1086/308692}

\bibitem[{M. Cantiello {et~al.}(2025)Cantiello, Hassan, Perna, Armitage, Begelman, Jiang, Ryu, \& Townsend}]{CantielloEtAl2025}
Cantiello, M., Hassan, J.~B., Perna, R., {et~al.} 2025, Pulsational {{Instability}} of {{Quasi-Stars}}: {{Interpreting}} the {{Variability}} of {{Little Red Dots}}, arXiv, \dodoi{10.48550/arXiv.2512.17997}

\bibitem[{A.~C. Carnall {et~al.}(2023)Carnall, Begley, McLeod, Hamadouche, Donnan, McLure, Dunlop, {Milvang-Jensen}, Bondestam, Cullen, Jewell, \& Pollock}]{CarnallEtAl2023}
Carnall, A.~C., Begley, R., McLeod, D.~J., {et~al.} 2023, \bibinfo{title}{A First Look at the {{SMACS0723 JWST ERO}}: Spectroscopic Redshifts, Stellar Masses, and Star-Formation Histories,} Monthly Notices of the Royal Astronomical Society: Letters, 518, L45, \dodoi{10.1093/mnrasl/slac136}

\bibitem[{C.~M. Casey {et~al.}(2025)Casey, Akins, Finkelstein, Franco, Fujimoto, Liu, Long, Magdis, Manning, McKinney, Shuntov, \& Tanaka}]{CaseyEtAl2025}
Casey, C.~M., Akins, H.~B., Finkelstein, S.~L., {et~al.} 2025, \bibinfo{title}{An {{Upper Limit}} of 106 {{M}}{$\odot$} in {{Dust}} from {{ALMA Observations}} in 60 {{Little Red Dots}},} The Astrophysical Journal, 990, L61, \dodoi{10.3847/2041-8213/adfa91}

\bibitem[{Q.~O. {Casey} {et~al.}(2026){Casey}, {Hickox}, {Cleri}, {Cohn}, {Alexander}, {Durodola}, {Whalen}, {Hviding}, \& {Ananna}}]{CaseyEtAl2026}
{Casey}, Q.~O., {Hickox}, R.~C., {Cleri}, N.~J., {et~al.} 2026, \bibinfo{title}{{A Population of Little Red Dot-like Quasars in SDSS},} arXiv e-prints, arXiv:2606.26098, \dodoi{10.48550/arXiv.2606.26098}

\bibitem[{S.-J. Chang {et~al.}(2025)Chang, Gronke, Matthee, \& Mason}]{ChangEtAl2025}
Chang, S.-J., Gronke, M., Matthee, J., \& Mason, C. 2025, \bibinfo{title}{Impact of {{Resonance}}, {{Raman}}, and {{Thomson Scattering}} on {{Hydrogen Line Formation}} in {{Little Red Dots}},} Monthly Notices of the Royal Astronomical Society, \dodoi{10.1093/mnras/staf2131}

\bibitem[{C.-H. Chen {et~al.}(2025)Chen, Ho, Li, \& Zhuang}]{ChenEtAl2025}
Chen, C.-H., Ho, L.~C., Li, R., \& Zhuang, M.-Y. 2025, \bibinfo{title}{The {{Host Galaxy}} ({{If Any}}) of the {{Little Red Dots}},} The Astrophysical Journal, 983, 60, \dodoi{10.3847/1538-4357/ada93a}

\bibitem[{C.-H. {Chen} {et~al.}(2026){Chen}, {Shangguan}, {Ho}, {Zhang}, {Inayoshi}, \& {Li}}]{ChenEtAl2026}
{Chen}, C.-H., {Shangguan}, J., {Ho}, L.~C., {et~al.} 2026, \bibinfo{title}{{ABCD: The Nuclear Structure of the Little Red Dots Revealted through Absorption, Break, Continuum, and Decrement},} arXiv e-prints, arXiv:2606.04711, \dodoi{10.48550/arXiv.2606.04711}

\bibitem[{K. {Chen} {et~al.}(2025){Chen}, {Li}, {Inayoshi}, \& {Ho}}]{ChenEtAl2025c}
{Chen}, K., {Li}, Z., {Inayoshi}, K., \& {Ho}, L.~C. 2025, \bibinfo{title}{{Dust Budget Crisis in Little Red Dots},} \apjl, 994, L42, \dodoi{10.3847/2041-8213/ae1955}

\bibitem[{A. {de Graaff} {et~al.}(2024){de Graaff}, {Rix}, {Carniani}, {Suess}, {Charlot}, {Curtis-Lake}, {Arribas}, {Baker}, {Boyett}, {Bunker}, {Cameron}, {Chevallard}, {Curti}, {Eisenstein}, {Franx}, {Hainline}, {Hausen}, {Ji}, {Johnson}, {Jones}, {Maiolino}, {Maseda}, {Nelson}, {Parlanti}, {Rawle}, {Robertson}, {Tacchella}, {{\"U}bler}, {Williams}, {Willmer}, \& {Willott}}]{DeGraaffEtAl2024}
{de Graaff}, A., {Rix}, H.-W., {Carniani}, S., {et~al.} 2024, \bibinfo{title}{{Ionised gas kinematics and dynamical masses of z {\ensuremath{\gtrsim}} 6 galaxies from JADES/NIRSpec high-resolution spectroscopy},} \aap, 684, A87, \dodoi{10.1051/0004-6361/202347755}

\bibitem[{A. de~Graaff {et~al.}(2025)de~Graaff, Rix, Naidu, Labb{\'e}, Wang, Leja, Matthee, Katz, Greene, Hviding, Baggen, Bezanson, Boogaard, Brammer, Dayal, van Dokkum, Goulding, Hirschmann, Maseda, McConachie, Miller, Nelson, Oesch, Setton, Shivaei, Weibel, Whitaker, \& Williams}]{DeGraaffEtAl2025a}
de~Graaff, A., Rix, H.-W., Naidu, R.~P., {et~al.} 2025, \bibinfo{title}{A Remarkable Ruby: {{Absorption}} in Dense Gas, Rather than Evolved Stars, Drives the Extreme {{Balmer}} Break of a Little Red Dot at z = 3.5,} Astronomy \& Astrophysics, 701, A168, \dodoi{10.1051/0004-6361/202554681}

\bibitem[{A. {de Graaff} {et~al.}(2025){de Graaff}, Hviding, Naidu, Greene, Miller, Leja, Matthee, Brammer, Katz, Bezanson, Boogaard, Bose, Chisholm, Cleri, Dayal, Feldmann, Fudamoto, Fujimoto, Furtak, Glazebrook, Gottumukkala, Heintz, Kokorev, Labbe, Maseda, McConachie, Nanayakkara, Nelson, Nowaczyk, Oesch, Rix, Setton, Torralba, Walter, Wang, Weibel, \& {van der Wel}}]{DeGraaffEtAl2025}
{de Graaff}, A., Hviding, R.~E., Naidu, R.~P., {et~al.} 2025, Little {{Red Dots}} Host {{Black Hole Stars}}: {{A}} Unified Family of Gas-Reddened {{AGN}} Revealed by {{JWST}}/{{NIRSpec}} Spectroscopy, arXiv, \dodoi{10.48550/arXiv.2511.21820}

\bibitem[{I. {Delvecchio} {et~al.}(2025){Delvecchio}, {Daddi}, {Magnelli}, {Elbaz}, {Giavalisco}, {Traina}, {Lanzuisi}, {Akins}, {Belli}, {Casey}, {Gentile}, {Gruppioni}, {Pozzi}, \& {Zamorani}}]{DelvecchioEtAl2025}
{Delvecchio}, I., {Daddi}, E., {Magnelli}, B., {et~al.} 2025, \bibinfo{title}{{Active galactic nuclei-heated dust revealed in ``little red dots''},} \aap, 704, A313, \dodoi{10.1051/0004-6361/202557164}

\bibitem[{ {DESI Collaboration} {et~al.}(2026){DESI Collaboration}, {Abdul Karim}, {Adame}, {Aguado}, {Aguilar}, {Ahlen}, {Alam}, {Aldering}, {Alexander}, {Alfarsy}, {Allen}, {Allende Prieto}, {Alves}, {Anand}, {Andrade}, {Armengaud}, {Avila}, {Aviles}, {Awan}, {Bailey}, {Baleato Lizancos}, {Ballester}, {Bault}, {Bautista}, {Bean}, {Behera}, {BenZvi}, {Beraldo e Silva}, {Bermejo-Climent}, {Beutler}, {Bianchi}, {Blake}, {Blum}, {Bolton}, {Bonici}, {Brieden}, {Brodzeller}, {Brooks}, {Buckley-Geer}, {Burtin}, {Bystr{\"o}m}, {Canning}, {Carnero Rosell}, {Carr}, {Carrilho}, {Casas}, {Castander}, {Cereskaite}, {Cervantes-Cota}, {Chaussidon}, {Chaves-Montero}, {Chen}, {Chen}, {Circosta}, {Claybaugh}, {Cole}, {Cooper}, {Cousinou}, {Cuceu}, {Davis}, {Dawson}, {de Belsunce}, {de la Cruz}, {de la Macorra}, {de Mattia}, {Deiosso}, {Della Costa}, {Demina}, {Demirbozan}, {DeRose}, {Dey}, {Dey}, {Ding}, {Ding}, {Doel}, {Douglass}, {Dowicz}, {Ebina}, {Edelstein}, {Eisenstein}, {Elbers}, {Emas}, {Escoffier}, {Fagrelius},
  {Fan}, {Fanning}, {Favole}, {Fawcett}, {Fern{\'a}ndez-Garc{\'\i}a}, {Ferraro}, {Findlay}, {Font-Ribera}, {Forero-Romero}, {Forero-S{\'a}nchez}, {Frenk}, {G{\"a}nsicke}, {Galbany}, {Garc{\'\i}a-Bellido}, {Garcia-Quintero}, {Garrison}, {Gazta{\~n}aga}, {Gil-Mar{\'\i}n}, {Gloudemans}, {Gnedin}, {Gontcho A Gontcho}, {Gonzalez}, {Gonzalez-Morales}, {Gonzalez-Perez}, {Gordon}, {Graur}, {Green}, {Gruen}, {Gsponer}, {Guandalin}, {Gutierrez}, {Guy}, {Hahn}, {Han}, {Han}, {He}, {Herrera-Alcantar}, {Heydenreich}, {Honscheid}, {Hou}, {Howlett}, {Huterer}, {Ir{\v{s}}i{\v{c}}}, {Ishak}, {Jacques}, {Jiang}, {Jimenez}, {Jing}, {Joachimi}, {Joudaki}, {Joyce}, {Jullo}, {Juneau}, {Kara{\c{c}}ayl{\i}}, {Karim}, {Kehoe}, {Kent}, {Khederlarian}, {Kirkby}, {Kisner}, {Kitaura}, {Kizhuprakkat}, {Kong}, {Koposov}, {Kremin}, {Krolewski}, {Lahav}, {Lai}, {Lamman}, {Lan}, {Landriau}, {Lang}, {Lange}, {Lasker}, {Le Goff}, {Le Guillou}, {Leauthaud}, {Levi}, {Li}, {Li}, {Liu}, {Lodha}, {Lokken}, {Luo}, {Magneville}, {Manera}, {Manser},
  {Margala}, {Martini}, {Maus}, {McCullough}, {McDonald}, {Medina}, {Medina-Varela}, {Meisner}, {Mena-Fern{\'a}ndez}, {Menegas}, {Meneses-Rizo}, {Mezcua}, {Miquel}, {Montero-Camacho}, {Moon}, {Moustakas}, {Mu{\~n}oz-Guti{\'e}rrez}, {Mu noz-Santos}, {Myers}, {Myles}, {Nadathur}, {Najita}, {Napolitano}, {Newman}, {Nikakhtar}, {Nikutta}, {Niz}, {Noriega}, \& {Nugent}}]{DESI_DR1}
{DESI Collaboration}, {Abdul Karim}, M., {Adame}, A.~G., {et~al.} 2026, \bibinfo{title}{{Data Release 1 of the Dark Energy Spectroscopic Instrument},} \aj, 171, 285, \dodoi{10.3847/1538-3881/ae4c43}

\bibitem[{F. {D'Eugenio} {et~al.}(2025){D'Eugenio}, {Nelson}, {Ji}, {Baggen}, {Greene}, {Labb{\'e}}, {Pezzulli}, {Brown}, {Maiolino}, {Matthee}, {Terlevich}, {Terlevich}, {Torralba}, \& {Carniani}}]{DEugenioEtAl2025}
{D'Eugenio}, F., {Nelson}, E., {Ji}, X., {et~al.} 2025, \bibinfo{title}{{Irony at z=6.68: a bright AGN with forbidden Fe emission and multi-component Balmer absorption},} arXiv e-prints, arXiv:2510.00101, \dodoi{10.48550/arXiv.2510.00101}

\bibitem[{F. D'Eugenio {et~al.}(2026)D'Eugenio, Juod{\v z}balis, Ji, Scholtz, Maiolino, Carniani, Perna, Mazzolari, {\"U}bler, Arribas, Bhatawdekar, Bunker, Cresci, {Curtis-Lake}, Hainline, Inayoshi, Isobe, Ji, Johnson, Jones, Looser, Nelson, Parlanti, Pusk{\'a}s, Rinaldi, Robertson, Rodr{\'i}guez Del~Pino, Shivaei, Sun, Tacchella, Venturi, Volonteri, Williams, Willmer, Willott, \& Witstok}]{DEugenioEtAl2026}
D'Eugenio, F., Juod{\v z}balis, I., Ji, X., {et~al.} 2026, \bibinfo{title}{{{JADES}} and {{BlackTHUNDER}}: Rest-Frame {{Balmer-line}} Absorption and the Local Environment in a {{Little Red Dot}} at z = 5,} Monthly Notices of the Royal Astronomical Society, 545, staf2117, \dodoi{10.1093/mnras/staf2117}

\bibitem[{F. {D'Eugenio} {et~al.}(2026){D'Eugenio}, {Maiolino}, {Perna}, {{\"U}bler}, {Ji}, {McClymont}, {Koudmani}, {Sijacki}, {Juod{\v{z}}balis}, {Scholtz}, {Bennett}, {Bunker}, {Carniani}, {Charlot}, {Cresci}, {Curtis-Lake}, {Bont{\`a}}, {Inayoshi}, {Jones}, {Lyu}, {Marconi}, {Mazzolari}, {Nelson}, {Parlanti}, {Robertson}, {Schneider}, {Simmonds}, {Tacchella}, {Venturi}, {Willott}, {Witstok}, \& {Witten}}]{DEugenioEtAl2026b}
{D'Eugenio}, F., {Maiolino}, R., {Perna}, M., {et~al.} 2026, \bibinfo{title}{{BlackTHUNDER strikes twice: Balmer-line absorption in an overmassive Little Red Dot at z = 7.04},} \mnras, 547, stag401, \dodoi{10.1093/mnras/stag401}

\bibitem[{P. {Du} {et~al.}(2016){Du}, {Lu}, {Zhang}, {Huang}, {Wang}, {Hu}, {Qiu}, {Li}, {Fan}, {Fang}, {Bai}, {Bian}, {Yuan}, {Ho}, {Wang}, \& {SEAMBH Collaboration}}]{DuEtAl2016}
{Du}, P., {Lu}, K.-X., {Zhang}, Z.-X., {et~al.} 2016, \bibinfo{title}{{Supermassive Black Holes with High Accretion Rates in Active Galactic Nuclei. V. A New Size-Luminosity Scaling Relation for the Broad-line Region},} \apj, 825, 126, \dodoi{10.3847/0004-637X/825/2/126}

\bibitem[{P. {Du} {et~al.}(2018){Du}, {Zhang}, {Wang}, {Huang}, {Zhang}, {Lu}, {Hu}, {Li}, {Bai}, {Bian}, {Yuan}, {Ho}, {Wang}, \& {SEAMBH Collaboration}}]{DuEtAl2018}
{Du}, P., {Zhang}, Z.-X., {Wang}, K., {et~al.} 2018, \bibinfo{title}{{Supermassive Black Holes with High Accretion Rates in Active Galactic Nuclei. IX. 10 New Observations of Reverberation Mapping and Shortened H{\ensuremath{\beta}} Lags},} \apj, 856, 6, \dodoi{10.3847/1538-4357/aaae6b}

\bibitem[{D. {Foreman-Mackey} {et~al.}(2013){Foreman-Mackey}, {Hogg}, {Lang}, \& {Goodman}}]{emcee}
{Foreman-Mackey}, D., {Hogg}, D.~W., {Lang}, D., \& {Goodman}, J. 2013, \bibinfo{title}{emcee: The MCMC Hammer,} PASP, 125, 306, \dodoi{10.1086/670067}

\bibitem[{L.~B. {Fries} {et~al.}(2023){Fries}, {Trump}, {Davis}, {Grier}, {Shen}, {Anderson}, {Dwelly}, {Eracleous}, {Homayouni}, {Horne}, {Krumpe}, {Morrison}, {Runnoe}, {Trakhtenbrot}, {Assef}, {Brandt}, {Brownstein}, {Dabbieri}, {Fix}, {Fonseca Alvarez}, {Frederick}, {Hall}, {Koekemoer}, {Li}, {Liu}, {Mart{\'\i}nez-Aldama}, {Ricci}, {Schneider}, {Sharp}, {Temple}, {Yang}, {Zeltyn}, \& {Bizyaev}}]{FriesEtAl2023}
{Fries}, L.~B., {Trump}, J.~R., {Davis}, M.~C., {et~al.} 2023, \bibinfo{title}{{The SDSS-V Black Hole Mapper Reverberation Mapping Project: Unusual Broad-line Variability in a Luminous Quasar},} \apj, 948, 5, \dodoi{10.3847/1538-4357/acbfb7}

\bibitem[{S. Fujimoto {et~al.}(2024)Fujimoto, Wang, Weaver, Kokorev, Atek, Bezanson, Labbe, Brammer, Greene, Chemerynska, Dayal, {de Graaff}, Furtak, Oesch, Setton, Price, Miller, Williams, Whitaker, Zitrin, Cutler, Leja, Pan, Coe, {van Dokkum}, Feldmann, Fudamoto, Goulding, Khullar, Marchesini, Maseda, Nanayakkara, Nelson, Smit, Stefanon, \& Weibel}]{FujimotoEtAl2024}
Fujimoto, S., Wang, B., Weaver, J.~R., {et~al.} 2024, \bibinfo{title}{{{UNCOVER}}: {{A NIRSpec Census}} of {{Lensed Galaxies}} at z = 8.50--13.08 {{Probing}} a {{High-AGN Fraction}} and {{Ionized Bubbles}} in the {{Shadow}},} The Astrophysical Journal, 977, 250, \dodoi{10.3847/1538-4357/ad9027}

\bibitem[{L.~J. Furtak {et~al.}(2023)Furtak, Zitrin, Plat, Fujimoto, Wang, Nelson, Labb{\'e}, Bezanson, Brammer, {van Dokkum}, Endsley, Glazebrook, Greene, Leja, Price, Smit, Stark, Weaver, Whitaker, Atek, Chevallard, {Curtis-Lake}, Dayal, Feltre, Franx, Fudamoto, Marchesini, Mowla, Pan, Suess, {Vidal-Garc{\'i}a}, \& Williams}]{FurtakEtAl2023}
Furtak, L.~J., Zitrin, A., Plat, A., {et~al.} 2023, \bibinfo{title}{{{JWST UNCOVER}}: {{Extremely Red}} and {{Compact Object}} at Zphot {$\simeq$} 7.6 {{Triply Imaged}} by {{A2744}},} The Astrophysical Journal, 952, 142, \dodoi{10.3847/1538-4357/acdc9d}

\bibitem[{L.~J. Furtak {et~al.}(2024)Furtak, Labb{\'e}, Zitrin, Greene, Dayal, Chemerynska, Kokorev, Miller, Goulding, {de Graaff}, Bezanson, Brammer, Cutler, Leja, Pan, Price, Wang, Weaver, Whitaker, Atek, Bogd{\'a}n, Charlot, {Curtis-Lake}, {van Dokkum}, Endsley, Feldmann, Fudamoto, Fujimoto, Glazebrook, Juneau, Marchesini, Maseda, Nelson, Oesch, Plat, Setton, Stark, \& Williams}]{FurtakEtAl2024}
Furtak, L.~J., Labb{\'e}, I., Zitrin, A., {et~al.} 2024, \bibinfo{title}{A High Black-Hole-to-Host Mass Ratio in a Lensed {{AGN}} in the Early {{Universe}},} Nature, 628, 57, \dodoi{10.1038/s41586-024-07184-8}

\bibitem[{L.~J. Furtak {et~al.}(2025)Furtak, Secunda, Greene, Zitrin, Labb{\'e}, Golubchik, Bezanson, Kokorev, Atek, Brammer, Chemerynska, Cutler, Dayal, Feldmann, Fujimoto, Glazebrook, Leja, Ma, Matthee, Naidu, Nelson, Oesch, Pan, Price, Suess, Wang, Weaver, \& Whitaker}]{FurtakEtAl2025}
Furtak, L.~J., Secunda, A.~R., Greene, J.~E., {et~al.} 2025, \bibinfo{title}{Investigating Photometric and Spectroscopic Variability in the Multiply Imaged Little Red Dot {{A2744-QSO1}},} Astronomy and Astrophysics, 698, A227, \dodoi{10.1051/0004-6361/202554110}

\bibitem[{C.~M. {Gaskell} \& A.~J. {Benker}(2007){Gaskell} \& {Benker}}]{GaskellBenker2007}
{Gaskell}, C.~M., \& {Benker}, A.~J. 2007, \bibinfo{title}{{AGN Reddening and Ultraviolet Extinction Curves from Hubble Space Telescope Spectra},} arXiv e-prints, arXiv:0711.1013, \dodoi{10.48550/arXiv.0711.1013}

\bibitem[{K.~D. Gordon(2024)Gordon}]{Gordon2024}
Gordon, K.~D. 2024, \bibinfo{title}{Dust\_extinction: {{Interstellar Dust Extinction Models}},} Journal of Open Source Software, 9, 7023, \dodoi{10.21105/joss.07023}

\bibitem[{K.~D. Gordon {et~al.}(2023)Gordon, Clayton, Decleir, Fitzpatrick, Massa, Misselt, \& Tollerud}]{GordonEtAl2023}
Gordon, K.~D., Clayton, G.~C., Decleir, M., {et~al.} 2023, \bibinfo{title}{One {{Relation}} for {{All Wavelengths}}: {{The Far-ultraviolet}} to {{Mid-infrared Milky Way Spectroscopic R}}({{V}})-Dependent {{Dust Extinction Relationship}},} The Astrophysical Journal, 950, 86, \dodoi{10.3847/1538-4357/accb59}

\bibitem[{K.~D. Gordon {et~al.}(2003)Gordon, Clayton, Misselt, Landolt, \& Wolff}]{GordonEtAl2003}
Gordon, K.~D., Clayton, G.~C., Misselt, K.~A., Landolt, A.~U., \& Wolff, M.~J. 2003, \bibinfo{title}{A {{Quantitative Comparison}} of the {{Small Magellanic Cloud}}, {{Large Magellanic Cloud}}, and {{Milky Way Ultraviolet}} to {{Near-Infrared Extinction Curves}},} The Astrophysical Journal, 594, 279, \dodoi{10.1086/376774}

\bibitem[{K.~D. Gordon {et~al.}(2024)Gordon, Fitzpatrick, Massa, Bohlin, Chastenet, Murray, Clayton, Lennon, Misselt, \& Sandstrom}]{GordonEtAl2024}
Gordon, K.~D., Fitzpatrick, E.~L., Massa, D., {et~al.} 2024, \bibinfo{title}{Expanded {{Sample}} of {{Small Magellanic Cloud Ultraviolet Dust Extinction Curves}}: {{Correlations}} between the 2175 {{\AA{} Bump}}, q {{PAH}}, {{Ultraviolet Extinction Shape}}, and {{N}}({{H I}})/{{A}}({{V}}),} The Astrophysical Journal, 970, 51, \dodoi{10.3847/1538-4357/ad4be1}

\bibitem[{J.~E. Greene {et~al.}(2024)Greene, Labbe, Goulding, Furtak, Chemerynska, Kokorev, Dayal, Volonteri, Williams, Wang, Setton, Burgasser, Bezanson, Atek, Brammer, Cutler, Feldmann, Fujimoto, Glazebrook, {de Graaff}, Khullar, Leja, Marchesini, Maseda, Matthee, Miller, Naidu, Nanayakkara, Oesch, Pan, Papovich, Price, {van Dokkum}, Weaver, Whitaker, \& Zitrin}]{GreeneEtAl2024}
Greene, J.~E., Labbe, I., Goulding, A.~D., {et~al.} 2024, \bibinfo{title}{{{UNCOVER Spectroscopy Confirms}} the {{Surprising Ubiquity}} of {{Active Galactic Nuclei}} in {{Red Sources}} at z {$>$} 5,} The Astrophysical Journal, 964, 39, \dodoi{10.3847/1538-4357/ad1e5f}

\bibitem[{C.~J. {Grier} {et~al.}(2017){Grier}, {Pancoast}, {Barth}, {Fausnaugh}, {Brewer}, {Treu}, \& {Peterson}}]{GrierEtAl2017}
{Grier}, C.~J., {Pancoast}, A., {Barth}, A.~J., {et~al.} 2017, \bibinfo{title}{{The Structure of the Broad-line Region in Active Galactic Nuclei. II. Dynamical Modeling of Data From the AGN10 Reverberation Mapping Campaign},} \apj, 849, 146, \dodoi{10.3847/1538-4357/aa901b}

\bibitem[{Y. Harikane {et~al.}(2023)Harikane, Zhang, Nakajima, Ouchi, Isobe, Ono, Hatano, Xu, \& Umeda}]{HarikaneEtAl2023}
Harikane, Y., Zhang, Y., Nakajima, K., {et~al.} 2023, \bibinfo{title}{A {{JWST}}/{{NIRSpec First Census}} of {{Broad-line AGNs}} at z = 4--7: {{Detection}} of 10 {{Faint AGNs}} with {{MBH}} {$\sim$} 106--{{108M}}{$\odot$} and {{Their Host Galaxy Properties}},} The Astrophysical Journal, 959, 39, \dodoi{10.3847/1538-4357/ad029e}

\bibitem[{C.~R. Harris {et~al.}(2020)Harris, Millman, {van der Walt}, Gommers, Virtanen, Cournapeau, Wieser, Taylor, Berg, Smith, Kern, Picus, Hoyer, {van Kerkwijk}, Brett, Haldane, {del R{\'i}o}, Wiebe, Peterson, {G{\'e}rard-Marchant}, Sheppard, Reddy, Weckesser, Abbasi, Gohlke, \& Oliphant}]{HarrisEtAl2020b}
Harris, C.~R., Millman, K.~J., {van der Walt}, S.~J., {et~al.} 2020, \bibinfo{title}{Array Programming with {{NumPy}},} Nature, 585, 357, \dodoi{10.1038/s41586-020-2649-2}

\bibitem[{C. {Hu} {et~al.}(2021){Hu}, {Li}, {Yang}, {Yang}, {Guo}, {Bao}, {Jiang}, {Du}, {Li}, {Xiao}, {Songsheng}, {Yu}, {Bai}, {Ho}, {Brotherton}, {Aceituno}, {Winkler}, {Wang}, \& {Seambh Collaboration}}]{HuEtAl2021}
{Hu}, C., {Li}, S.-S., {Yang}, S., {et~al.} 2021, \bibinfo{title}{{Supermassive Black Holes with High Accretion Rates in Active Galactic Nuclei. XII. Reverberation Mapping Results for 15 PG Quasars from a Long-duration High-cadence Campaign},} \apjs, 253, 20, \dodoi{10.3847/1538-4365/abd774}

\bibitem[{H. {Hu} {et~al.}(2022){Hu}, {Inayoshi}, {Haiman}, {Quataert}, \& {Kuiper}}]{HuEtAl2022}
{Hu}, H., {Inayoshi}, K., {Haiman}, Z., {Quataert}, E., \& {Kuiper}, R. 2022, \bibinfo{title}{{Long-term Evolution of Supercritical Black Hole Accretion with Outflows: A Subgrid Feedback Model for Cosmological Simulations},} \apj, 934, 132, \dodoi{10.3847/1538-4357/ac75d8}

\bibitem[{J.~D. Hunter(2007)Hunter}]{Hunter2007b}
Hunter, J.~D. 2007, \bibinfo{title}{Matplotlib: {{A 2D Graphics Environment}},} Computing in Science \& Engineering, 9, 90, \dodoi{10.1109/MCSE.2007.55}

\bibitem[{J.~B. {Hutchings} {et~al.}(2002){Hutchings}, {Crenshaw}, {Kraemer}, {Gabel}, {Kaiser}, {Weistrop}, \& {Gull}}]{HutchingsEtAl2002}
{Hutchings}, J.~B., {Crenshaw}, D.~M., {Kraemer}, S.~B., {et~al.} 2002, \bibinfo{title}{{Balmer and He I Absorption in the Nuclear Spectrum of NGC 4151},} \aj, 124, 2543, \dodoi{10.1086/344080}

\bibitem[{K. {Inayoshi} \& K. {Ichikawa}(2024){Inayoshi} \& {Ichikawa}}]{InayoshiIchikawa2024}
{Inayoshi}, K., \& {Ichikawa}, K. 2024, \bibinfo{title}{{Birth of Rapidly Spinning, Overmassive Black Holes in the Early Universe},} \apjl, 973, L49, \dodoi{10.3847/2041-8213/ad74e2}

\bibitem[{K. Inayoshi {et~al.}(2025)Inayoshi, Kimura, \& Noda}]{InayoshiEtAl2025a}
Inayoshi, K., Kimura, S.~S., \& Noda, H. 2025, \bibinfo{title}{Weakness of {{X-rays}} and Variability in High-Redshift Active Galactic Nuclei with Super-{{Eddington}} Accretion,} Publications of the Astronomical Society of Japan, 77, 811, \dodoi{10.1093/pasj/psaf050}

\bibitem[{X. Ji {et~al.}(2025)Ji, Maiolino, {\"U}bler, Scholtz, D'Eugenio, Sun, Perna, Turner, Carniani, Arribas, Bennett, Bunker, Charlot, Cresci, Curti, Egami, Fabian, Inayoshi, Isobe, Jones, Juod{\v z}balis, Kumari, Lyu, Mazzolari, Parlanti, Robertson, Rodr{\'i}guez Del~Pino, Schneider, Sijacki, Tacchella, Trinca, Valiante, Venturi, Volonteri, Willott, Witten, \& Witstok}]{JiEtAl2025}
Ji, X., Maiolino, R., {\"U}bler, H., {et~al.} 2025, \bibinfo{title}{{{BlackTHUNDER}} - {{A}} Non-Stellar {{Balmer}} Break in a Black Hole-Dominated Little Red Dot at z = 7.04,} Monthly Notices of the Royal Astronomical Society, 544, 3900, \dodoi{10.1093/mnras/staf1867}

\bibitem[{X. Ji {et~al.}(2026)Ji, D'Eugenio, Juod{\v z}balis, Walton, Fabian, Maiolino, Ramos~Almeida, Acosta~Pulido, Belokurov, Isobe, Jones, Maraston, Scholtz, Simmonds, Tacchella, Terlevich, \& Terlevich}]{JiEtAl2026}
Ji, X., D'Eugenio, F., Juod{\v z}balis, I., {et~al.} 2026, \bibinfo{title}{Lord of {{LRDs}}: Insights into a '{{Little Red Dot}}' with a Low-Ionization Spectrum at z = 0.1,} Monthly Notices of the Royal Astronomical Society, 545, staf2235, \dodoi{10.1093/mnras/staf2235}

\bibitem[{Y.-F. {Jiang} {et~al.}(2014){Jiang}, {Stone}, \& {Davis}}]{JiangEtAl2014}
{Jiang}, Y.-F., {Stone}, J.~M., \& {Davis}, S.~W. 2014, \bibinfo{title}{{A Global Three-dimensional Radiation Magneto-hydrodynamic Simulation of Super-Eddington Accretion Disks},} \apj, 796, 106, \dodoi{10.1088/0004-637X/796/2/106}

\bibitem[{I. {Juod{\v{z}}balis} {et~al.}(2026){Juod{\v{z}}balis}, {Maiolino}, {Baker}, {Lake}, {Scholtz}, {D'Eugenio}, {Trefoloni}, {Isobe}, {Tacchella}, {Bunker}, {Carniani}, {Charlot}, {Jones}, {Parlanti}, {Perna}, {Rinaldi}, {Robertson}, {{\"U}bler}, {Venturi}, \& {Willott}}]{JuodzbalisEtAl2026}
{Juod{\v{z}}balis}, I., {Maiolino}, R., {Baker}, W.~M., {et~al.} 2026, \bibinfo{title}{{JADES: comprehensive census of broad-line AGN from reionization to cosmic noon revealed by JWST},} \mnras, 546, stag086, \dodoi{10.1093/mnras/stag086}

\bibitem[{I. Juod{\v z}balis {et~al.}(2024)Juod{\v z}balis, Ji, Maiolino, D'Eugenio, Scholtz, Risaliti, Fabian, Mazzolari, Gilli, Prandoni, Arribas, Bunker, Carniani, Charlot, {Curtis-Lake}, {de~Graaff}, Hainline, Parlanti, Perna, {P{\'e}rez-Gonz{\'a}lez}, Robertson, Tacchella, {\"U}bler, Williams, Willott, \& Witstok}]{JuodzbalisEtAl2024}
Juod{\v z}balis, I., Ji, X., Maiolino, R., {et~al.} 2024, \bibinfo{title}{{{JADES}} -- the {{Rosetta}} Stone of {{JWST-discovered AGN}}: Deciphering the Intriguing Nature of Early {{AGN}},} Monthly Notices of the Royal Astronomical Society, 535, 853, \dodoi{10.1093/mnras/stae2367}

\bibitem[{B.~C. {Kelly} {et~al.}(2009){Kelly}, {Bechtold}, \& {Siemiginowska}}]{KellyEtAl2009}
{Kelly}, B.~C., {Bechtold}, J., \& {Siemiginowska}, A. 2009, \bibinfo{title}{{Are the Variations in Quasar Optical Flux Driven by Thermal Fluctuations?},} \apj, 698, 895, \dodoi{10.1088/0004-637X/698/1/895}

\bibitem[{B.~C. {Kelly} {et~al.}(2014){Kelly}, {Becker}, {Sobolewska}, {Siemiginowska}, \& {Uttley}}]{KellyEtAl2014}
{Kelly}, B.~C., {Becker}, A.~C., {Sobolewska}, M., {Siemiginowska}, A., \& {Uttley}, P. 2014, \bibinfo{title}{{Flexible and Scalable Methods for Quantifying Stochastic Variability in the Era of Massive Time-domain Astronomical Data Sets},} \apj, 788, 33, \dodoi{10.1088/0004-637X/788/1/33}

\bibitem[{D. Kido {et~al.}(2025)Kido, Ioka, Hotokezaka, Inayoshi, \& Irwin}]{KidoEtAl2025}
Kido, D., Ioka, K., Hotokezaka, K., Inayoshi, K., \& Irwin, C.~M. 2025, \bibinfo{title}{Black Hole Envelopes in {{Little Red Dots}},} Monthly Notices of the Royal Astronomical Society, 544, 3407, \dodoi{10.1093/mnras/staf1898}

\bibitem[{D. {Kim} {et~al.}(2010){Kim}, {Im}, \& {Kim}}]{KimEtAl2010}
{Kim}, D., {Im}, M., \& {Kim}, M. 2010, \bibinfo{title}{{New Estimators of Black Hole Mass in Active Galactic Nuclei with Hydrogen Paschen Lines},} \apj, 724, 386, \dodoi{10.1088/0004-637X/724/1/386}

\bibitem[{D.~D. Kocevski {et~al.}(2023)Kocevski, Onoue, Inayoshi, Trump, Arrabal~Haro, Grazian, Dickinson, Finkelstein, Kartaltepe, Hirschmann, Aird, Holwerda, Fujimoto, Juneau, Amor{\'i}n, Backhaus, Bagley, Barro, Bell, Bisigello, Calabr{\`o}, Cleri, Cooper, Ding, Grogin, Ho, Hutchison, Inoue, Jiang, Jones, Koekemoer, Li, Li, McGrath, Molina, Papovich, {P{\'e}rez-Gonz{\'a}lez}, Pirzkal, Wilkins, Yang, \& Yung}]{KocevskiEtAl2023a}
Kocevski, D.~D., Onoue, M., Inayoshi, K., {et~al.} 2023, \bibinfo{title}{Hidden {{Little Monsters}}: {{Spectroscopic Identification}} of {{Low-mass}}, {{Broad-line AGNs}} at z {$>$} 5 with {{CEERS}},} The Astrophysical Journal, 954, L4, \dodoi{10.3847/2041-8213/ace5a0}

\bibitem[{D.~D. Kocevski {et~al.}(2025)Kocevski, Finkelstein, Barro, Taylor, Calabr{\`o}, Laloux, Buchner, Trump, Leung, Yang, Dickinson, {P{\'e}rez-Gonz{\'a}lez}, Pacucci, Inayoshi, Somerville, McGrath, Akins, Bagley, Bowler, Bisigello, Carnall, Casey, Cheng, Cleri, Costantin, Cullen, Davis, Donnan, Dunlop, Ellis, Ferguson, Fujimoto, Fontana, Giavalisco, Grazian, Grogin, Hathi, Hirschmann, {Huertas-Company}, Holwerda, Illingworth, Juneau, Kartaltepe, Koekemoer, Li, Lucas, Magee, Mason, McLeod, McLure, Napolitano, Papovich, Pirzkal, Rodighiero, Santini, Wilkins, \& Yung}]{KocevskiEtAl2025}
Kocevski, D.~D., Finkelstein, S.~L., Barro, G., {et~al.} 2025, \bibinfo{title}{The {{Rise}} of {{Faint}}, {{Red Active Galactic Nuclei}} at z {$>$} 4: {{A Sample}} of {{Little Red Dots}} in the {{JWST Extragalactic Legacy Fields}},} The Astrophysical Journal, 986, 126, \dodoi{10.3847/1538-4357/adbc7d}

\bibitem[{V. Kokorev {et~al.}(2023)Kokorev, Fujimoto, Labbe, Greene, Bezanson, Dayal, Nelson, Atek, Brammer, Caputi, Chemerynska, Cutler, Feldmann, Fudamoto, Furtak, Goulding, {de Graaff}, Leja, Marchesini, Miller, Nanayakkara, Oesch, Pan, Price, Setton, Smit, Stefanon, Wang, Weaver, Whitaker, Williams, \& Zitrin}]{KokorevEtAl2023}
Kokorev, V., Fujimoto, S., Labbe, I., {et~al.} 2023, \bibinfo{title}{{{UNCOVER}}: {{A NIRSpec Identification}} of a {{Broad-line AGN}} at z = 8.50,} The Astrophysical Journal, 957, L7, \dodoi{10.3847/2041-8213/ad037a}

\bibitem[{V. Kokorev {et~al.}(2024)Kokorev, Caputi, Greene, Dayal, Trebitsch, Cutler, Fujimoto, Labb{\'e}, Miller, Iani, {Navarro-Carrera}, \& Rinaldi}]{KokorevEtAl2024}
Kokorev, V., Caputi, K.~I., Greene, J.~E., {et~al.} 2024, \bibinfo{title}{A {{Census}} of {{Photometrically Selected Little Red Dots}} at 4 {$<$} z {$<$} 9 in {{JWST Blank Fields}},} The Astrophysical Journal, 968, 38, \dodoi{10.3847/1538-4357/ad4265}

\bibitem[{V. Kokorev {et~al.}(2025)Kokorev, Chisholm, Naidu, Fujimoto, Atek, Brammer, Finkelstein, Akins, Berg, Furtak, Fei, Hsiao, Labb{\'e}, Matthee, Mu{\~n}oz, Oesch, Pan, Rinaldi, {Saldana-Lopez}, Schaerer, Volonteri, \& Zitrin}]{KokorevEtAl2025}
Kokorev, V., Chisholm, J., Naidu, R.~P., {et~al.} 2025, The {{Deepest GLIMPSE}} of a {{Dense Gas Cocoon Enshrouding}} a {{Little Red Dot}}, arXiv, \dodoi{10.48550/arXiv.2511.07515}

\bibitem[{M. {Kokubo}(2024){Kokubo}}]{Kokubo2024}
{Kokubo}, M. 2024, \bibinfo{title}{{Rayleigh and Raman scattering cross-sections and phase matrices of the ground-state hydrogen atom, and their astrophysical implications},} \mnras, 529, 2131, \dodoi{10.1093/mnras/stae515}

\bibitem[{M. Kokubo \& Y. Harikane(2025)Kokubo \& Harikane}]{KokuboHarikane2025}
Kokubo, M., \& Harikane, Y. 2025, \bibinfo{title}{Challenging the {{Active Galactic Nucleus Scenario}} for {{JWST}}/{{NIRSpec Little Red Dot}} and {{Non}}--{{Little Red Dot Broad H$\alpha$ Emitters}} in {{Light}} of {{Nondetection}} of {{NIRCam Photometric Variability}} and {{X-Ray}},} The Astrophysical Journal, 995, 24, \dodoi{10.3847/1538-4357/ae119e}

\bibitem[{S. {Koz{\l}owski}(2016){Koz{\l}owski}}]{Kozlowski2016}
{Koz{\l}owski}, S. 2016, \bibinfo{title}{{Revisiting Stochastic Variability of AGNs with Structure Functions},} \apj, 826, 118, \dodoi{10.3847/0004-637X/826/2/118}

\bibitem[{J. {Kwan} \& J.~H. {Krolik}(1981){Kwan} \& {Krolik}}]{KwanKrolik1981}
{Kwan}, J., \& {Krolik}, J.~H. 1981, \bibinfo{title}{{The formation of emission lines in quasars and Seyfert nuclei},} \apj, 250, 478, \dodoi{10.1086/159395}

\bibitem[{I. Labb{\'e} {et~al.}(2023)Labb{\'e}, {van Dokkum}, Nelson, Bezanson, Suess, Leja, Brammer, Whitaker, Mathews, Stefanon, \& Wang}]{LabbeEtAl2023}
Labb{\'e}, I., {van Dokkum}, P., Nelson, E., {et~al.} 2023, \bibinfo{title}{A Population of Red Candidate Massive Galaxies 600 {{Myr}} after the {{Big Bang}},} Nature, 616, 266, \dodoi{10.1038/s41586-023-05786-2}

\bibitem[{I. Labbe {et~al.}(2024)Labbe, Greene, Matthee, Treiber, Kokorev, Miller, Kramarenko, Setton, Ma, Goulding, Bezanson, Naidu, Williams, Atek, Brammer, Cutler, Chemerynska, Cloonan, Dayal, {de Graaff}, Fudamoto, Fujimoto, Furtak, Glazebrook, Heintz, Leja, Marchesini, Nanayakkara, Nelson, Oesch, Pan, Price, Shivaei, Sobral, Suess, {van Dokkum}, Wang, Weaver, Whitaker, \& Zitrin}]{LabbeEtAl2024}
Labbe, I., Greene, J.~E., Matthee, J., {et~al.} 2024, An Unambiguous {{AGN}} and a {{Balmer}} Break in an {{Ultraluminous Little Red Dot}} at Z=4.47 from {{Ultradeep UNCOVER}} and {{All}} the {{Little Things Spectroscopy}}, arXiv, \dodoi{10.48550/arXiv.2412.04557}

\bibitem[{I. Labbe {et~al.}(2025)Labbe, Greene, Bezanson, Fujimoto, Furtak, Goulding, Matthee, Naidu, Oesch, Atek, Brammer, Chemerynska, Coe, Cutler, Dayal, Feldmann, Franx, Glazebrook, Leja, Maseda, Marchesini, Nanayakkara, Nelson, Pan, Papovich, Price, Suess, Wang, Weaver, Whitaker, Williams, \& Zitrin}]{LabbeEtAl2025}
Labbe, I., Greene, J.~E., Bezanson, R., {et~al.} 2025, \bibinfo{title}{{{UNCOVER}}: {{Candidate Red Active Galactic Nuclei}} at 3 \&lt; z \&lt; 7 with {{JWST}} and {{ALMA}},} The Astrophysical Journal, 978, 92, \dodoi{10.3847/1538-4357/ad3551}

\bibitem[{S.~K. Lam {et~al.}(2015)Lam, Pitrou, \& Seibert}]{LamEtAl2015a}
Lam, S.~K., Pitrou, A., \& Seibert, S. 2015, \bibinfo{title}{Numba: A {{LLVM-based Python JIT}} Compiler,} in Proceedings of the {{Second Workshop}} on the {{LLVM Compiler Infrastructure}} in {{HPC}}, {{LLVM}} '15 (New York, NY, USA: Association for Computing Machinery), 1--6, \dodoi{10.1145/2833157.2833162}

\bibitem[{E. Lambrides {et~al.}(2024)Lambrides, Garofali, Larson, Ptak, Chiaberge, Long, Hutchison, Norman, McKinney, Akins, Berg, Chisholm, Civano, Cloonan, Endsley, Faisst, Gilli, Gillman, Hirschmann, Kartaltepe, Kocevski, Kokorev, Pacucci, Richardson, Stiavelli, \& Whalen}]{LambridesEtAl2024}
Lambrides, E., Garofali, K., Larson, R., {et~al.} 2024, The {{Case}} for {{Super-Eddington Accretion}}: {{Connecting Weak X-ray}} and {{UV Line Emission}} in {{JWST Broad-Line AGN During}} the {{First Gyr}} of {{Cosmic Time}}, arXiv, \dodoi{10.48550/arXiv.2409.13047}

\bibitem[{E. {Lambrides} {et~al.}(2026){Lambrides}, {Hutchison}, {Larson}, {Arrabal Haro}, {Papovich}, {Hu}, {Cleri}, {Finkelstein}, {Trump}, {Perez-Gonzalez}, {Wang}, {Kocevski}, {Chisholm}, {Secunda}, {Bosman}, {Akins}, {Karmen}, {Dickinson}, {Bromm}, {Backhaus}, {Chiaberge}, {Cooper}, {Ajay}, {Barro}, {Berg}, {Cann}, {Cooper}, {Grogin}, {Hirschmann}, {Huertas-Company}, {Kartaltepe}, {Koekemoer}, {Lucas}, {Long}, {Gilli}, {Norman}, {Ptak}, {Richardson}, {Rigby}, {Vanderhoof}, {Yung}, \& {Zavala}}]{GlimmIr}
{Lambrides}, E., {Hutchison}, T.~A., {Larson}, R.~L., {et~al.} 2026, \bibinfo{title}{{The GlimmIr: Spectroscopic Variability in a z\raisebox{-0.5ex}\textasciitilde7 LRD Indicates Rapid Changes in Both the Narrow and Broad Line Regions},} arXiv e-prints, arXiv:2604.25991, \dodoi{10.48550/arXiv.2604.25991}

\bibitem[{I. {Lamperti} {et~al.}(2017){Lamperti}, {Koss}, {Trakhtenbrot}, {Schawinski}, {Ricci}, {Oh}, {Landt}, {Riffel}, {Rodr{\'\i}guez-Ardila}, {Gehrels}, {Harrison}, {Masetti}, {Mushotzky}, {Treister}, {Ueda}, \& {Veilleux}}]{LampertiEtAl2017}
{Lamperti}, I., {Koss}, M., {Trakhtenbrot}, B., {et~al.} 2017, \bibinfo{title}{{BAT AGN Spectroscopic Survey - IV: Near-Infrared Coronal Lines, Hidden Broad Lines, and Correlation with Hard X-ray Emission},} \mnras, 467, 540, \dodoi{10.1093/mnras/stx055}

\bibitem[{H. {Landt} {et~al.}(2008){Landt}, {Bentz}, {Ward}, {Elvis}, {Peterson}, {Korista}, \& {Karovska}}]{LandtEtAl2008}
{Landt}, H., {Bentz}, M.~C., {Ward}, M.~J., {et~al.} 2008, \bibinfo{title}{{The Near-Infrared Broad Emission Line Region of Active Galactic Nuclei. I. The Observations},} \apjs, 174, 282, \dodoi{10.1086/522373}

\bibitem[{R.~L. Larson {et~al.}(2023)Larson, Finkelstein, Kocevski, Hutchison, Trump, Arrabal~Haro, Bromm, Cleri, Dickinson, Fujimoto, Kartaltepe, Koekemoer, Papovich, Pirzkal, Tacchella, Zavala, Bagley, Behroozi, Champagne, Cole, Jung, Morales, Yang, Zhang, Zitrin, Amor{\'i}n, Burgarella, Casey, Ch{\'a}vez~Ortiz, Cox, Chworowsky, Fontana, Gawiser, Grazian, Grogin, Harish, Hathi, Hirschmann, Holwerda, Juneau, Leung, Lucas, McGrath, {P{\'e}rez-Gonz{\'a}lez}, Rigby, Seill{\'e}, Simons, {de La Vega}, Weiner, Wilkins, Yung, \& {Ceers Team}}]{LarsonEtAl2023}
Larson, R.~L., Finkelstein, S.~L., Kocevski, D.~D., {et~al.} 2023, \bibinfo{title}{A {{CEERS Discovery}} of an {{Accreting Supermassive Black Hole}} 570 {{Myr}} after the {{Big Bang}}: {{Identifying}} a {{Progenitor}} of {{Massive}} z {$>$} 6 {{Quasars}},} The Astrophysical Journal, 953, L29, \dodoi{10.3847/2041-8213/ace619}

\bibitem[{R. {Lin} {et~al.}(2026){Lin}, {Zheng}, {Wang}, {Ho}, {Zavala}, {Zhang}, {Jiang}, {Lin}, {Yuan}, {Jiang}, {Wang}, \& {Zhang}}]{LinR_2026}
{Lin}, R., {Zheng}, Z.-Y., {Wang}, J., {et~al.} 2026, \bibinfo{title}{{Local Analogs of Little Red Dots: Optical Variability and Evidence for an Active Galactic Nucleus Origin},} \apj, 1005, 150, \dodoi{10.3847/1538-4357/ae74c6}

\bibitem[{X. Lin {et~al.}(2026)Lin, Fan, Wang, Sun, Champagne, Egami, Kakiichi, Lyu, Tee, Yang, Bian, Bosman, Cai, Casey, Decarli, Faisst, Finkelstein, Fujimoto, Harish, Ilbert, Inoue, Jin, Kartaltepe, Kocevski, Li, Liu, Liu, Schindler, Shuntov, Tanaka, Vestergaard, Wu, Zhang, \& Zhang}]{LinEtAl2026}
Lin, X., Fan, X., Wang, F., {et~al.} 2026, \bibinfo{title}{Bridging {{Quasars}} and {{Little Red Dots}}: {{Insights}} into {{Broad-line Active Galactic Nuclei}} at z = 5-8 from the {{First JWST COSMOS-3D Dataset}},} The Astrophysical Journal, 996, 93, \dodoi{10.3847/1538-4357/ae1b9b}

\bibitem[{X. {Lin} {et~al.}(2026){Lin}, {Fan}, {Cai}, {Liu}, {Sun}, {Bian}, {Li}, {Mao}, {Greene}, {Liu}, \& et~al.}]{LinEtAl2026b}
{Lin}, X., {Fan}, X., {Cai}, Z., {et~al.} 2026, \bibinfo{title}{{(LRDs)$^2$: The Low-ReDshift Little Red Dots Survey. II. DESI DR1 Sample},} arXiv e-prints, arXiv:2605.21574, \dodoi{10.48550/arXiv.2605.21574}

\bibitem[{H. Liu {et~al.}(2025)Liu, Jiang, Quataert, Greene, \& Ma}]{LiuEtAl2025b}
Liu, H., Jiang, Y.-F., Quataert, E., Greene, J.~E., \& Ma, Y. 2025, \bibinfo{title}{The {{Balmer Break}} and {{Optical Continuum}} of {{Little Red Dots}} from {{Super-Eddington Accretion}},} The Astrophysical Journal, 994, 113, \dodoi{10.3847/1538-4357/ae0c19}

\bibitem[{J.-R. {Liu} {et~al.}(2026){Liu}, {Feng}, \& {Ho}}]{Liu_etal_2026}
{Liu}, J.-R., {Feng}, H., \& {Ho}, L.~C. 2026, \bibinfo{title}{{Optically Thick Outflow Driven by Supercritical Accretion May Explain Little Red Dots},} arXiv e-prints, arXiv:2607.17448.
\newblock \doarXiv{2607.17448}

\bibitem[{Z. {Liu} {et~al.}(2026){Liu}, {Naidu}, {Secunda}, {Greene}, {Matthee}, {Chisholm}, {de Graaff}, {Robbins}, {Antwi-Danso}, {Brammer}, {Sun}, {Eilers}, {Fujimoto}, {Furtak}, {Kara}, {Kokorev}, {Marchesini}, {Oesch}, {Pierel}, {Shen}, {Simcoe}, {Torralba}, \& {Vogelsberger}}]{LiuEtAl2026}
{Liu}, Z., {Naidu}, R.~P., {Secunda}, A., {et~al.} 2026, \bibinfo{title}{{How I Wonder What You Are -- JWST's Little Red Dots do not TWINKLE},} arXiv e-prints, arXiv:2604.13000, \dodoi{10.48550/arXiv.2604.13000}

\bibitem[{K.-X. {Lu} {et~al.}(2019){Lu}, {Huang}, {Zhang}, {Wang}, {Du}, {Hu}, {Xiao}, {Li}, {Bai}, {Bian}, {Yuan}, {Ho}, {Wang}, \& {SEAMBH Collaboration}}]{LuEtAl2019}
{Lu}, K.-X., {Huang}, Y.-K., {Zhang}, Z.-X., {et~al.} 2019, \bibinfo{title}{{Supermassive Black Holes with High Accretion Rates in Active Galactic Nuclei. X. Optical Variability Characteristics},} \apj, 877, 23, \dodoi{10.3847/1538-4357/ab16e8}

\bibitem[{Y. {Ma} {et~al.}(2025){Ma}, {Greene}, {Setton}, {Volonteri}, {Leja}, {Wang}, {Bezanson}, {Brammer}, {Cutler}, {Dayal}, {van Dokkum}, {Furtak}, {Glazebrook}, {Goulding}, {de Graaff}, {Kokorev}, {Labbe}, {Pan}, {Price}, {Weaver}, {Williams}, {Whitaker}, \& {Zitrin}}]{MaEtAl2025}
{Ma}, Y., {Greene}, J.~E., {Setton}, D.~J., {et~al.} 2025, \bibinfo{title}{{UNCOVER: 404 Error{\textemdash}Models Not Found for the Triply Imaged Little Red Dot A2744-QSO1},} \apj, 981, 191, \dodoi{10.3847/1538-4357/ada613}

\bibitem[{C.~L. {MacLeod} {et~al.}(2010){MacLeod}, {Ivezi{\'c}}, {Kochanek}, {Koz{\l}owski}, {Kelly}, {Bullock}, {Kimball}, {Sesar}, {Westman}, {Brooks}, {Gibson}, {Becker}, \& {de Vries}}]{MacLeodEtAl2010}
{MacLeod}, C.~L., {Ivezi{\'c}}, {\v{Z}}., {Kochanek}, C.~S., {et~al.} 2010, \bibinfo{title}{{Modeling the Time Variability of SDSS Stripe 82 Quasars as a Damped Random Walk},} \apj, 721, 1014, \dodoi{10.1088/0004-637X/721/2/1014}

\bibitem[{C.~L. {MacLeod} {et~al.}(2012){MacLeod}, {Ivezi{\'c}}, {Sesar}, {de Vries}, {Kochanek}, {Kelly}, {Becker}, {Lupton}, {Hall}, {Richards}, {Anderson}, \& {Schneider}}]{MacLeodEtAl2012}
{MacLeod}, C.~L., {Ivezi{\'c}}, {\v{Z}}., {Sesar}, B., {et~al.} 2012, \bibinfo{title}{{A Description of Quasar Variability Measured Using Repeated SDSS and POSS Imaging},} \apj, 753, 106, \dodoi{10.1088/0004-637X/753/2/106}

\bibitem[{P. Madau \& F. Haardt(2024)Madau \& Haardt}]{MadauHaardt2024}
Madau, P., \& Haardt, F. 2024, \bibinfo{title}{X-{{Ray Weak Active Galactic Nuclei}} from {{Super-Eddington Accretion}} onto {{Infant Black Holes}},} The Astrophysical Journal Letters, 976, L24, \dodoi{10.3847/2041-8213/ad90e1}

\bibitem[{P. {Madau} \& R. {Maiolino}(2026){Madau} \& {Maiolino}}]{MadauMaiolino2026}
{Madau}, P., \& {Maiolino}, R. 2026, \bibinfo{title}{{Little Red Dots as Obscured Little Blue Dots: A Super-Eddington Unification Model},} arXiv e-prints, arXiv:2602.22386, \dodoi{10.48550/arXiv.2602.22386}

\bibitem[{P. {Madau} {et~al.}(2026){Madau}, {Maiolino}, {Scholtz}, \& {D'Eugenio}}]{MadauEtAl2026}
{Madau}, P., {Maiolino}, R., {Scholtz}, J., \& {D'Eugenio}, F. 2026, \bibinfo{title}{{Wings of little dots: Exponential broad lines from a stratified BLR},} arXiv e-prints, arXiv:2604.04216, \dodoi{10.48550/arXiv.2604.04216}

\bibitem[{R. Maiolino {et~al.}(2024)Maiolino, Scholtz, {Curtis-Lake}, Carniani, Baker, {de Graaff}, Tacchella, {\"U}bler, D'Eugenio, Witstok, Curti, Arribas, Bunker, Charlot, Chevallard, Eisenstein, Egami, Ji, Jones, Lyu, Rawle, Robertson, Rujopakarn, Perna, Sun, Venturi, Williams, \& Willott}]{MaiolinoEtAl2024}
Maiolino, R., Scholtz, J., {Curtis-Lake}, E., {et~al.} 2024, \bibinfo{title}{{{JADES}}: {{The}} Diverse Population of Infant Black Holes at 4 {$<$} z {$<$} 11: {{Merging}}, Tiny, Poor, but Mighty,} Astronomy and Astrophysics, 691, A145, \dodoi{10.1051/0004-6361/202347640}

\bibitem[{R. Maiolino {et~al.}(2025)Maiolino, Risaliti, Signorini, Trefoloni, Juod{\v z}balis, Scholtz, {\"U}bler, D'Eugenio, Carniani, Fabian, Ji, Mazzolari, Bertola, Brusa, Bunker, Charlot, Comastri, Cresci, DeCoursey, Egami, Fiore, Gilli, Perna, Tacchella, \& Venturi}]{MaiolinoEtAl2025}
Maiolino, R., Risaliti, G., Signorini, M., {et~al.} 2025, \bibinfo{title}{{{JWST}} Meets {{Chandra}}: A Large Population of {{Compton}} Thick, Feedback-Free, and Intrinsically {{X-ray}} Weak {{AGN}}, with a Sprinkle of {{SNe}},} Monthly Notices of the Royal Astronomical Society, 538, 1921, \dodoi{10.1093/mnras/staf359}

\bibitem[{J. Matthee {et~al.}(2024)Matthee, Naidu, Brammer, Chisholm, Eilers, Goulding, Greene, Kashino, Labbe, Lilly, Mackenzie, Oesch, Weibel, Wuyts, Xiao, Bordoloi, Bouwens, {van Dokkum}, Illingworth, Kramarenko, Maseda, Mason, Meyer, Nelson, Reddy, Shivaei, Simcoe, \& Yue}]{MattheeEtAl2024}
Matthee, J., Naidu, R.~P., Brammer, G., {et~al.} 2024, \bibinfo{title}{Little {{Red Dots}}: {{An Abundant Population}} of {{Faint Active Galactic Nuclei}} at z {$\sim$} 5 {{Revealed}} by the {{EIGER}} and {{FRESCO JWST Surveys}},} The Astrophysical Journal, 963, 129, \dodoi{10.3847/1538-4357/ad2345}

\bibitem[{J. {Matthee} {et~al.}(2026){Matthee}, {Torralba}, {Pezzulli}, {Naidu}, {Chisholm}, {Mascia}, {Greene}, {Ishikawa}, {Gronke}, {Wuyts}, {Bordoloi}, {Brammer}, {Chang}, {Eilers}, {de Graaff}, {Hviding}, {Iani}, {Illingworth}, {Kashino}, {Labbe}, {Ma}, {Maseda}, {Meyer}, {Nelson}, {Oesch}, \& {Xiao}}]{MattheeEtAl2026}
{Matthee}, J., {Torralba}, A., {Pezzulli}, G., {et~al.} 2026, \bibinfo{title}{{The Engine and its Flows: Little Red Dot spectra are shaped by the column densities of their gas envelopes},} arXiv e-prints, arXiv:2603.17667, \dodoi{10.48550/arXiv.2603.17667}

\bibitem[{R.~M. {M{\'e}rida} {et~al.}(2026){M{\'e}rida}, {Sawicki}, {Willott}, {Gaspar}, \& {Iyer}}]{MeridaEtAl2026}
{M{\'e}rida}, R.~M., {Sawicki}, M., {Willott}, C.~J., {Gaspar}, G., \& {Iyer}, K.~G. 2026, \bibinfo{title}{{Between Degeneracy and Evolution: UV-to-optical Insights into the BH$^*$ Model in Little Red Dots},} arXiv e-prints, arXiv:2606.12355, \dodoi{10.48550/arXiv.2606.12355}

\bibitem[{R.~P. Naidu {et~al.}(2025)Naidu, Matthee, Katz, {de Graaff}, Oesch, Smith, Greene, Brammer, Weibel, Hviding, Chisholm, Labb{\textbackslash}'e, Simcoe, Witten, Atek, Baggen, Belli, Bezanson, Boogaard, Bose, {Covelo-Paz}, Dayal, Fudamoto, Furtak, Giovinazzo, Goulding, Gronke, Heintz, Hirschmann, Illingworth, Inoue, Johnson, Leja, Leonova, McConachie, Maseda, Natarajan, Nelson, Setton, Shivaei, Sobral, Stefanon, Tacchella, Toft, Torralba, {van Dokkum}, {van der Wel}, Volonteri, Walter, Wang, \& Watson}]{NaiduEtAl2025}
Naidu, R.~P., Matthee, J., Katz, H., {et~al.} 2025, A "{{Black Hole Star}}" {{Reveals}} the {{Remarkable Gas-Enshrouded Hearts}} of the {{Little Red Dots}}, arXiv, \dodoi{10.48550/arXiv.2503.16596}

\bibitem[{R.~P. {Naidu} {et~al.}(2026){Naidu}, {Matthee}, {de Graaff}, {Torralba}, {Ashall}, {Katz}, {Chisholm}, {Brammer}, {Dessart}, {Eilers}, \& et~al.}]{NaiduEtAl2026}
{Naidu}, R.~P., {Matthee}, J., {de Graaff}, A., {et~al.} 2026, \bibinfo{title}{{Little Red Dots as Intermediate Mass, Super-Eddington Engines: Insights from Type IIn Supernovae and The 1837-1856 Great Eruption of $η$ Carinae},} arXiv e-prints, arXiv:2606.30711, \dodoi{10.48550/arXiv.2606.30711}

\bibitem[{H. {Netzer}(2019){Netzer}}]{Netzer2019}
{Netzer}, H. 2019, \bibinfo{title}{{Bolometric correction factors for active galactic nuclei},} \mnras, 488, 5185, \dodoi{10.1093/mnras/stz2016}

\bibitem[{M. Newville {et~al.}(2025)Newville, Otten, Nelson, Stensitzki, Ingargiola, Allan, Fox, Carter, \& Rawlik}]{NewvilleEtAl2025}
Newville, M., Otten, R., Nelson, A., {et~al.} 2025, {{LMFIT}}: {{Non-Linear Least-Squares Minimization}} and {{Curve-Fitting}} for {{Python}}, \dodoi{10.5281/zenodo.16175987}

\bibitem[{G.~P. Nikopoulos {et~al.}(2025)Nikopoulos, Watson, Sneppen, Rusakov, Heintz, Witstok, \& Brammer}]{NikopoulosEtAl2025}
Nikopoulos, G.~P., Watson, D., Sneppen, A., {et~al.} 2025, Evidence of Violation of {{Case B}} Recombination in {{Little Red Dots}}, arXiv, \dodoi{10.48550/arXiv.2510.06362}

\bibitem[{F. Pacucci {et~al.}(2026)Pacucci, Ferrara, \& Kocevski}]{PacucciEtAl2026}
Pacucci, F., Ferrara, A., \& Kocevski, D.~D. 2026, The {{Little Red Dots Are Direct Collapse Black Holes}}, arXiv, \dodoi{10.48550/arXiv.2601.14368}

\bibitem[{F. Pacucci \& R. Narayan(2024)Pacucci \& Narayan}]{PacucciNarayan2024}
Pacucci, F., \& Narayan, R. 2024, \bibinfo{title}{Mildly {{Super-Eddington Accretion}} onto {{Slowly Spinning Black Holes Explains}} the {{X-Ray Weakness}} of the {{Little Red Dots}},} The Astrophysical Journal, 976, 96, \dodoi{10.3847/1538-4357/ad84f7}

\bibitem[{Z. {Pan} {et~al.}(2026){Pan}, {Zhuang}, {Shen}, {Wang}, {Greene}, {Burgasser}, {Li}, {Stone}, \& {Venkatraman}}]{PanEtAl2026}
{Pan}, Z., {Zhuang}, M.-Y., {Shen}, Y., {et~al.} 2026, \bibinfo{title}{{NEXUS: Abundance, Environments, and Spectral Diversity of Little Red Dots from the NIRSpec MSA Sample},} arXiv e-prints, arXiv:2606.09721.
\newblock \doarXiv{2606.09721}

\bibitem[{Y. {Pang} {et~al.}(2026){Pang}, {Wang}, {Cheng}, {Wang}, {Zhou}, {Zhou}, {Wu}, \& {Glazebrook}}]{PangEtAl2026}
{Pang}, Y., {Wang}, X., {Cheng}, C., {et~al.} 2026, \bibinfo{title}{{The Structure and Evolution of LRDs: Insights from JWST NIRSpec Medium- and High-resolution Spectroscopy at z {\ensuremath{\sim}} 4},} \apj, 1004, 185, \dodoi{10.3847/1538-4357/ae6e3f}

\bibitem[{K. {Park} {et~al.}(2026){Park}, {Torralba}, {Matthee}, {Mascia}, {Haiman}, {Naidu}, \& {de Graaff}}]{ParkEtAl2026}
{Park}, K., {Torralba}, A., {Matthee}, J., {et~al.} 2026, \bibinfo{title}{{A new sample of Little Red Dots at $z<0.45$ in DESI DR1: Broad Balmer lines, low ionization spectrum and no variability},} arXiv e-prints, arXiv:2605.14233, \dodoi{10.48550/arXiv.2605.14233}

\bibitem[{B.~M. Peterson(2001)Peterson}]{Peterson2001a}
Peterson, B.~M. 2001, Variability of {{Active Galactic Nuclei}} (eprint: arXiv:astro-ph/0109495), 3, \dodoi{10.1142/9789812811318_0002}

\bibitem[{S. {Rakshit} \& C.~S. {Stalin}(2017){Rakshit} \& {Stalin}}]{RakshitStalin2017}
{Rakshit}, S., \& {Stalin}, C.~S. 2017, \bibinfo{title}{{Optical Variability of Narrow-line and Broad-line Seyfert 1 Galaxies},} \apj, 842, 96, \dodoi{10.3847/1538-4357/aa72f4}

\bibitem[{P. Rinaldi {et~al.}(2025)Rinaldi, Bonaventura, Rieke, Alberts, Caputi, Baker, Baum, Bhatawdekar, Bunker, Carniani, {Curtis-Lake}, D'Eugenio, Egami, Ji, Johnson, Hainline, Helton, Lin, Lyu, Ma, Maiolino, {P{\'e}rez-Gonz{\'a}lez}, Rieke, Robertson, Shivaei, Stone, Sun, Tacchella, {\"U}bler, Williams, Willmer, Willott, Zhang, \& Zhu}]{RinaldiEtAl2025}
Rinaldi, P., Bonaventura, N., Rieke, G.~H., {et~al.} 2025, \bibinfo{title}{Not {{Just}} a {{Dot}}: {{The Complex UV Morphology}} and {{Underlying Properties}} of {{Little Red Dots}},} The Astrophysical Journal, 992, 71, \dodoi{10.3847/1538-4357/adfa10}

\bibitem[{A. {Sadowski} {et~al.}(2015){Sadowski}, {Narayan}, {Tchekhovskoy}, {Abarca}, {Zhu}, \& {McKinney}}]{SkadowskiEtAl2015}
{Sadowski}, A., {Narayan}, R., {Tchekhovskoy}, A., {et~al.} 2015, \bibinfo{title}{{Global simulations of axisymmetric radiative black hole accretion discs in general relativity with a mean-field magnetic dynamo},} \mnras, 447, 49, \dodoi{10.1093/mnras/stu2387}

\bibitem[{P. {S{\'a}nchez-S{\'a}ez} {et~al.}(2018){S{\'a}nchez-S{\'a}ez}, {Lira}, {Mej{\'\i}a-Restrepo}, {Ho}, {Ar{\'e}valo}, {Kim}, {Cartier}, \& {Coppi}}]{SanchezSaezEtAl2018}
{S{\'a}nchez-S{\'a}ez}, P., {Lira}, P., {Mej{\'\i}a-Restrepo}, J., {et~al.} 2018, \bibinfo{title}{{The QUEST-La Silla AGN Variability Survey: Connection between AGN Variability and Black Hole Physical Properties},} \apj, 864, 87, \dodoi{10.3847/1538-4357/aad7f9}

\bibitem[{J. {Scholtz} {et~al.}(2026){Scholtz}, {D'Eugenio}, {Maiolino}, {Brazzini}, {{\"U}bler}, {Ji}, {Perna}, {Sun}, {Brocchi}, {Carniani}, {Cresci}, {Ivey}, {Juod{\v{z}}balis}, {Marconi}, {Mazzolari}, {Risaliti}, \& {Trefoloni}}]{ScholtzEtAl2026}
{Scholtz}, J., {D'Eugenio}, F., {Maiolino}, R., {et~al.} 2026, \bibinfo{title}{{Little Red and Blue Dots: simply stratified Broad Line Regions},} arXiv e-prints, arXiv:2603.22277, \dodoi{10.48550/arXiv.2603.22277}

\bibitem[{A. Secunda {et~al.}(2025)Secunda, Somerville, Jiang, Greene, Furtak, \& Zitrin}]{SecundaEtAl2025b}
Secunda, A., Somerville, R.~S., Jiang, Y.-F., {et~al.} 2025, \bibinfo{title}{Do {{Little Red Dots Vary}}?} The Astrophysical Journal, 996, 6, \dodoi{10.3847/1538-4357/ae1f08}

\bibitem[{D.~J. Setton {et~al.}(2025)Setton, Greene, Spilker, Williams, Labb{\'e}, Ma, Wang, Whitaker, Leja, {de Graaff}, Alberts, Bezanson, Boogaard, Brammer, Cutler, Cleri, Cooper, Dayal, Fujimoto, Furtak, Goulding, Hirschmann, Kokorev, Maseda, McConachie, Matthee, Miller, Naidu, Oesch, Pan, Price, Suess, Weaver, Xiao, Zhang, \& Zitrin}]{SettonEtAl2025}
Setton, D.~J., Greene, J.~E., Spilker, J.~S., {et~al.} 2025, \bibinfo{title}{A {{Confirmed Deficit}} of {{Hot}} and {{Cold Dust Emission}} in the {{Most Luminous Little Red Dots}},} The Astrophysical Journal Letters, 991, L10, \dodoi{10.3847/2041-8213/ade78b}

\bibitem[{A.~J. {Shajib} {et~al.}(2025){Shajib}, {Treu}, {Melo}, {Roberts-Borsani}, {Knabel}, {Cappellari}, \& {Frieman}}]{ShajibEtAl2025}
{Shajib}, A.~J., {Treu}, T., {Melo}, A., {et~al.} 2025, \bibinfo{title}{{An accurate measurement of the spectral resolution of the JWST Near Infrared Spectrograph},} \aap, 702, L12, \dodoi{10.1051/0004-6361/202556281}

\bibitem[{Y. Shen {et~al.}(2015)Shen, Brandt, Dawson, Hall, McGreer, Anderson, Chen, Denney, Eftekharzadeh, Fan, Gao, Green, Greene, Ho, Horne, Jiang, Kelly, Kinemuchi, Kochanek, P{\^a}ris, Peters, Peterson, Petitjean, Ponder, Richards, Schneider, Seth, Smith, Strauss, Tao, Trump, {Wood-Vasey}, Zu, Eisenstein, Pan, Bizyaev, Malanushenko, Malanushenko, \& Oravetz}]{ShenEtAl2015}
Shen, Y., Brandt, W.~N., Dawson, K.~S., {et~al.} 2015, \bibinfo{title}{The {{Sloan Digital Sky Survey Reverberation Mapping Project}}: {{Technical Overview}},} The Astrophysical Journal Supplement Series, 216, 4, \dodoi{10.1088/0067-0049/216/1/4}

\bibitem[{Y. Shen {et~al.}(2019)Shen, Hall, Horne, Zhu, McGreer, Simm, Trump, Kinemuchi, Brandt, Green, Grier, Guo, Ho, Homayouni, Jiang, {I-Hsiu Li}, Morganson, Petitjean, Richards, Schneider, Starkey, Wang, Chambers, Kaiser, Kudritzki, Magnier, \& Waters}]{ShenEtAl2019}
Shen, Y., Hall, P.~B., Horne, K., {et~al.} 2019, \bibinfo{title}{The {{Sloan Digital Sky Survey Reverberation Mapping Project}}: {{Sample Characterization}},} The Astrophysical Journal Supplement Series, 241, 34, \dodoi{10.3847/1538-4365/ab074f}

\bibitem[{Y. {Shen} {et~al.}(2024){Shen}, {Grier}, {Horne}, {Stone}, {Li}, {Yang}, {Homayouni}, {Trump}, {Anderson}, {Brandt}, {Hall}, {Ho}, {Jiang}, {Petitjean}, {Schneider}, {Tao}, {Donnan}, {AlSayyad}, {Bershady}, {Blanton}, {Bizyaev}, {Bundy}, {Chen}, {Davis}, {Dawson}, {Fan}, {Greene}, {Gr{\"o}ller}, {Guo}, {Ibarra-Medel}, {Jiang}, {Keenan}, {Kollmeier}, {Lejoly}, {Li}, {de la Macorra}, {Moe}, {Nie}, {Rossi}, {Smith}, {Tee}, {Weijmans}, {Xu}, {Yue}, {Zhou}, {Zhou}, \& {Zou}}]{ShenEtAl2024_rm}
{Shen}, Y., {Grier}, C.~J., {Horne}, K., {et~al.} 2024, \bibinfo{title}{{The Sloan Digital Sky Survey Reverberation Mapping Project: Key Results},} \apjs, 272, 26, \dodoi{10.3847/1538-4365/ad3936}

\bibitem[{Y. Shen {et~al.}(2024)Shen, Zhuang, Li, Burgasser, Fan, Greene, Narayan, Shapley, Sun, Wang, \& Yang}]{ShenEtAl2024a}
Shen, Y., Zhuang, M.-Y., Li, J., {et~al.} 2024, {{NEXUS}}: The {{North}} Ecliptic Pole {{EXtragalactic Unified Survey}}, arXiv, \dodoi{10.48550/arXiv.2408.12713}

\bibitem[{T. {Simm} {et~al.}(2016){Simm}, {Salvato}, {Saglia}, {Ponti}, {Lanzuisi}, {Trakhtenbrot}, {Nandra}, \& {Bender}}]{SimmEtAl2016}
{Simm}, T., {Salvato}, M., {Saglia}, R., {et~al.} 2016, \bibinfo{title}{{Pan-STARRS1 variability of XMM-COSMOS AGN. II. Physical correlations and power spectrum analysis},} \aap, 585, A129, \dodoi{10.1051/0004-6361/201527353}

\bibitem[{N. {Smith}(2013){Smith}}]{Smith2013}
{Smith}, N. 2013, \bibinfo{title}{{A model for the 19th century eruption of Eta Carinae: CSM interaction like a scaled-down Type IIn Supernova},} \mnras, 429, 2366, \dodoi{10.1093/mnras/sts508}

\bibitem[{N. {Smith}(2026){Smith}}]{Smith2026}
{Smith}, N. 2026, \bibinfo{title}{{Luminous blue variables},} in Encyclopedia of Astrophysics, Volume 2, Vol.~2, 508--532, \dodoi{10.1016/B978-0-443-21439-4.00147-4}

\bibitem[{A. {Sneppen} {et~al.}(2026){Sneppen}, {Watson}, {Matthews}, \& {Nikopoulos}}]{SneppenEtAl2026c}
{Sneppen}, A., {Watson}, D., {Matthews}, J.~H., \& {Nikopoulos}, G. 2026, \bibinfo{title}{{TBD LBD: The nature of `little blue dots'},} arXiv e-prints, arXiv:2606.12509.
\newblock \doarXiv{2606.12509}

\bibitem[{A. Sneppen {et~al.}(2026)Sneppen, Watson, Matthews, Nikopoulos, Allen, Brammer, Damgaard, Heintz, Knigge, Long, Rusakov, Sim, \& Witstok}]{SneppenEtAl2026}
Sneppen, A., Watson, D., Matthews, J.~H., {et~al.} 2026, Inside the Cocoon: A Comprehensive Explanation of the Spectra of {{Little Red Dots}}, arXiv

\bibitem[{M. {Stepney} {et~al.}(2026){Stepney}, {Banerji}, {Tang}, {Temple}, \& {Hewett}}]{StepneyEtAl2026}
{Stepney}, M., {Banerji}, M., {Tang}, S., {Temple}, M.~J., \& {Hewett}, P.~C. 2026, \bibinfo{title}{{The rest-ultraviolet to infrared spectral energy distributions of heavily reddened quasars are 'V-shaped' and hot-dust poor},} \mnras, 546, stag191, \dodoi{10.1093/mnras/stag191}

\bibitem[{J. {Stern} \& A. {Laor}(2012){Stern} \& {Laor}}]{SternLaor2012}
{Stern}, J., \& {Laor}, A. 2012, \bibinfo{title}{{Type 1 AGN at low z - II. The relative strength of narrow lines and the nature of intermediate type AGN},} \mnras, 426, 2703, \dodoi{10.1111/j.1365-2966.2012.21772.x}

\bibitem[{Z. Stone {et~al.}(2022)Stone, Shen, Burke, Chen, Yang, Liu, Gruendl, Adam{\'o}w, {Andrade-Oliveira}, Annis, Bacon, Bertin, Bocquet, Brooks, Burke, Carnero~Rosell, Carrasco~Kind, Carretero, {da Costa}, Pereira, De~Vicente, Desai, Diehl, Doel, Ferrero, Friedel, Frieman, {Garc{\'i}a-Bellido}, Gaztanaga, Gruen, Gutierrez, Hinton, Hollowood, Honscheid, James, Kuehn, Kuropatkin, Lidman, Maia, Menanteau, Miquel, Morgan, {Paz-Chinch{\'o}n}, Pieres, Plazas~Malag{\'o}n, {Rodriguez-Monroy}, Sanchez, Scarpine, Serrano, {Sevilla-Noarbe}, Smith, Suchyta, Swanson, Tarl{\'e}, To, \& {DES Collaboration}}]{StoneEtAl2022b}
Stone, Z., Shen, Y., Burke, C.~J., {et~al.} 2022, \bibinfo{title}{Optical Variability of Quasars with 20-Yr Photometric Light Curves,} Monthly Notices of the Royal Astronomical Society, 514, 164, \dodoi{10.1093/mnras/stac1259}

\bibitem[{Z. Stone {et~al.}(2025)Stone, Shen, Zhuang, Hu, Pierel, Li, Burgasser, Greene, Pan, Shapley, Sun, Venkatraman, \& Wang}]{StoneEtAl2025a}
Stone, Z., Shen, Y., Zhuang, M.-Y., {et~al.} 2025, {{NEXUS}}: {{A Search}} for {{Nuclear Variability}} with the {{First Two JWST NIRCam Epochs}}, arXiv, \dodoi{10.48550/arXiv.2509.19585}

\bibitem[{Z. {Stone} {et~al.}(2025){Stone}, {Shen}, {Anderson}, {Bauer}, {Brandt}, {Chakraborty}, {Davis}, {Fries}, {Grier}, {Hall}, {Horne}, {Koekemoer}, {Mart{\'\i}nez-Aldama}, {Long}, {Morrison}, {Ricci}, {Schneider}, {Temple}, \& {Trump}}]{StoneEtAl2025b}
{Stone}, Z., {Shen}, Y., {Anderson}, S.~F., {et~al.} 2025, \bibinfo{title}{{The SDSS-V Black Hole Mapper Reverberation Mapping Project: Multiline Dynamical Modeling of a Highly Variable Active Galactic Nucleus with Decade-long Light Curves},} \apj, 991, 218, \dodoi{10.3847/1538-4357/adfd4c}

\bibitem[{A.~J. Taylor {et~al.}(2025{\natexlab{a}})Taylor, Kokorev, Kocevski, Akins, Cullen, Dickinson, Finkelstein, Arrabal~Haro, Bromm, Giavalisco, Inayoshi, Juneau, Leung, {P{\'e}rez-Gonz{\'a}lez}, Somerville, Trump, Amor{\'i}n, Barro, Burgarella, Brooks, Carnall, Casey, Cheng, Chisholm, Chworowsky, Davis, Donnan, Dunlop, Ellis, Fern{\'a}ndez, Fujimoto, Grogin, Gupta, Hathi, Jung, Hirschmann, Kartaltepe, Koekemoer, Larson, Leung, Llerena, Lucas, McLeod, McLure, Napolitano, Papovich, Stanton, Tripodi, Wang, Wilkins, Yung, \& Zavala}]{TaylorEtAl2025}
Taylor, A.~J., Kokorev, V., Kocevski, D.~D., {et~al.} 2025{\natexlab{a}}, \bibinfo{title}{{{CAPERS-LRD-z9}}: {{A Gas-enshrouded Little Red Dot Hosting}} a {{Broad-line Active Galactic Nucleus}} at z = 9.288,} The Astrophysical Journal Letters, 989, L7, \dodoi{10.3847/2041-8213/ade789}

\bibitem[{A.~J. Taylor {et~al.}(2025{\natexlab{b}})Taylor, Finkelstein, Kocevski, Jeon, Bromm, Amor{\'i}n, Arrabal~Haro, Backhaus, Bagley, Banados, Bhatawdekar, Brooks, Calabr{\`o}, Ch{\'a}vez~Ortiz, Cheng, Cleri, Cole, Davis, Dickinson, Donnan, Dunlop, Ellis, Fern{\'a}ndez, Fontana, Fujimoto, Giavalisco, Grazian, Guo, Hathi, Holwerda, Hirschmann, Inayoshi, Kartaltepe, Khusanova, Koekemoer, Kokorev, Larson, Leung, Lucas, McLeod, Napolitano, Onoue, Pacucci, Papovich, {P{\'e}rez-Gonz{\'a}lez}, Pirzkal, Somerville, Trump, Wilkins, Yung, \& Zhang}]{TaylorEtAl2025a}
Taylor, A.~J., Finkelstein, S.~L., Kocevski, D.~D., {et~al.} 2025{\natexlab{b}}, \bibinfo{title}{Broad-Line {{AGNs}} at 3.5 {$<$} z {$<$} 6: {{The Black Hole Mass Function}} and a {{Connection}} with {{Little Red Dots}},} The Astrophysical Journal, 986, 165, \dodoi{10.3847/1538-4357/add15b}

\bibitem[{W.~L. Tee {et~al.}(2025)Tee, Fan, Wang, \& Yang}]{TeeEtAl2025}
Tee, W.~L., Fan, X., Wang, F., \& Yang, J. 2025, \bibinfo{title}{Lack of {{Rest-frame Ultraviolet Variability}} in {{Little Red Dots Based}} on {{HST}} and {{JWST Observations}},} The Astrophysical Journal Letters, 983, L26, \dodoi{10.3847/2041-8213/adc5e3}

\bibitem[{A. Torralba {et~al.}(2025)Torralba, Matthee, Pezzulli, Naidu, Ishikawa, Brammer, Chang, Chisholm, {de Graaff}, D'Eugenio, Di~Cesare, Eilers, Greene, Gronke, Iani, Kokorev, Kotiwale, Kramarenko, Ma, Mascia, Navarrete, Nelson, Oesch, Simcoe, \& Wuyts}]{TorralbaEtAl2025}
Torralba, A., Matthee, J., Pezzulli, G., {et~al.} 2025, The Warm Outer Layer of a {{Little Red Dot}} as the Source of [{{Fe II}}] and Collisional {{Balmer}} Lines with Scattering Wings, arXiv, \dodoi{10.48550/arXiv.2510.00103}

\bibitem[{A. Trinca {et~al.}(2024)Trinca, Valiante, Schneider, Juod{\v z}balis, Maiolino, Graziani, Lupi, Natarajan, Volonteri, \& Zana}]{TrincaEtAl2024}
Trinca, A., Valiante, R., Schneider, R., {et~al.} 2024, Episodic Super-{{Eddington}} Accretion as a Clue to {{Overmassive Black Holes}} in the Early {{Universe}}, arXiv, \dodoi{10.48550/arXiv.2412.14248}

\bibitem[{H. {\"U}bler {et~al.}(2023){\"U}bler, Maiolino, {Curtis-Lake}, {P{\'e}rez-Gonz{\'a}lez}, Curti, Perna, Arribas, Charlot, Marshall, D'Eugenio, Scholtz, Bunker, Carniani, Ferruit, Jakobsen, Rix, Rodr{\'i}guez Del~Pino, Willott, Boeker, Cresci, Jones, Kumari, \& Rawle}]{UblerEtAl2023}
{\"U}bler, H., Maiolino, R., {Curtis-Lake}, E., {et~al.} 2023, \bibinfo{title}{{{GA-NIFS}}: {{A}} Massive Black Hole in a Low-Metallicity {{AGN}} at z {$\sim$} 5.55 Revealed by {{JWST}}/{{NIRSpec IFS}},} Astronomy and Astrophysics, 677, A145, \dodoi{10.1051/0004-6361/202346137}

\bibitem[{P. Virtanen {et~al.}(2020)Virtanen, Gommers, Oliphant, Haberland, Reddy, Cournapeau, Burovski, Peterson, Weckesser, Bright, {van der Walt}, Brett, Wilson, Millman, Mayorov, Nelson, Jones, Kern, Larson, Carey, Polat, Feng, Moore, {VanderPlas}, Laxalde, Perktold, Cimrman, Henriksen, Quintero, Harris, Archibald, Ribeiro, Pedregosa, {van Mulbregt}, \& {SciPy 1.0 Contributors}}]{scipy}
Virtanen, P., Gommers, R., Oliphant, T.~E., {et~al.} 2020, \bibinfo{title}{{{SciPy} 1.0: Fundamental Algorithms for Scientific Computing in Python},} Nature Methods, 17, 261, \dodoi{10.1038/s41592-019-0686-2}

\bibitem[{B. {Wang} {et~al.}(2024){Wang}, {Leja}, {de Graaff}, {Brammer}, {Weibel}, {van Dokkum}, {Baggen}, {Suess}, {Greene}, {Bezanson}, {Cleri}, {Hirschmann}, {Labb{\'e}}, {Matthee}, {McConachie}, {Naidu}, {Nelson}, {Oesch}, {Setton}, \& {Williams}}]{WangEtAl2024b}
{Wang}, B., {Leja}, J., {de Graaff}, A., {et~al.} 2024, \bibinfo{title}{{RUBIES: Evolved Stellar Populations with Extended Formation Histories at z {\ensuremath{\sim}} 7─8 in Candidate Massive Galaxies Identified with JWST/NIRSpec},} \apjl, 969, L13, \dodoi{10.3847/2041-8213/ad55f7}

\bibitem[{B. Wang {et~al.}(2025)Wang, {de Graaff}, Davies, Greene, Leja, Brammer, Goulding, Miller, Suess, Weibel, Williams, Bezanson, Boogaard, Cleri, Hirschmann, Katz, Labb{\'e}, Maseda, Matthee, McConachie, Naidu, Oesch, Rix, Setton, \& Whitaker}]{WangEtAl2025}
Wang, B., {de Graaff}, A., Davies, R.~L., {et~al.} 2025, \bibinfo{title}{{{RUBIES}}: {{JWST}}/{{NIRSpec Confirmation}} of an {{Infrared-luminous}}, {{Broad-line Little Red Dot}} with an {{Ionized Outflow}},} The Astrophysical Journal, 984, 121, \dodoi{10.3847/1538-4357/adc1ca}

\bibitem[{J. {Wang} \& D.~W. {Xu}(2015){Wang} \& {Xu}}]{WangXu2015}
{Wang}, J., \& {Xu}, D.~W. 2015, \bibinfo{title}{{Identifying AGN Balmer absorptions and stratified narrow emission-line region kinematics in SDSS J112611.63+425246.4},} \aap, 573, A15, \dodoi{10.1051/0004-6361/201424848}

\bibitem[{W. Wang {et~al.}(2024)Wang, Wylezalek, Breuck, Vernet, Rupke, Zakamska, Vayner, Lehnert, Nesvadba, \& Stern}]{WangEtAl2024}
Wang, W., Wylezalek, D., Breuck, C.~D., {et~al.} 2024, \bibinfo{title}{{{JWST}} Discovers an {{AGN}} Ionization Cone but Only Weak Radiatively Driven Feedback in a Powerful z {$\approx$} 3.5 Radio-Loud {{AGN}},} Astronomy \& Astrophysics, 683, A169, \dodoi{10.1051/0004-6361/202348531}

\bibitem[{M.~L. Waskom(2021)Waskom}]{Waskom2021}
Waskom, M.~L. 2021, \bibinfo{title}{Seaborn: Statistical Data Visualization,} Journal of Open Source Software, 6, 3021, \dodoi{10.21105/joss.03021}

\bibitem[{C.~C. Williams {et~al.}(2024)Williams, Alberts, Ji, Hainline, Lyu, Rieke, Endsley, Suess, Sun, Johnson, Florian, Shivaei, Rujopakarn, Baker, Bhatawdekar, Boyett, Bunker, Cameron, Carniani, Charlot, {Curtis-Lake}, DeCoursey, {de Graaff}, Egami, Eisenstein, Gibson, Hausen, Helton, Maiolino, Maseda, Nelson, {P{\'e}rez-Gonz{\'a}lez}, Rieke, Robertson, Saxena, Tacchella, Willmer, \& Willott}]{WilliamsEtAl2024}
Williams, C.~C., Alberts, S., Ji, Z., {et~al.} 2024, \bibinfo{title}{The {{Galaxies Missed}} by {{Hubble}} and {{ALMA}}: {{The Contribution}} of {{Extremely Red Galaxies}} to the {{Cosmic Census}} at 3 {$<$} z {$<$} 8,} The Astrophysical Journal, 968, 34, \dodoi{10.3847/1538-4357/ad3f17}

\bibitem[{Q. Wu \& Y. Shen(2022)Wu \& Shen}]{WuShen2022}
Wu, Q., \& Shen, Y. 2022, \bibinfo{title}{A {{Catalog}} of {{Quasar Properties}} from {{Sloan Digital Sky Survey Data Release}} 16,} The Astrophysical Journal Supplement Series, 263, 42, \dodoi{10.3847/1538-4365/ac9ead}

\bibitem[{M. Xiao {et~al.}(2025)Xiao, Oesch, Bing, Elbaz, Matthee, Fudamoto, Fujimoto, {Marques-Chaves}, Williams, {Dessauges-Zavadsky}, Valentino, Brammer, {Covelo-Paz}, Daddi, Fynbo, Gillman, Ginolfi, Giovinazzo, Greene, Gu, Illingworth, Inayoshi, Kokorev, Meyer, Naidu, Reddy, Schaerer, Shapley, Stefanon, Steinhardt, Setton, Vestergaard, \& Wang}]{XiaoEtAl2025}
Xiao, M., Oesch, P.~A., Bing, L., {et~al.} 2025, \bibinfo{title}{No [{{C II}}] or Dust Detection in Two {{Little Red Dots}} at Zspec{$>$} 7,} Astronomy and Astrophysics, 700, A231, \dodoi{10.1051/0004-6361/202554361}

\bibitem[{W. {Yu} {et~al.}(2025){Yu}, {Richards}, {Ruan}, {Vogeley}, {Bauer}, \& {Graham}}]{YuEtAl2025}
{Yu}, W., {Richards}, G.~T., {Ruan}, J.~J., {et~al.} 2025, \bibinfo{title}{{Examining Active Galactic Nucleus UV/Optical Variability beyond the Simple Damped Random Walk. II. Insights from 22 yr Observations of SDSS, PS1, and ZTF},} \apj, 992, 130, \dodoi{10.3847/1538-4357/adfdd2}

\bibitem[{M. Yue {et~al.}(2024)Yue, Eilers, Ananna, Panagiotou, Kara, \& Miyaji}]{YueEtAl2024}
Yue, M., Eilers, A.-C., Ananna, T.~T., {et~al.} 2024, \bibinfo{title}{Stacking {{X-Ray Observations}} of ``{{Little Red Dots}}'': {{Implications}} for {{Their Active Galactic Nucleus Properties}},} The Astrophysical Journal Letters, 974, L26, \dodoi{10.3847/2041-8213/ad7eba}

\bibitem[{Z. {Zhang} {et~al.}(2026){Zhang}, {Inayoshi}, {Oguri}, {Jiang}, {Sun}, {Li}, \& {Lin}}]{ZhangEtAl2026}
{Zhang}, Z., {Inayoshi}, K., {Oguri}, M., {et~al.} 2026, \bibinfo{title}{{Little red dots as a cosmological probe: constraining $H_0$ with quasi-periodic pulsations},} arXiv e-prints, arXiv:2606.05281, \dodoi{10.48550/arXiv.2606.05281}

\bibitem[{Z. Zhang {et~al.}(2025{\natexlab{a}})Zhang, Jiang, Liu, \& Ho}]{ZhangEtAl2025c}
Zhang, Z., Jiang, L., Liu, W., \& Ho, L.~C. 2025{\natexlab{a}}, \bibinfo{title}{Analysis of {{Multi-epoch JWST Images}} of {$\sim$}300 {{Little Red Dots}}: {{Tentative Detection}} of {{Variability}} in a {{Minority}} of {{Sources}},} The Astrophysical Journal, 985, 119, \dodoi{10.3847/1538-4357/adcb3e}

\bibitem[{Z. Zhang {et~al.}(2025{\natexlab{b}})Zhang, Jiang, Liu, Ho, \& Inayoshi}]{ZhangEtAl2025a}
Zhang, Z., Jiang, L., Liu, W., Ho, L.~C., \& Inayoshi, K. 2025{\natexlab{b}}, {{JWST Insights}} into {{Narrow-line Little Red Dots}}, arXiv, \dodoi{10.48550/arXiv.2506.04350}

\bibitem[{Z. Zhang {et~al.}(2025{\natexlab{c}})Zhang, Li, Oguri, Lin, Inayoshi, Cerny, Coe, Diego, Fujimoto, Jiang, Mahler, Matthee, Naidu, Sharon, Shen, Zitrin, {Abdurro'uf}, Akins, Allingham, Amor{\'i}n, Asada, Atek, Bauer, Brada{\v c}, Bradley, Cai, Cantalupo, Conselice, Dai, Dayal, Egami, Eisenstein, Faisst, Fan, Fei, Frye, Fudamoto, Furtak, Golubchik, {Gonz{\'a}lez-Otero}, Harikane, Hsiao, {Jim{\'e}nez-Teja}, Kartaltepe, Kiyota, Koekemoer, Kohno, Kokorev, Kumari, Labbe, Lagos, Larison, Liang, Lucas, Lyu, Martis, Magdis, Messa, Nakane, Noirot, Ortiz, Ouchi, Pierel, Postman, Reddy, Ricotti, Schaerer, Schneider, Steidel, Tee, Tripodi, Trussler, Umeda, Valentino, Vanzella, Wang, Windhorst, Wu, Wu, Yanagisawa, Yang, \& Sun}]{ZhangEtAl2025d}
Zhang, Z., Li, M., Oguri, M., {et~al.} 2025{\natexlab{c}}, Little Red Dot Variability over a Century Reveals Black Hole Envelope via a Giant {{Einstein}} Cross, arXiv, \dodoi{10.48550/arXiv.2512.05180}

\bibitem[{S. Zhou {et~al.}(2025)Zhou, Sun, Zhang, Chen, \& Ho}]{ZhouEtAl2025b}
Zhou, S., Sun, M., Zhang, Z., Chen, J., \& Ho, L.~C. 2025, \bibinfo{title}{On the {{Variability Features}} of {{Active Galactic Nuclei}} in {{Little Red Dots}},} The Astrophysical Journal, 991, 137, \dodoi{10.3847/1538-4357/adfd5f}

\bibitem[{M.-Y. Zhuang {et~al.}(2025)Zhuang, Li, Shen, Lin, Shapley, Wang, Wu, \& Yang}]{ZhuangEtAl2025}
Zhuang, M.-Y., Li, J., Shen, Y., {et~al.} 2025, {{NEXUS}}: {{A Spectroscopic Census}} of {{Broad-line AGNs}} and {{Little Red Dots}} at \$3\textbackslash lesssim Z\textbackslash lesssim 6\$, arXiv, \dodoi{10.48550/arXiv.2505.20393}

\bibitem[{M.-Y. {Zhuang} {et~al.}(2026{\natexlab{a}}){Zhuang}, {Wang}, {Sun}, {Shen}, {Li}, {Burgasser}, {Fan}, {Greene}, {Narayan}, {Shapley}, \& {Yang}}]{nexus_edr}
{Zhuang}, M.-Y., {Wang}, F., {Sun}, F., {et~al.} 2026{\natexlab{a}}, \bibinfo{title}{{NEXUS Early Data Release: NIRCam Imaging and WFSS Spectroscopy from the First (Partial) Wide Epoch},} \apjs, 282, 54, \dodoi{10.3847/1538-4365/ae2d05}

\bibitem[{M.-Y. {Zhuang} {et~al.}(2026{\natexlab{b}}){Zhuang}, {Shen}, {Pan}, {Hu}, {Burgasser}, {Coulter}, {Greene}, {Li}, \& {Wang}}]{nexus-qdr}
{Zhuang}, M.-Y., {Shen}, Y., {Pan}, Z., {et~al.} 2026{\natexlab{b}}, \bibinfo{title}{{NEXUS: Quick Release Notes},} arXiv e-prints, arXiv:2603.04586, \dodoi{10.48550/arXiv.2603.04586}

\end{thebibliography}
\bibliographystyle{aasjournalv7}



\end{document}